\documentclass[numberedappendix,apj,twocolumn]{emulateapj}
\usepackage[colorlinks=true, citecolor=blue, linkcolor=black,breaklinks]{hyperref}
\usepackage{amsmath,amsthm, amsfonts,amssymb} % For AMS Beautification

\newcommand{\lya}{Ly$\alpha$}

\newcommand{\xhi}{$\overline{x}_{\mathrm{HI}}$}
\newcommand\emissionline[2]{#1$\;${\scshape{#2}}}%                       % ion, i.e., CII = \ion{C}{ii}
\newcommand{\oiii}{[\emissionline{O}{iii}]}
\newcommand{\oii}{[\emissionline{O}{ii}]}

\newcommand{\HST}{\emph{HST}}
\newcommand{\spitzer}{\emph{Spitzer}}
\newcommand{\Spitzer}{\emph{Spitzer}}

\newcommand{\civ}{C\,{\sc iv}}
\newcommand{\ciii}{C\,{\sc iii}]}

\newcommand{\ncluster}{12}
\newcommand{\macssampleid}{MACS0744-064}
\newcommand{\rxjsampleid}{RXJ1347-018}

\newcommand{\nhighz}{68}

\newcommand{\xhibest}{$\overline{x}_{\mathrm{HI}}=0.88^{+0.05}_{-0.10}$}
\newcommand{\xhione}{$\overline{x}_{\mathrm{HI}}=0.89^{+0.04}_{-0.10}$}
\newcommand{\xhithree}{$\overline{x}_{\mathrm{HI}}=0.88^{+0.05}_{-0.09}$}
\def\frac#1#2{{\textstyle{{#1}\over {#2}}}}

\def\lsim{\mathrel{\rlap{\lower4pt\hbox{\hskip1pt$\sim$}}
    \raise1pt\hbox{$<$}}}
\def\gsim{\mathrel{\rlap{\lower4pt\hbox{\hskip1pt$\sim$}}
    \raise1pt\hbox{$>$}}}
\def\sqr#1#2{{\vcenter{\vbox{\hrule height.#2pt
         \hbox{\vrule width.#2pt height#1pt \kern#1pt
         \vrule width.#2pt}
         \hrule height.#2pt}}}}

\def\frac#1#2{{\textstyle{{#1}\over {#2}}}}

\def\abs#1{\left|{#1}\right|}

\def\lsim{\mathrel{\rlap{\lower4pt\hbox{\hskip1pt$\sim$}}
    \raise1pt\hbox{$<$}}}
\def\gsim{\mathrel{\rlap{\lower4pt\hbox{\hskip1pt$\sim$}}
    \raise1pt\hbox{$>$}}}
\def\sqr#1#2{{\vcenter{\vbox{\hrule height.#2pt
         \hbox{\vrule width.#2pt height#1pt \kern#1pt
         \vrule width.#2pt}
         \hrule height.#2pt}}}}

\newcommand{\bea}{\begin{eqnarray}}
\newcommand{\eea}{\end{eqnarray}}
\newcommand{\bit}{\begin{itemize}}
\newcommand{\eit}{\end{itemize}}

\def\picture #1 by #2 (#3){
  \vbox to #2{
    \hrule width #1 height 0pt depth 0pt
    \vfill
    \special{picture #3} % this is the low-level interface
    }
  }

\def\scaledpicture #1 by #2 (#3 scaled #4){{
  \dimen0=#1 \dimen1=#2
  \divide\dimen0 by 1000 \multiply\dimen0 by #4
  \divide\dimen1 by 1000 \multiply\dimen1 by #4
  \picture \dimen0 by \dimen1 (#3 scaled #4)}
  }

\begin{document}

\title{Constraining the neutral fraction of hydrogen in the IGM at redshift 7.5}

\author{A. Hoag\altaffilmark{1},
M. Brada\v{c}\altaffilmark{2}, 
K. Huang\altaffilmark{2},
C. Mason\altaffilmark{3$\dagger$},
T. Treu\altaffilmark{1},
K. B. Schmidt\altaffilmark{4},
M. Trenti\altaffilmark{5,6}
V. Strait\altaffilmark{2}, 
B. C. Lemaux\altaffilmark{2},
E. Finney\altaffilmark{2},
M. Paddock\altaffilmark{2}}

\altaffiltext{1}{Department of Physics and Astronomy, University of California, Los Angeles, CA 90095-1547, USA; \email{athoag@astro.ucla.edu}}
\altaffiltext{2}{Department of Physics, University of California, Davis, 1 Shields Ave, Davis, CA 95616, USA}
\altaffiltext{3}{Harvard-Smithsonian Center for Astrophysics, 60 Garden St, Cambridge, MA, 02138, USA}
\altaffiltext{4}{Leibniz-Institut f\"{u}r Astrophysik Potsdam (AIP), An der Sternwarte 16, 14482 Potsdam, Germany}
\altaffiltext{5}{School of Physics, University of Melbourne, Parkville 3010, VIC, Australia}
\altaffiltext{6}{ARC Centre of Excellence for All Sky Astrophysics in 3 Dimensions (ASTRO 3D), Australia}
\altaffiltext{$\dagger$}{Hubble Fellow}

% ArXiv friendly
\begin{abstract} 
We present a large spectroscopic campaign with Keck/MOSFIRE targeting Lyman-alpha emission (Ly$\alpha$) from intrinsically faint Lyman-break Galaxies (LBGs) behind 12 efficient galaxy cluster lenses. Gravitational lensing allows us to probe the more abundant faint galaxy population to sensitive Ly$\alpha$ equivalent width limits. During the campaign we targeted 70 LBG candidates with MOSFIRE Y-band, selected photometrically to cover Ly$\alpha$ over the range $7<z<8.2$. We detect $S/N>5$ emission lines in 2 of these galaxies and find that they are likely Ly$\alpha$ at $z=7.148\pm0.001$ and $z=7.161\pm0.001$. We present new lens models for 4 of the galaxy clusters, using our previously published lens models for the remaining clusters to determine the magnification factors for the source galaxies. Using a Bayesian framework that employs large scale reionization simulations of the intergalactic medium (IGM) as well as realistic properties of the interstellar medium and circumgalactic medium, we infer the volume-averaged neutral hydrogen fraction, $\overline{x}_{\mathrm{HI}}$, in the IGM during reionization to be $\overline{x}_{\mathrm{HI}}=0.88^{+0.05}_{-0.10}$ at $z=7.6\pm0.6$. Our result is consistent with a late and rapid reionization scenario inferred by \emph{Planck}. 
 
\end{abstract}


\vspace*{0.4truecm}

\section{Introduction}
\label{sec:intro}
The Epoch of Reionization is a significant gap in our understanding of cosmic history. In recent decades it has been the subject of intensive observational campaigns spanning the entire electromagnetic spectrum. A first step in understanding reionization is constraining the timeline, which will provide a link between the well-studied ionized universe at $z\lesssim6$ with the Dark Ages, the period of time after Recombination ($z\lesssim1100$) but before the first sources of light emerged ($z\gtrsim20$). Knowledge of the timeline will help to constrain the abundance and properties of the sources driving reionization, potentially solving the longstanding debate as to whether quasars or galaxies are the primary sources \citep[e.g.][]{Haiman+98,Madau+99}. If quasars are primarily responsible, they must exist in greater numbers than the current high-redshift quasar luminosity function suggests \citep{Onoue+17,McGreer+18}. 

Galaxies are more promising sources as they exist in far greater numbers than quasars. The faint-end slope of the galaxy luminosity function suggests that faint galaxies may provide far more ionizing photons beyond the current detection limit, provided ionizing photon production efficiencies and escape fractions do not decline with luminosity \citep[e.g.][]{Finkelstein+15,Bouwens+15b,Livermore+16,Bouwens+17}. Furthermore, \citet{Bouwens+15b} found that the redshift evolution of the ionizing emissivity required to produce the Thomson optical depth measured by \citet{Planck2015} approximately matched the redshift evolution of the galaxy luminosity density. 

The first observational evidence for reionization was the detection of a \citet{Gunn+1965} trough in a quasar at $z=6.28$ \citep{Becker+01}. By compiling observations of $\sim20$ quasars at $z\sim6$, \citet{Fan+06} found that at $z\lesssim6$, reionization had likely concluded. The ``instantaneous'' redshift of reionization has also been constrained via the Cosmic Microwave Background anisotropy from both WMAP \citep{Hinshaw+13} and Planck \citep{Planck+16}, with some tension between the two. \citet{Planck+16} measured this redshift to lie in the interval $z=7.8-8.8$, lower than the WMAP measurement of $z=9.5-11.7$ \citep{Hinshaw+13}. 

In the last decade, many techniques have been developed to probe the state of reionization via the volume-averaged neutral hydrogen fraction (\xhi{}) in the inter-galactic medium (IGM), which must evolve from \xhi{}=$1$ before reionization to \xhi{}$\simeq0$ after reionization is complete. Quasars are often used in this approach due to their intrinsic brightness, which allows them to be observed at very high redshift \cite[e.g.][]{Mcgreer+15,Greig+17a,Banados+18}. However, their rarity is a limitation in this application, especially considering each quasar only probes a single line of sight through the universe.

A complementary approach is to observe the Lyman-alpha (\lya{}, rest-frame $1216$\AA{}) line from a large sample of galaxies. \lya{} is a strong rest-frame ultraviolet (UV) emission line that is redshifted to near-infrared wavelengths for sources at the Epoch of Reionization ($z\gtrsim6$),  \lya{} photons are easily absorbed by neutral hydrogen, making the line a sensitive probe of the state of the IGM during reionization. The ``\lya{} fraction'' test \cite[e.g.][]{Stark+10,Pentericci+11}, i.e. measuring the fraction of Lyman Break Galaxies with detected \lya{} as a way to constrain the \lya{} optical depth, is one such technique. This kind of test has been carried out by several teams, confirming a decline in the fraction of LBGs that show \lya{} emission above a rest-frame equivalent width of $\mathrm{EW_{Ly\alpha}}=25$ \AA{} from $z=6$ to $z=7$, consistent with an increase in \lya{} optical depth \cite[e.g.][]{Fontana+10,Stark+11,Pentericci+11,Schenker+12,Ono+12,Treu+12,Tilvi+14,Pentericci+14,Schenker+14}, and thus an increase in the neutral hydrogen fraction over the same interval. An important difference between the \lya{} fraction test and the Gunn-Peterson trough constraint is the fact that the \lya{} transmission at $z>6$ is small but non-zero. This allows the test to be meaningfully extended to higher redshift. Due to the large abundance of galaxies, the test can also be binned more finely, enabling a more detailed evolutionary study and therefore insight into the evolution of the properties of the sources of ionizing photons.

Despite these advantages, measuring the \lya{} optical depth does not directly constrain \xhi{}. One must make assumptions regarding the distribution of the neutral and ionized gas within the interstellar media (ISM) and circumgalactic media (CGM) of the galaxies themselves because \lya{} can be highly attenuated before it enters the IGM. Furthermore, the large-scale structure of the IGM during reionization can affect the inference on the neutral fraction inferred from \lya{} surveys. 

\citet{Treu+12,Treu+13} suggested that using the full \lya{} equivalent width distribution, rather then the fraction of detections at some fixed arbitrary threshold, vastly improves the information content on the distribution of \lya{} optical depth, and thus in turn on the neutral fraction, provided that the effects of the ISM and IGM can be isolated. \citet{Mesinger+15} put forward a model that ties observations of \lya{} to large-scale reionization simulations of the IGM. \citet{Mason+18a} built on the Bayesian framework introduced by \citet{Treu+12,Treu+13} and the simulations by \citet{Mesinger+15}, adding empirical models of the effects of the ISM and CGM, to develop a complete end-to-end inference from \lya{} observations to the hydrogen neutral fraction. 

In this paper, we make use of the \citet{Mason+18a} models to infer the neutral hydrogen fraction at $z\sim7.5$ from a \lya{} spectroscopic campaign with the Multi-Object Spectrometer for InfraRed Exploration \citep[MOSFIRE;][]{mosfire}. Our spectroscopic campaign was designed to target \ncluster{} galaxy clusters, comprising some of the best gravitational lenses in terms of number of magnified high-$z$ galaxies. The galaxy clusters act as cosmic telescopes, magnifying the background galaxies. This allows us to probe the sub-$L_{\star}$ galaxy population, which likely comprises more typical regions of the universe at $z>7$ than the $\gtrsim L_{\star}$ population. Using \ncluster{} galaxy cluster fields also allows us to mitigate cosmic variance by probing multiple independent lines of sight.

In Section~\ref{sec:data}, we present our imaging and spectroscopic data.  We present our main results in Section~\ref{sec:results}, including the sample of high-z galaxies in Section~\ref{sec:targ_selection}, gravitational lens models in Section~\ref{sec:lensmodels}, the search for \lya{} in Sections~\ref{sec:lya_search}, \lya{} detections in Section~\ref{sec:detections}, analysis  of non-detections in Section~\ref{sec:nondetections}, and the neutral fraction inference in Section~\ref{sec:neutral_fraction}. We discuss the neutral fraction result and its implication on the reionization timeline in Section~\ref{sec:discussion}. We summarize our results in Section~\ref{sec:summary}. Throughout this work, we adopt the following cosmology: $\Omega_m=0.3$, $\Omega_{\Lambda}=0.7$ and $h=0.7$. All magnitudes are given in the AB system, all dates are given in UTC, and all uncertainties are $68\%$ confidence unless specified otherwise. 

\section{Data}
\label{sec:data}
Here we describe the spectroscopic and imaging data that we used in this work. 
We also describe the photometric pipeline we use to create source catalogs from the reduced images.

\subsection{Spectroscopic Data Reduction and Calibration}
\label{sec:specdata}
\begin{deluxetable*}{| ccccccc |}
				\tablecaption{MOSFIRE slit-masks targeted}
				\tablecolumns{7}
				\tablewidth{0pt}
				\tabletypesize{\tiny}
				\tablehead{Cluster & mask name & date of observation & $t_{\mathrm{exp}}$ & $<\mathrm{seeing}>$ & $<\mathrm{attenuation}>$ & $<\mathrm{airmass}>$
				\\ & & (UTC) & (sec.) & ($''$) & (mag.) & }  
\startdata
A2744 & A2744\_MR & 2015nov07 & 4320 & 0.73 & 0.05 & 1.13 \\ 
A2744 & A2744\_MR & 2015nov08 & 2520 & 1.18 & 0.05 & 1.17 \\ 
\hline \\ 
A370 & A370\_M & 2013dec16* & 3060 & 1.13 & \nodata & 1.10 \\ 
A370 & A370\_M & 2013dec17* & 1800 & 1.03 & 0.05 & 1.12 \\ 
A370 & A370\_M & 2013dec18* & 3600 & 0.89 & 0.1 & 1.09 \\ 
A370 & A370\_M & 2017oct01 & 14580 & 0.67 & 0.05 & 1.18 \\ 
\hline \\ 
MACS0416 & MACS0416\_M & 2015nov07 & 9000 & 0.67 & 0.04 & 1.12 \\ 
MACS0416 & MACS0416\_M & 2015nov08 & 7200 & 0.84 & 0.05 & 1.12 \\ 
\hline \\ 
MACS0454 & macs0454\_M & 2013dec15* & 3240 & 0.64 & 0.05 & 1.19 \\ 
\hline \\ 
MACS0744 & MACS0744\_M\_backup & 2015nov08 & 7200 & 0.60 & 0.08 & 1.08 \\ 
MACS0744 & MACS0744\_M & 2016feb22 & 14220 & 1.06 & 0.03 & 1.14 \\ 
MACS0744 & MACS0744\_M & 2016feb23* & 13680 & 0.66 & 0.05 & 1.12 \\ 
MACS0744 & MACS0744\_ATH4Yband\_mt\_M & 2016mar20 & 5040 & 0.52 & 0.08 & 1.40 \\ 
\hline \\ 
MACS1149 & MACS1149 & 2014feb14 & 2520 & 0.60 & 0.9 & 1.15 \\ 
MACS1149 & MACS1149\_rot180 & 2016feb22 & 10800 & 1.40 & \nodata & 1.15 \\ 
\hline \\ 
MACS1423 & miki14M & 2013jun13* & 5040 & 0.90 & 0.07 & 1.44 \\ 
MACS1423 & MACS1423\_M & 2015apr27* & 7560 & 0.68 & 0.06 & 1.31 \\ 
MACS1423 & MACS1423\_052715\_M & 2015may28 & 17280 & 0.69 & 0.06 & 1.20 \\ 
MACS1423 & MACS1423\_20160319\_M & 2016mar20 & 6660 & 0.85 & 0.05 & 1.16 \\ 
\hline \\ 
MACS2129 & MACS2129\_M & 2015apr27* & 1440 & 0.80 & 0.04 & 1.55 \\ 
MACS2129 & MACS2129\_052715\_M & 2015may28 & 7560 & 0.55 & 0.1 & 1.27 \\ 
MACS2129 & MACS2129\_M & 2015nov07 & 8460 & 0.53 & 0.05 & 1.17 \\ 
MACS2129 & MACS2129\_M & 2015nov08 & 9540 & 0.84 & 0.04 & 1.20 \\ 
\hline \\ 
MACS2214 & MACS2214\_M & 2017oct01 & 14400 & 0.84 & 0.04 & 1.16 \\ 
\hline \\ 
RCS2327 & rcs22327\_M & 2013dec15* & 3420 & 0.51 & 0.06 & 1.13 \\ 
RCS2327 & rcs22327\_M & 2013dec17* & 3600 & 0.71 & 0.07 & 1.13 \\ 
RCS2327 & rcs22327\_M\_1 & 2013dec18* & 3600 & 1.14 & 0.08 & 1.13 \\ 
\hline \\ 
RXJ1347 & RXJ1347\_v4 & 2018jun01 & 11700 & 0.60 & 0.18 & 1.13 \hspace{-0.2cm}
\tablecomments{* Indicates a half night of observation. Attentuation values marked by ``\nodata'' mean that attenutation data were not available on these dates. }
\enddata
\label{tab:obs_conditions}
\end{deluxetable*}
The spectroscopic data used in this work were obtained entirely using the MOSFIRE instrument on the Keck I telescope. The majority of the data (14/15 nights) come from the program  ``Dawn of the Galaxies: Spectroscopy of Sources at $z\gtrsim7$'' (PI Bradac; 2013: project codes: U032M, U004M, 2014: U004M, 2015: U005M, U031M, 2016: U004M, 2017: U027, U026, 2018: U005). However, we did not use 5/14 nights in our analysis due to poor weather. In addition, we use data from 1 night (2016 Mar 20) from project code Z054M (PI M. Trenti). 

We designed MOSFIRE slit-masks using the publicly available MAGMA software\footnote{\url{https://www2.keck.hawaii.edu/inst/mosfire/magma.html}}. The names and exposure times of the targeted slit-masks are listed in Table~\ref{tab:obs_conditions}. In designing the masks, we assigned $z\gtrsim7$ galaxies the highest priority, followed by $z\gtrsim6$ galaxies (Section~\ref{sec:photom}). We filled the remaining slits with, in decreasing order of priority, gravitationally lensed arcs, line-emitting galaxies and cluster members, depending on the availability of pre-existing spectroscopy for each cluster. There was an average of $\sim6$ of the highest priority ($z\gtrsim7$) galaxies on each mask, with each mask containing a total of $\sim30$ objects.

All MOSFIRE data were obtained with $0\farcs7$-wide slits in the Y-band, which covers \lya{} over the redshift range $7\lesssim z \lesssim 8.2$. We observed with an ABBA dither pattern, using $1\farcs{25}$ amplitude dithers, which results in a spacing of $2\farcs5$ between positive signal and each negative shadow. The average full-width at half maximum (FWHM) of the atmospheric seeing was generally $\lesssim1''$ and average attenuation of the data we used (measured in the V-band) was generally small ($\lesssim0.05$ mag). Seeing was either measured directly from stars on the science slit mask or from pre-science alignment images taken in the J-band. 

We reduced the MOSFIRE data using the publicly available data reduction pipeline (DRP\footnote{\url{https://www2.keck.hawaii.edu/inst/mosfire/drp.html}}). The DRP produces wavelength calibrated, rectified, background- and skyline-subtracted 2D signal and noise spectra for each slit on the mask. We extracted 1D signal and noise spectra from the 2D spectra for each slit, using uniform weights on all spatial pixels for the extraction. The spatial aperture we used for extraction was twice the FWHM of the atmospheric seeing. We found that for the majority of our observations, the pipeline-produced noise spectra underestimates the uncertainty derived directly from signal spectra in regions where no objects or sky emission were present. To remedy this, we scaled all 1D noise spectra so that the $S/N$ spectra in regions free from object traces and sky line emission had a distribution with standard deviation equal to 1. We then divided the 1D signal spectra by the scaled 1D noise spectra to obtain properly behaved 1D $S/N$ spectra for each slit on each slit-mask. We opted to use the re-scaled noise spectrum rather than, e.g., the standard deviation within the flux spectrum itself because of potential contamination in the flux spectrum from nearby objects.

Flux calibration of the 1D signal and noise spectra was performed following the steps outlined by \citet{Hoag+15}. For all masks observed before September 2016, we used a spectral type AOV star we observed on 2013 Dec 15 to flux calibrate our spectra. MOSFIRE was repaired during the interval between September 2016 and February 2017, potentially altering the response function of the instrument. Therefore, we used an AOV star observed after the repair (2017 Oct 01) to flux calibrate the masks observed after the repair. We account for differences in the extinction and airmass of the calibration star compared to each mask observation. This is done by de-reddening and applying an airmass correction using the average airmass value during observations to both the calibration star and each science spectrum before applying the telescope correction to the science spectra. This step effectively applies a slit-loss correction to each science spectrum assuming our targets are all point sources. Most sources are indeed not extended, so we do not apply an extended source slit-loss term to our spectra. While \lya{} is known to be on average more extended than the rest-frame UV continuum \citep[e.g.][]{Steidel+11,Wisotzki+16, Leclercq+17}, an analysis with the KMOS IFU down to similar flux limits as obtained in this work (Mason et al. 2019, submitted) found no evidence that would suggest we are missing \lya{} due to these effects. These effects may still, however, decrease our sensitivity, so our flux limits presented in Section~\ref{sec:nondetections} are effectively lower limits.

Many of the same objects were observed on multiple masks. In these cases, we stacked the calibrated 1D signal and noise spectra from all available individual masks using a simple mean and produced full-depth $S/N$ spectra from these stacks. While masks with somewhat poor conditions were included, this did not result in the loss of candidate emission lines in the final stacks; we inspected all individual spectra for emission lines in addition to our search described in Section~\ref{sec:lya_search}. 

\subsection{Imaging data}
\label{sec:image_data}
%%%%%%% Table 1: Clusters %%%%%%%
\begin{deluxetable*}{| ccccccc |}
				\tablecaption{Galaxy Clusters Targeted}
				\tablecolumns{7}
				\tablewidth{0pt}
				\tabletypesize{\tiny}
				\tablehead{Cluster name & Short Name &  $\alpha_{\mathrm{J2000}}$ & $\delta_{\mathrm{J2000}}$ & Redshift &  \HST{} imaging & \Spitzer{} imaging  
				\\ & & (deg.) & (deg.) &  & }  
\startdata
Abell 2744       & A2744    &  3.5975000  & -30.39056 & 0.308 & HFF/GLASS & SFF  \\ 
Abell 370        & A370     &  39.970000  & -1.576666 & 0.375 & HFF/GLASS & SFF \\ 
M0416.1-2403     & MACS0416 &  64.039167  & -24.06778 & 0.420 & HFF/CLASH/GLASS & SFF  \\ 
MACSJ0454.1-0300 & MACS0454 &  73.545417  & -3.018611 & 0.540 & HST-GO-11591/GO-9836/GO-9722 & SURFSUP \\ 
MACSJ0717.1-0300 & MACS0717 &  109.38167  & 37.755000  & 0.548 & HFF/CLASH/GLASS & SFF \\ 
MACSJ0744.8+3927 & MACS0744 &  116.215833 & 39.459167 & 0.686 & CLASH/GLASS & SURFSUP \\ 
MACSJ1149.5+2223 & MACS1149 &  177.392917 & 22.395000 & 0.544 & HFF/CLASH/GLASS & SFF/SURFSUP \\ 
MACSJ1423.8+2404 & MACS1423 &  215.951250 & 24.079722 & 0.545 & CLASH/GLASS & SURFSUP \\ 
MACSJ2129.4-0741 & MACS2129 &  322.359208 & -7.690611 & 0.570 & CLASH/GLASS & SURFSUP \\ 
MACSJ2214.9-1359 & MACS2214 &  333.739208 & -14.00300 & 0.500 & SURFSUP & SURFSUP \\ 
RCS2-2327.4-0204 & RCS2327  &  351.867500 & -2.073611 & 0.699 & SURFSUP/HST-GO-10846 & SURFSUP  \\
RXJ1347.5-1145   & RXJ1347  &  206.87750  & -11.75278 & 0.451 & CLASH/GLASS & SURFSUP
\tablecomments{SFF = \emph{Spitzer} Frontier Fields (PI T. Soifer, P. Capak). Clusters are referred to by their short names throughout this work. }
\enddata
\label{tab:clusters_ongoing}
\end{deluxetable*}

%%% Figures 1 and 2: HST images, critical curves and high-z objects %%%

 \begin{figure*}[htb]
    \centering
    \includegraphics[width=\linewidth]{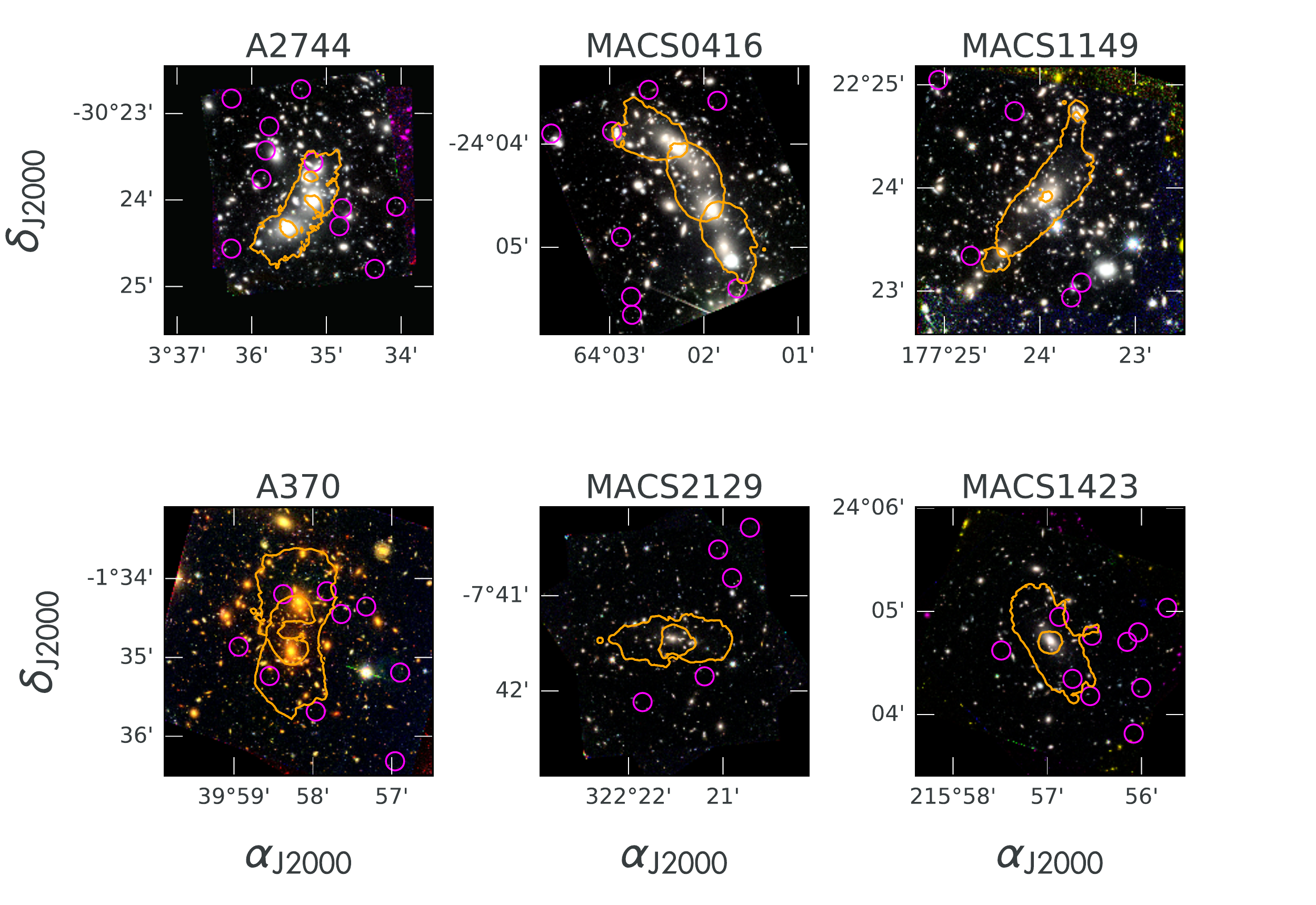}
    \caption{6/11 Galaxy clusters targeted in this work, including 4 Hubble Frontier Fields clusters. Critical curves (orange lines) from our lens models are shown at $z=7.6$ along with the $z\gtrsim7$ LBGs (magenta circles) we targeted with MOSFIRE. Magnification can be as large as $\sim50$ near the critical curve, but falls off quickly, reaching typical values of $1.5-2$ near the edge of the shown fields, which are the \HST{} WFC3/IR footprints.  }
 \label{fig:critical_curves1}
 \end{figure*}
 
  \begin{figure*}[htb]
    \centering
    \includegraphics[width=\linewidth]{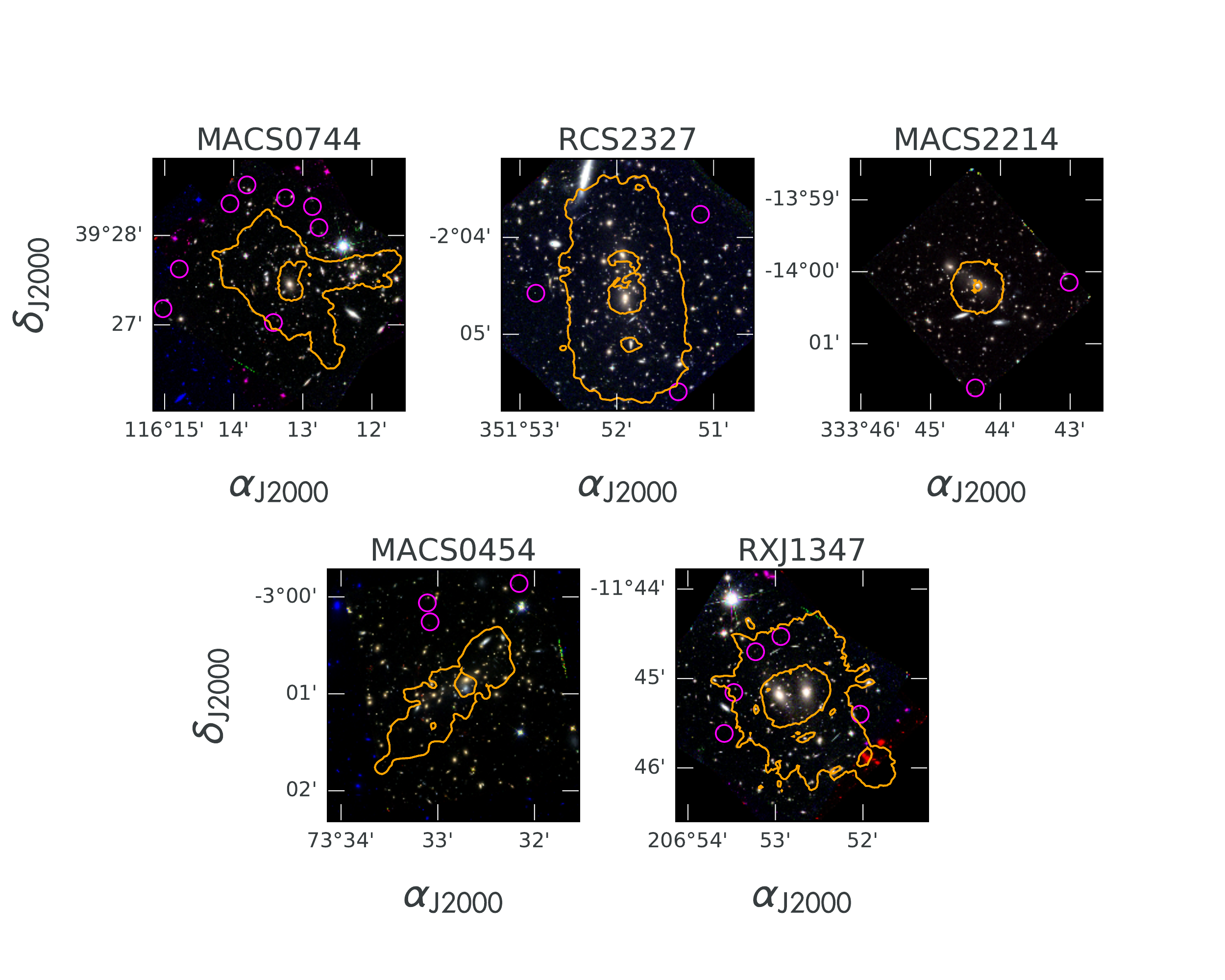}
    \caption{Same as Figure~\ref{fig:critical_curves1}, but for the remaining 5 galaxy clusters targeted in this work. }
 \label{fig:critical_curves2}
 \end{figure*}

%%%%%%%%%%%%%%%%%%%%% 

The imaging data were primarily used to measure photometric redshifts to select a sample of high-z galaxies for spectroscopic follow-up. The images were also used in constructing the lens models (Section~\ref{sec:lensmodels}). The data come from several \HST{} and \Spitzer{}/IRAC programs, summarized in Table~\ref{tab:clusters_ongoing}. \HST{} mosaics of the clusters used in this work are shown in Figures~\ref{fig:critical_curves1} and~\ref{fig:critical_curves2}. We targeted several clusters belonging to the Hubble Frontier Fields Initiative \citep[HFF;][]{Lotz+16}. These are Abell 2744 (A2744), M0416.1-2403 (MACS0416), MACSJ0717.5+3745 (MACS0717), MACSJ1149.5+2223 (MACS1149) and Abell 370 (A370). As explained in Section~\ref{sec:targ_selection}, MACS0717 had 0 targets in our final photometric selection, so it is not included in the rest of this work. For these clusters, we also use Keck/MOSFIRE $\mathrm{K_s}$-band photometry from the K-band Imaging of the Frontier Fields program \citep{Brammer+16}.

Another 4 clusters: MACS0744, MACS1423, MACS2129, and RXJ1347 are part of the Cluster Lensing And Supernova survey with Hubble \citep[CLASH;][]{Postman+12}. MACS0454 was observed by HST-GO-11591/GO-9836/GO-9722, and the \HST{} images for both MACS2214 and RCS2327 were obtained as part the Spitzer UltRa Faint SUrvey Program \citep[SURFSUP;][]{Bradac+14}. Many of these clusters were observed by the Grism Lens-Amplified Survey from Space \citep[GLASS;][]{Schmidt+14,Treu+15a,Schmidt+16}, which obtained \HST{} WFC3/IR imaging and grism spectroscopy of the prime cluster fields. Our target selection was mildly influenced by the \lya{} candidates assembled by \citet[][see Section~\ref{sec:targ_selection}]{Schmidt+16}.

The depths of the \HST{} images vary among the clusters. For example, the images of the HFF clusters that we targeted all reach limiting magnitudes of $\sim29$ AB mag ($5\sigma$, point source) per band, whereas the remaining clusters are $\sim2$ magnitudes shallower per band. The images of the HFF and CLASH clusters were obtained from the latest public mosaics available on the Mikulski Archive for Space Telescopes (MAST\footnote{\url{https://archive.stsci.edu/pub/hlsp/frontier/}}$^{,}$\footnote{\url{https://archive.stsci.edu/pub/hlsp/clash/}}). The HFF data span 3 Advanced Camera for Surveys (ACS) filters: F435W, F606W and F814W, and 4 Wide Field Camera 3 (WFC3) IR filters: F105W, F125W, F140W and F160W. The CLASH clusters share these filters but have 5 additional filters that we use: F475W, F625W, F775W, F850LP (ACS) and F110W (WFC3)\footnote{CLASH also observed clusters in 4 WFC3/UVIS filters, but we do not use these data in our analysis as they are not constraining.}. The three non-CLASH and non-HFF clusters: MACS0454, MACS2214 and RCS2327 were processed as described by \citet{Huang+16a} and have a subset of the CLASH filters. 

We also use deep \Spitzer{}/IRAC data in tandem with the \HST{} data. The primary purposes of these data are to more accurately select and characterize the high-$z$ sample \cite[e.g.][Strait et al. 2019, in prep.]{Huang+16a}. The IRAC images were obtained by the \emph{Spitzer} Ultra-Faint Survey Program \citep[SURFSUP; PID 90009;][]{Bradac+14} and the \emph{Spitzer Frontier Fields} (SFF; PIDs: 83, 137, 10171, 40652, 60034, 80168, 90257, 90258, 90259, 90260). The data have similar depths in all \ncluster{} clusters, reaching at least $\sim30~\mathrm{hr}$ in the $[3.6]~\mu m$ and $[4.5]~\mu m$ bands (hereafter $[3.6]$ and $[4.5]$). The processing of the \Spitzer{}/IRAC data is described by \citet{Ryan+14} and \citet{Huang+16a}. 

\subsection{HST and IRAC photometry}
\label{sec:photom}
While the \HST{} photometry procedure is described in detail by \citet{Huang+16a}, we briefly outline the primary steps we take to produce photometric catalogs. First, we produce point-spread function (PSF) matched \HST{} images with a 60mas pixel size, convolving all images (except F160W) to match the resolution of F160W, the lowest resolution \HST{} image. We produce a stacked near-infrared (NIR) image from the \HST{} WFC3 images to use for object detection. We then run Source Extractor \citep[SExtractor;][]{Bertin+96} in dual-image mode on each filter using the stacked WFC3 NIR image as the detection image. We correct each filter for galactic extiction using the IR dust maps obtained by \citet{Schlegel+98} and the coefficients in each filter from \citet{Postman+12}. We also find from simulations that the flux errors in each filter reported by SExtractor are underestimated. We correct the errors via source simulations using a similar method to \citet{Trenti+11}.

For some of the clusters, i.e. the HFF clusters, MACS1423, MACS2129 and RXJ1347, we also performed a subtraction of the intra-cluster light (ICL) and bright cluster members using the method developed by the ASTRODEEP collaboration \citep{Merlin+16,Castellano+16b}. This was done to obtain more accurate photometry and to improve number counts of lensed, high-redshift sources. It is important to perform on the deeper images, e.g. the HFF images, due to the stronger contamination from the ICL. However, the ICL is less of an issue for the CLASH-depth clusters, so the fact that this is was not performed on MACS0454, MACS0744, and RCS2327 is therefore of little concern. In particular, the sources that would be unveiled in these clusters from ICL and cluster member subtraction would be far too contaminated in the MOSFIRE slits to obtain constraining \lya{} flux limits. 

The IRAC photometry is described by \citet{Huang+16a}. In brief, we use T-PHOT \citep{Merlin+15} to measure colors between \HST{} and IRAC. We use priors based on the segmentation map from the \HST{} F160W image, the image closest in wavelength to the IRAC bands, to deblend objects in the lower resolution IRAC bands. In this way, we can assemble a consistent photometric catalog with IRAC for all of the sources identified in the \HST{} NIR detection image. The end result of the process is a merged \HST{} and IRAC photometric catalog for all detected sources in the field.

%%%%%%%%%%%%%%%%%%%%%%%%%%%%%%%%%%%%%%%%%%
  
\section{Results}
\label{sec:results}
\subsection{Selection of high-$z$ sample}
\label{sec:targ_selection}

The selection criteria for including high redshift galaxies on our slit-masks was very inclusive.  Where possible, we incorporated multiple selections from the literature to maximize the number of high-$z$ targets on the mask, using the compilation by \citet{Schmidt+16}. Since our observations began in December 2013, we have obtained significantly improved \HST{} and \Spitzer{} data for many of the clusters in this sample. In particular the HFF program and much of the SURFSUP and GLASS programs were performed during this time. 

In order to use a more consistent photometric selection across all clusters in this work and to make use of the much deeper current data, we performed the same photometric processing of all of the clusters in our sample after all of our spectroscopic observations were taken. We recalculated the photometry for each cluster using the pipeline described in Section~\ref{sec:photom} and re-fit all of the photometry to obtain photometric redshift probability distributions, $P(z)$s, and SEDs using EA$z$Y \citep{Brammer+08}, adopting the default \textit{v1.3} EA$z$Y spectral templates. As in \citet{Huang+16a}, we inspected the distribution of best-fit photo-$z$s for all clusters, ensuring that the distribution peaked at the cluster redshift. For the clusters in the GLASS sample (c.f. Table~\ref{tab:clusters_ongoing}), we were able to anchor the EA$z$Y photo-$z$ distribution by comparing our photo-$z$s to spectroscopic redshifts of intermediate-$z$ galaxies in the GLASS data set. We found good statistical agreement in all clusters \citep[e.g.][]{Hoag+16,Wang+15}. This was not possible for the non-GLASS clusters: MACS0454, MACS2214 and RCS2327. The objects in these clusters hold less weight in the neutral fraction inference because they generally have fewer objects and shallower observations. Nonetheless, their photo-$z$s are potentially less reliable than the photo-$z$s of targets in the GLASS clusters.

We used a flat prior on $z$, rather than adopting the default magnitude prior when deriving the $P(z)$s of all objects with EA$z$Y. We did not adopt the default EA$z$Y magnitude prior because our targets are lensed. From the $P(z)$ of each galaxy we calculated the probability that \lya{} falls in the MOSFIRE Y-band, i.e. $P(7<z<8.2)$. We then selected objects with $P(7<z<8.2)>0.01$, finding a total of 2828 such objects in all 11 clusters. A significant fraction of these were spurious. We cleaned the catalog of objects that were obvious parts of larger galaxies, diffraction spikes, bad pixels, or clearly originated from the edge effects at the boundary of the detector or overlap between epochs. This inspection was performed visually, so it is naturally subjective. However, it is more reliable than our current automatic methods to remove such artifacts. After cleaning the sample, we arrived at a sample of 530 objects that fulfill the $P(7<z<8.2)>0.01$ condition. Of these, we observed 70 in our Keck/MOSFIRE survey. 2 of these objects were spectroscopically confirmed to be at lower redshift ($6<z<7$) by \citet{Huang+16b} and S. Fuller et al. (in preparation) with Keck/DEIMOS during a simultaneous spectroscopic campaign. We note that the spectroscopic redshifts of both galaxies are consistent with the photometric redshifts we obtained for those objects with EA$z$Y. The remaining \nhighz{} objects will hereafter be referred to as the ``MOSFIRE'' sample. We note that while we observed several candidates from our original photometric catalogs in MACS0717 in December 2013 based on pre-HFF \HST{} and \Spitzer{}/IRAC imaging, the final photometric selection from our full-depth images yielded 0 candidates. As a result, MACS0717 is not part of the MOSFIRE sample, and is not discussed in the rest of this work. 

\begin{figure*}[htb]
    \centering
	    \includegraphics[width=\linewidth]{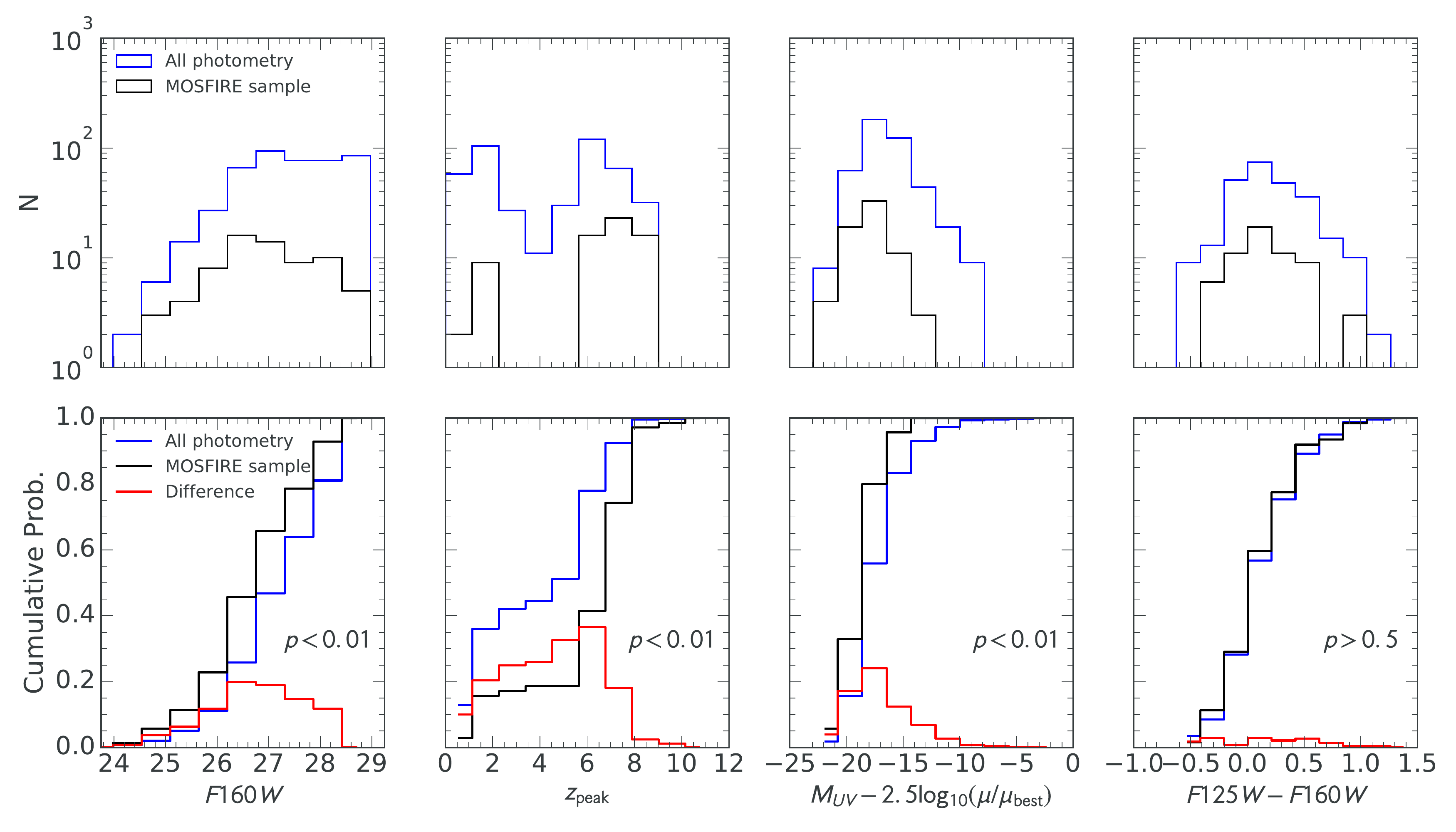}
    \caption{Comparison of the sample targeted with MOSFIRE and the parent photometric catalog of high-$z$ sources. \textbf{Top}: from left to right: distributions of the \HST{} F160W apparent magnitude, peak photometric redshift ($z_\mathrm{peak}$), absolute magnitude (corrected for lensing magnification) and F125W-F160W color (rest frame UV color) for the entire photometric high-$z$ sample (blue) and the MOSFIRE sub-sample. \textbf{Bottom}: The cumulative probability distributions of the above quantities, following the same color scheme. The red line in each panel is the difference between the two cumulative distributions, used in a two-sample Kolmogorov-Smirnov (KS) test. The KS tests indicate that the MOSFIRE sample is statistically brighter (both apparently and intrinsically) and higher redshift than the parent photometry sample. We describe how we account for these effects in Section~\ref{sec:neutral_fraction}. The distributions of rest-frame UV colors (observed IR colors: F125W - F160W) are statistically consistent between the two samples. P-values are shown in the bottom panels indicating the significance of the rejection or acceptance of the null hypothesis, i.e. that the two samples are drawn from identical distributions.}
 \label{fig:compare_phot}
 \end{figure*}

During some of our observing runs, we preferentially targeted GLASS \lya{} emitter candidates from \citet{Schmidt+16}. While these represented typically only 1-2 targets per cluster, we checked that this selection did not introduce a bias into our final result. We ran the neutral fraction inference (described in Section~\ref{sec:neutral_fraction}) with and without these targets and found that the difference was insignificant. We also compared the distributions of properties of the MOSFIRE sample with the parent photometric sample. Figure~\ref{fig:compare_phot} shows the comparison for the F160W apparent magnitude, peak photometric redshift ($z_{\mathrm{peak}}$), intrinsic absolute magnitude corrected for lensing ($M_{UV}-2.5\log_{10}(\mu/\mu_{\mathrm{best}})$), and the rest-frame UV color (measured near-IR color F125W-F160W). We performed a two-sample Kolmogorov-Smirov (KS) test for each of these properties, finding significant differences ($>99\%$ probability) between the distributions for the F160W magnitude, $z_{\mathrm{peak}}$, and the absolute magnitude. These differences illustrate our mask design strategy which was to put brighter candidates that had a higher probability of being high redshift on the masks. We correct for these selection effects when inferring the neutral fraction (see Section~\ref{sec:neutral_fraction}). The rest-UV color is statistically consistent between the two samples, which is important because we do not correct for any influence this color can have on the \lya{} properties, such as the EW distribution.

\begin{figure}[htb]
    \centering
    \includegraphics[width=\linewidth]{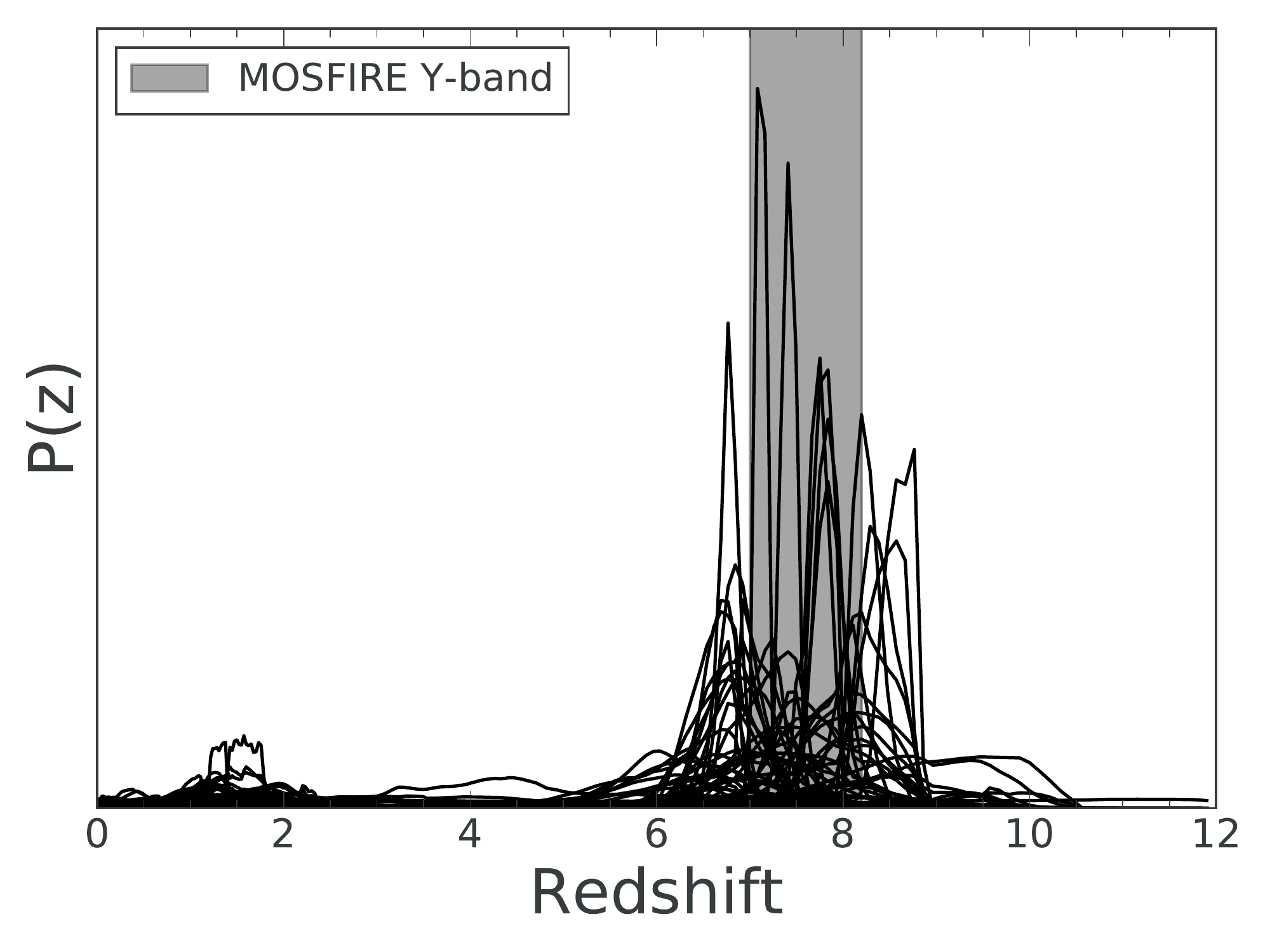}
    \caption{Photometric redshift distributions for all \nhighz{} galaxies in the MOSFIRE sample. The light gray band denotes the MOSFIRE Y-band redshift coverage of the \lya{} emission line, $7<z<8.2$. We account for the fact that not all of the probability distribution falls within the Y-band coverage redshift when inferring the neutral fraction (Section~\ref{sec:neutral_fraction}). }
 \label{fig:pzs}
 \end{figure}

We plot the $P(z)$s for all \nhighz{} galaxies in the MOSFIRE sample in Figure~\ref{fig:pzs}. Most of the probability from the \nhighz{} galaxies falls within the 
MOSFIRE Y-band coverage. We account for the photometric redshift impurity in the neutral fraction inference, as explained in Section~\ref{sec:neutral_fraction}. We list the targets along with some of their properties in Table~\ref{tab:targets}. Throughout this work, we refer to individual targets via their cluster and ID in the following manner: the first target in Table~\ref{tab:targets} is ID=A370-000 and the last is ID=MACS0744-069. We note that the median absolute magnitude of our sample is $M_{UV} = -18.25$,  where we account for magnification for each galaxy before computing the median. This is $\sim0.1L_{\star}$ at $z\sim7-8$ \citep{Bouwens+15a}, illustrating that the galaxies we are targeting are intrinsically very faint and probably more characteristic of the general galaxy population than $\sim L_{\star}$ galaxies. 

\subsection{Gravitational Lens Models}
\label{sec:lensmodels}
To link the observed LBGs to halos in reionization simulations for performing the inference on the hydrogen neutral fraction (see Section~\ref{sec:neutral_fraction}), we use the galaxy intrinsic luminosity. Because we observed LBGs in lensed fields, we must correct their observed luminosities by the magnification factor. To obtain the magnification factor for each galaxy, we produce lens models of all of the clusters in our sample. Lens modeling is performed using the free-from code developed by \citet{Bradac+05,Bradac+09}. Briefly, the code solves for the gravitational potential on an adaptive pixel grid via $\chi^2$ minimization. The lensing quantities such as convergence, shear and magnification are all derived from the best-fit gravitational potential. 

The lens models for all clusters were constructed using the \HST{} imaging data described in Section~\ref{sec:image_data}. \HST{} images are used to identify strongly lensed (multiply-imaged) galaxies as well as weakly-lensed (singly-imaged) galaxies, both of which are used as constraints to the lens model. We also used spectroscopic redshifts from GLASS \citep{Schmidt+14,Wang+15,Treu+16,Hoag+16}, from the CLASH-VLT program \citep[P.I.: P. Rosati;][]{Rosati+14}, and from \citet{Limousin+12,Johnson+14,Zitrin+15a,Grillo+16,Monna+17,Lagattuta+17} to improve the precision of the lens models. In general, we used only the set of mulitply-imaged systems with either secure spectroscopic redshifts or consistent photometric redshifts \citep[as in e.g.][]{Hoag+16}. To obtain uncertainties on the lensing quantities, including the magnification, we bootstrap resample the weak lensing catalog and re-run the lens modeling code 100 times. The number of weakly-lensed galaxies (typically $\sim500-1000$) is much larger than the number of strongly-lensed galaxies, so we opt to use the weak lensing catalog for the bootstraps. In cases where the number of multiply-imaged systems is sufficiently large ($\gtrsim15$), we simultaneously resample from the strongly-lensed and weakly-lensed galaxy catalogs to obtain uncertainties. For more details on this procedure see \citet{Hoag+16}. 

Some of the lens models used in this work are described in previous works: the RCS2327 model was obtained by \citet{Hoag+15}, A2744 by \citet{Wang+15}; MACS0416 by \citet{Hoag+16}, MACS2129 by \citet{Huang+16b}; MACS1423 by \citet{Hoag+17}, MACS1149 by \citet{Finney+18}, and A370 by \citet{Strait+18}. Lens models for MACS0454, MACS0744, MACS2214 and RXJ1347 are presented here for the first time. We show the critical curves from all of our lens models as well as the location of the high redshift objects targeted in the MOSFIRE campaign with non-zero probability of being at $7<z<8.2$ in Figures~\ref{fig:critical_curves1} and~\ref{fig:critical_curves2}.

%%%%%%%%%%%%%%%%%%%%%  

%%%%%%% Figure 3: Comparison with parent sample %%%%%%%

%%%%%%% Figure 4: All P(z)s %%%%%%%

%%%%%%%%%%%%%%%%%%%%%%%%%%%%%%%%%%%%%%%%%%    

%%%%%%%%%%%%%%%%%%%%%%%%%%%%%%%%%%%%%%%%%%

\subsection{Search for \lya{} emission}
\label{sec:lya_search}

After collecting all of our MOSFIRE observations and selecting the \nhighz{} targets that fulfill our high-$z$ selection criteria, we performed an automatic search for emission lines. For each target, we calculated $S/N_{int}$, the integrated $S/N$, at each wavelength in the full-depth 1D $S/N$ spectrum, assembling the integrated signal-to-noise spectrum. We used a spectral bandpass of $6$\AA{} ($\sim6$ pixels) for the integration because this is twice the FWHM spectral resolution of the instrument. For each object, we flagged any wavelengths where $S/N_{int}\geq5$, i.e. lines with high confidence. This resulted in a total of 22 candidate emission lines.

We visually inspected all 22 candidate lines identified by the automatic detection procedure. Real emission lines have several characteristics which can distinguish them from spurious features. These are:
\begin{enumerate}
\item Two negative residuals at symmetric positions flanking the central emission line and with approximately half of the intensity of the central emission line due to our ABBA dither pattern
\item At least as spectrally extended as the MOSFIRE resolution, i.e. $\sim3$ pixels in Y-band
\item At least as spatially extended as the atmospheric seeing at the time of observation 
\item Distinguishable from sky emission line features
\end{enumerate}

Because these tests were performed visually, they are inherently subjective. However, the candidate list was short (22 candidates) and most of the candidates failed multiple of these tests obviously. There was 1 ambiguous case which we describe in Section~\ref{sec:detections}.

\subsection{\lya{} detections}
\label{sec:detections}
\begin{figure*}[htb]
    \centering
    \includegraphics[width=\linewidth]{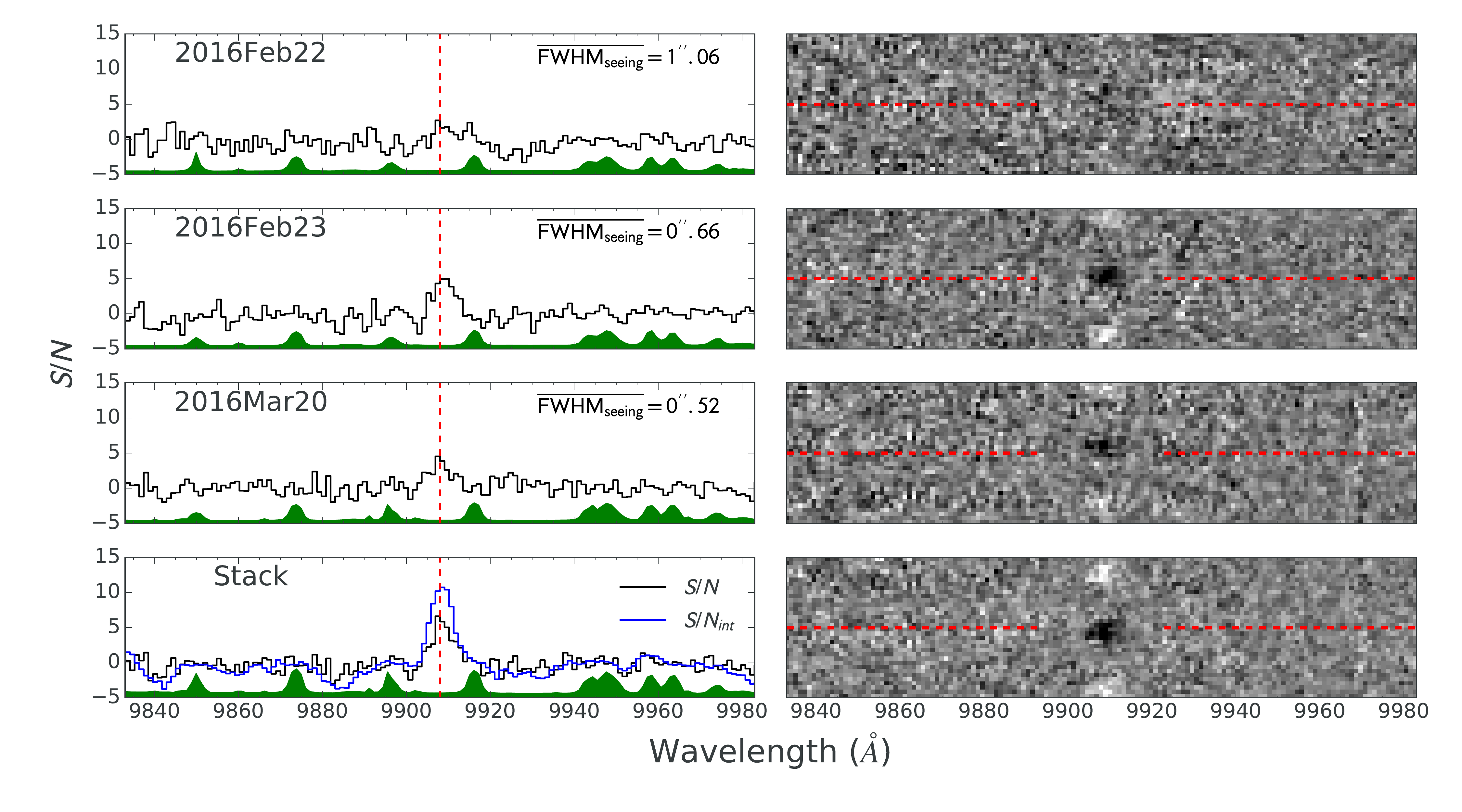}
    \caption{Detection of \lya{} at $9908$\AA{} ($z_{Ly\alpha}=7.148\pm0.001$) from ID=\macssampleid{}.  \textbf{Left:} 1D $S/N$ spectra (black histograms) from three different nights of observation and the stack of all three. The integrated $S/N$ is also shown (blue histogram) in the stacked panel, reaching $S/N_{int}=10.3$ at the line center. The vertical red dashed line in each panel is centered at $9908$\AA{}. An arbitrarily scaled noise spectrum is shown in green in each panel. \textbf{Right:} 2D $S/N$ spectra from which the 1D spectra shown on the left are extracted. The red horizontal dashed lines show the expected spatial position of the object in the slit. Black represents positive values, while white represents negative values. Negative residuals are clearly seen in white on either side of the central emission line in most panels. Note that the seeing was poor ($>1''$) on 2016 Feb 22 but favorable ($<0.7''$) on the other two dates, explaining the non-detection in the top panel. }
 \label{fig:m0744det}
 \end{figure*}

\begin{figure*}[htb]
    \centering
    \includegraphics[width=\linewidth]{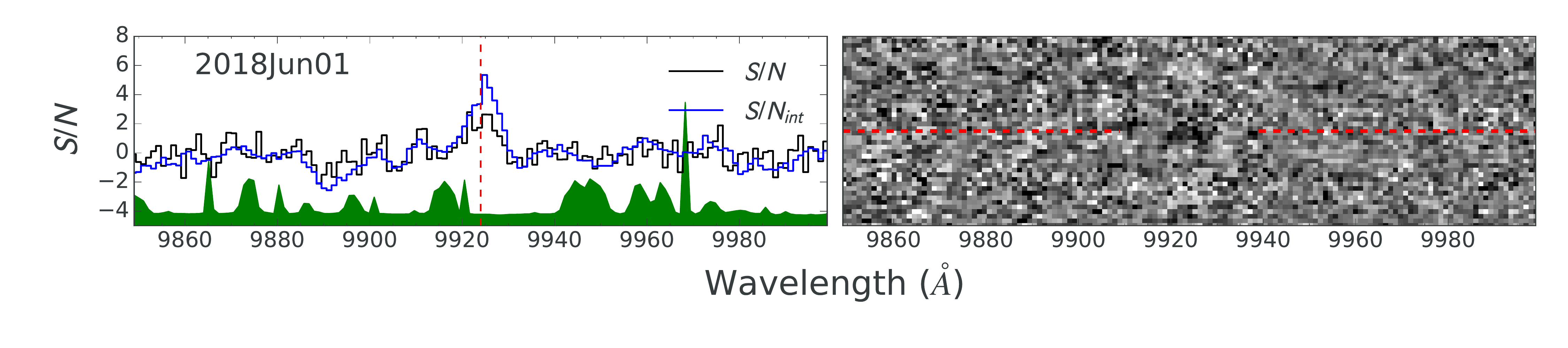}
    \caption{Detection of \lya{} at $9924$\AA{} ($z_{Ly\alpha}=7.161\pm0.001$) from sample ID=\rxjsampleid{}.  \textbf{Left:} 1D $S/N$ spectrum (black histogram) from the single night of observation. The integrated $S/N$ is also shown (blue histogram) reaching $S/N_{int}=5.1$ at the line center. The vertical red dashed line is centered at $9924$\AA{}. An arbitrarily scaled noise spectrum is shown in green. \textbf{Right:} 2D $S/N$ spectra from which the 1D spectra shown on the left are extracted. The red horizontal dashed lines show the expected spatial position of the object in the slit. Black represents positive values, while white represents negative values. Negative residuals are formally detected at the correct positions, although the low $S/N$ makes them hard to visually distinguish. }
 \label{fig:rxj1347det}
 \end{figure*}

We identified 2 emission lines that passed the tests described in Section~\ref{sec:lya_search}. The $S/N$ spectra for these two objects, \macssampleid{} and \rxjsampleid{}, are shown in Figures~\ref{fig:m0744det} and ~\ref{fig:rxj1347det}. 

We observed \macssampleid{} (Figure~\ref{fig:m0744det}) on 2016 Feb 22, 2016 Feb 23 and 2016 Mar 20 with slit-masks MACS0744\_M, MACS0744\_M, and MACS0744\_ATH4Yband\_mt\_M, respectively (see Table~\ref{tab:obs_conditions}). The emission line is significantly detected on 2016 Feb 23 and 2016 Mar 20, but hardly detected on 2016 Feb 22. While the exposure time was comparable for 2016 Feb 22 and 2016 Feb 23, the seeing was much worse on 2016 Feb 22, explaining the difference in $S/N$. The seeing and attenuation were both good on 2016 Feb 22 and 2016 Mar 20, the two dates where the emission line is significantly detected. ID=\macssampleid{} is confidently detected, with an integrated signal-to-noise ratio of $S/N_{int} = 10.3$ in the full-depth stack. The two negative residuals are also clearly detected in the stack at the correct locations and can also be seen on multiple nights of observation. The line is completely separated from sky emission lines. As a result, this line is considered secure.

ID=\rxjsampleid{} is detected at $S/N_{int}=5.1$. Unlike \macssampleid{}, we only observed this object during a single night of observation, so we do not have multiple nights to corroborate the detection. To test the robustness of the detection, we split the data in half, re-reduced each half, and re-extracted the spectra to see if the emission line was present in both halves\footnote{We split the data by using every even dither pair for one half and every odd pair for the other pair. This is opposed to dividing the dataset in half temporally, which could introduce differences in the two halves of the data due to variable weather.}. We found a peak in the integrated signal to noise in both halves at the same wavelength as in the full-depth spectrum. Using the same bandpass as in the full-depth spectrum, we found values of $S/N_{int}=3.82$ and $S/N_{int}=3.11$ in the two halves, consistent with a real line of $S/N_{int}\sim5$ in the combined dataset. 

Despite this evidence in support of the line, there are other aspects of the line that are concerning. A sky emission line is present just blueward of the line-peak emission wavelength of $9924$\AA{}. Part of the 6\AA{} spectral bandpass used to calculate the integrated signal-to-noise falls on this sky emission line. While the noise spectrum in principle takes into account the noise from the sky emission line, the sky subtraction is imperfect and could introduce spurious features at this level of $S/N$. The negative residuals are present for this object with ratios of $-0.9\pm0.3$ for both top and bottom residuals. These ratios should be consistent with $-0.5$ as a result of our dither pattern. However, because the central emission line is only detected at $S/N_{int}\sim5$, the negative residuals themselves are only expected to be detected at $\abs{S/N_{int}}\sim3.5$. At such low $S/N$, the ratios of the negative residuals to the central emission are less robust, and thus this test is less reliable. For example, the proximity to the sky line could influence the deviation of these ratios via imperfect sky line subtraction. Given that these factors reduce our confidence in the emission line, we carry out the neutral fraction inference with and without this detection and compare the results in Section~\ref{sec:neutral_fraction}, finding that its inclusion or exclusion does not change our result in a statistically significant way. Future follow up of this cluster is needed to confirm or deny the validity of this emission line. 

Both of the emission lines presented here are the only lines detected in the spectra of their respective targets. Here we provide additional evidence to properly identify them as \lya{}. For both objects, our main evidence in support of \lya{} is the photometric redshift distribution, $P(z)$. We show the $P(z)$ plots for \macssampleid{} and \rxjsampleid{} in Figures~\ref{fig:macs0744sedpz} and ~\ref{fig:rxj1347sedpz}, respectively. In both cases, the spectroscopic redshift is in good agreement of the photometric $P(z)$, falling within the $68\%$ confidence interval. The most common contaminant of \lya{} at $z\sim7$ is the \oii$\lambda \lambda 3726,3729$ doublet at $z\sim1.5$. The $P(z)$s suggest that \oii{} is very unlikely for both \macssampleid{} and \rxjsampleid{}. Furthermore, the \oii{} doublet would be resolved given the MOSFIRE resolution, with a peak separation of $\sim7$\AA{}. We do not detect any doublets with this separation in our spectra. Within the $95\%$ confidence interval of the $P(z)$s of both targets, the \civ{} and \ciii{} doublets could fall in MOSFIRE Y-band. However, these lines are typically much weaker in LBGs than the rest-frame equivalent widths that would be required to detect them given the faintness of these targets \cite[e.g.][]{Shapley+03,Stark+14,LeFevre+17}. Furthermore, in some cases we would expect to see both lines in the doublet. However, we cannot completely rule out that high equivalent width \civ{} or \ciii{}, such as values observed by \citet{Stark+15,Stark+16}, could explain the emission lines.

The RXJ1347 emission line looks marginally double peaked, but the separation of these two features is $\sim4$\AA{} and cannot be explained by any strong emission line doublets at redshifts consistent with the $P(z)$. The bluer peak falls on a narrow feature in the noise spectrum in between a bright sky line and the center of the emission line, but which is not associated with the sky line. The feature originates from two pixels in the 2D noise spectrum, which are identified by the DRP. We tested to see whether this feature was responsible for the bluer bump in $S/N$ spectrum. We masked out these two pixels and re-extracted the $S/N$  spectrum, finding that the bump did not disappear, and that the integrated $S/N$ was slightly larger than in the original case. We conclude that the blue bump originates from the flux spectrum, but it could be a result of imperfect sky subtraction of the bright sky line at $\sim9917$\AA{}. 

\subsection{Physical characteristics of \macssampleid{} and \rxjsampleid{} }
\begin{figure*}[htb]
    \centering
    \includegraphics[height=0.25\textheight]{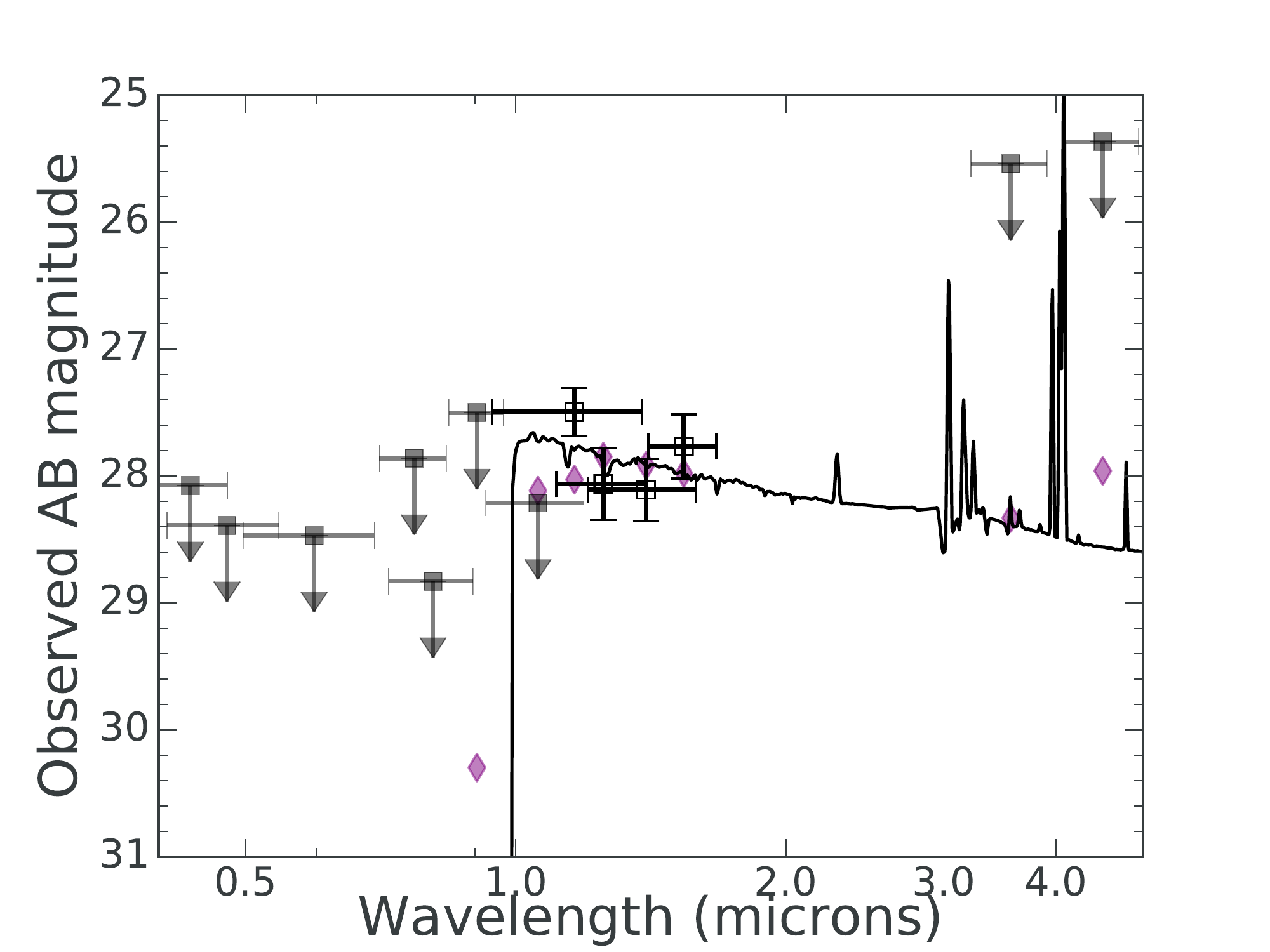}
    \includegraphics[height=0.25\textheight]{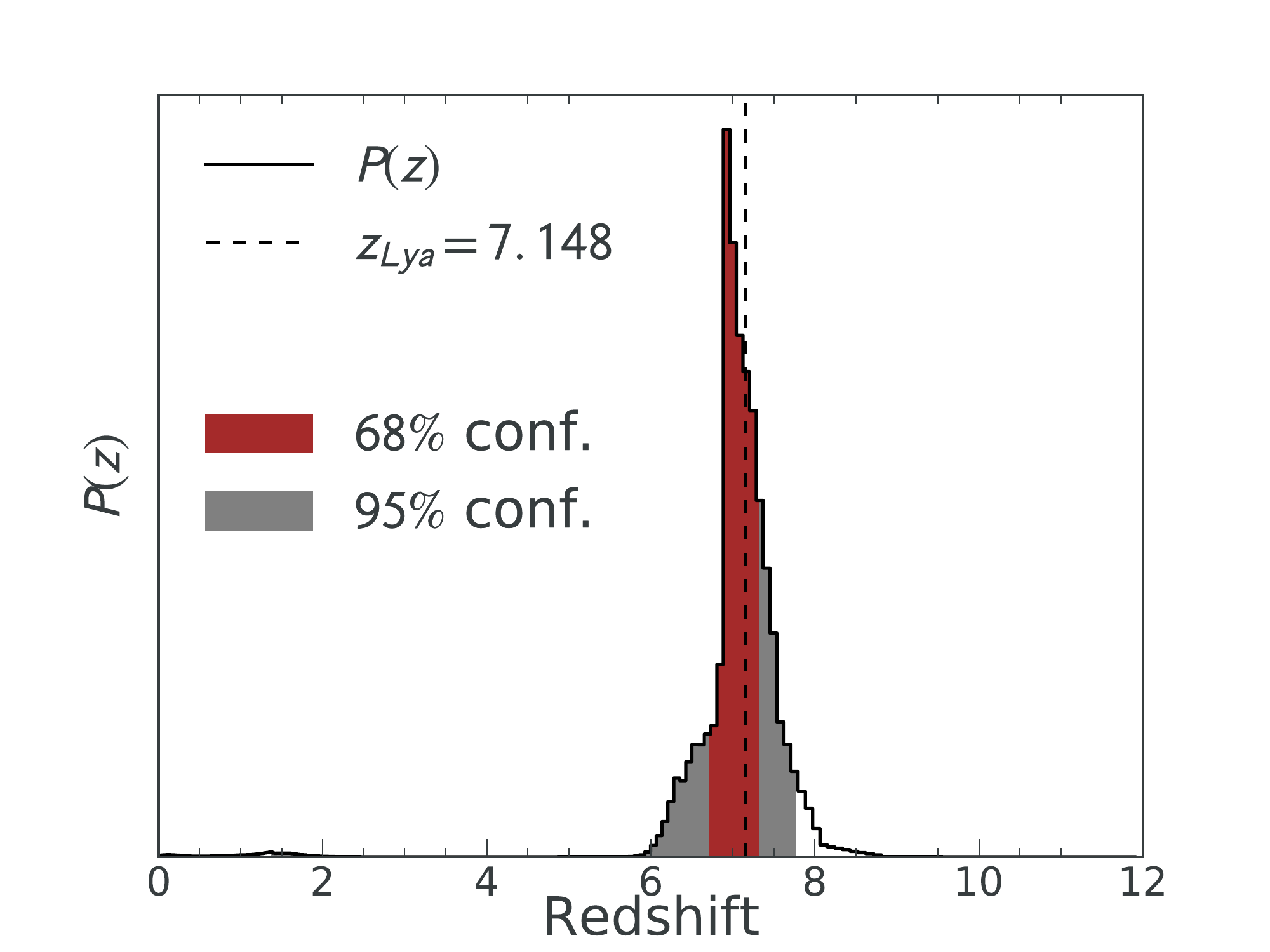}
    \caption{\textbf{Left}: Best-fit \citet{BC03} spectral energy distribution of the target \macssampleid{} (blue) fixed at the \lya{} redshift, $z=7.148$. Measured data points and $1\sigma$ error bars (black) and $3\sigma$ upper limits (gray) from \HST{} and \Spitzer{}/IRAC photometry are shown, along with the synthetic photometry from the best-fit SED (purple diamonds). \textbf{Right}: The photometric redshift distribution, $P(z)$ of the same galaxy obtained from EA$z$Y. The \lya{} spectroscopic redshift is in good agreement with the photometric redshift, and there is a very small probability of a lower-$z$ solution.}
 \label{fig:macs0744sedpz}
 \end{figure*}

\begin{figure*}[htb]
    \centering
    \includegraphics[height=0.25\textheight]{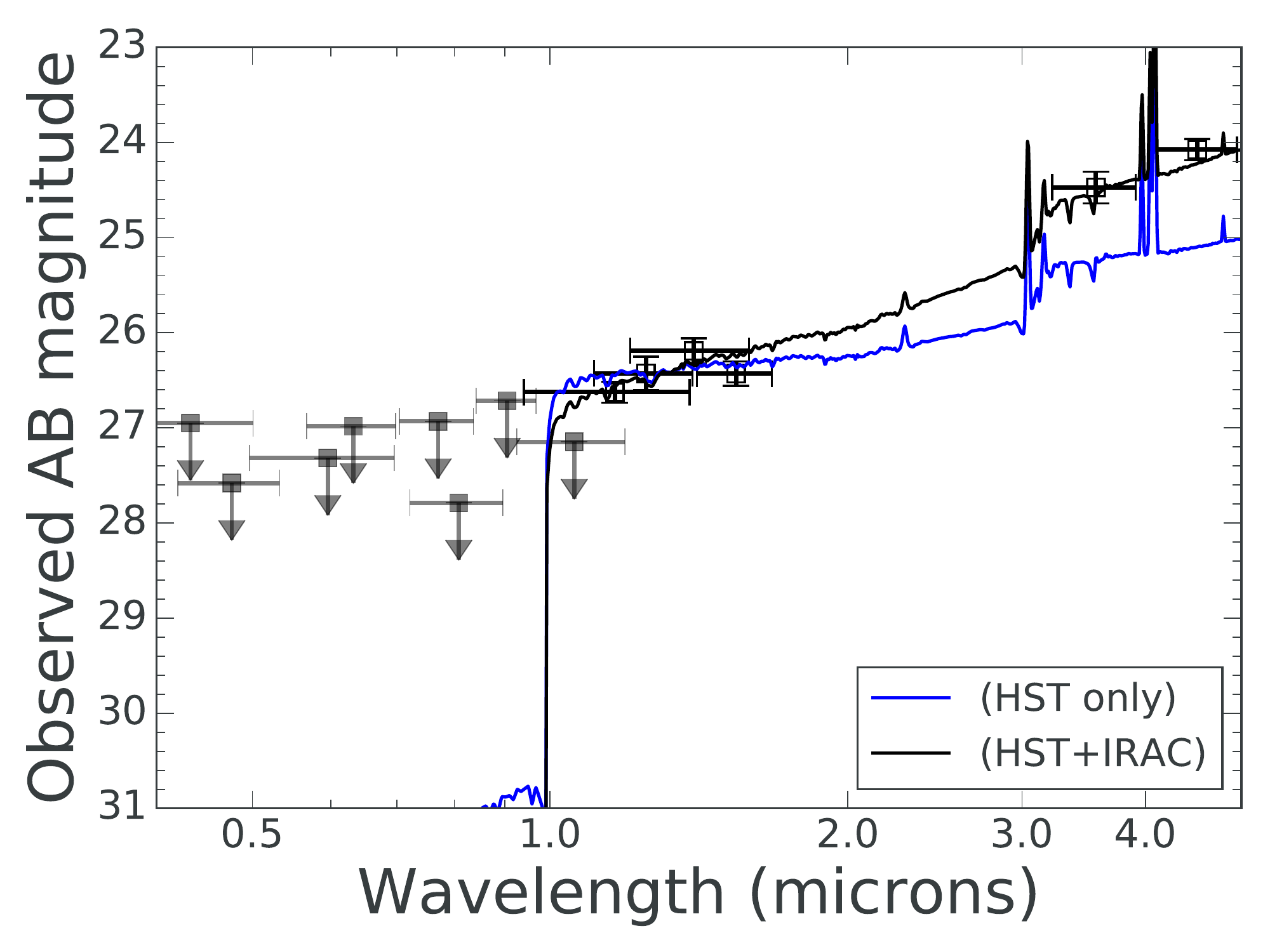}
    \includegraphics[height=0.25\textheight]{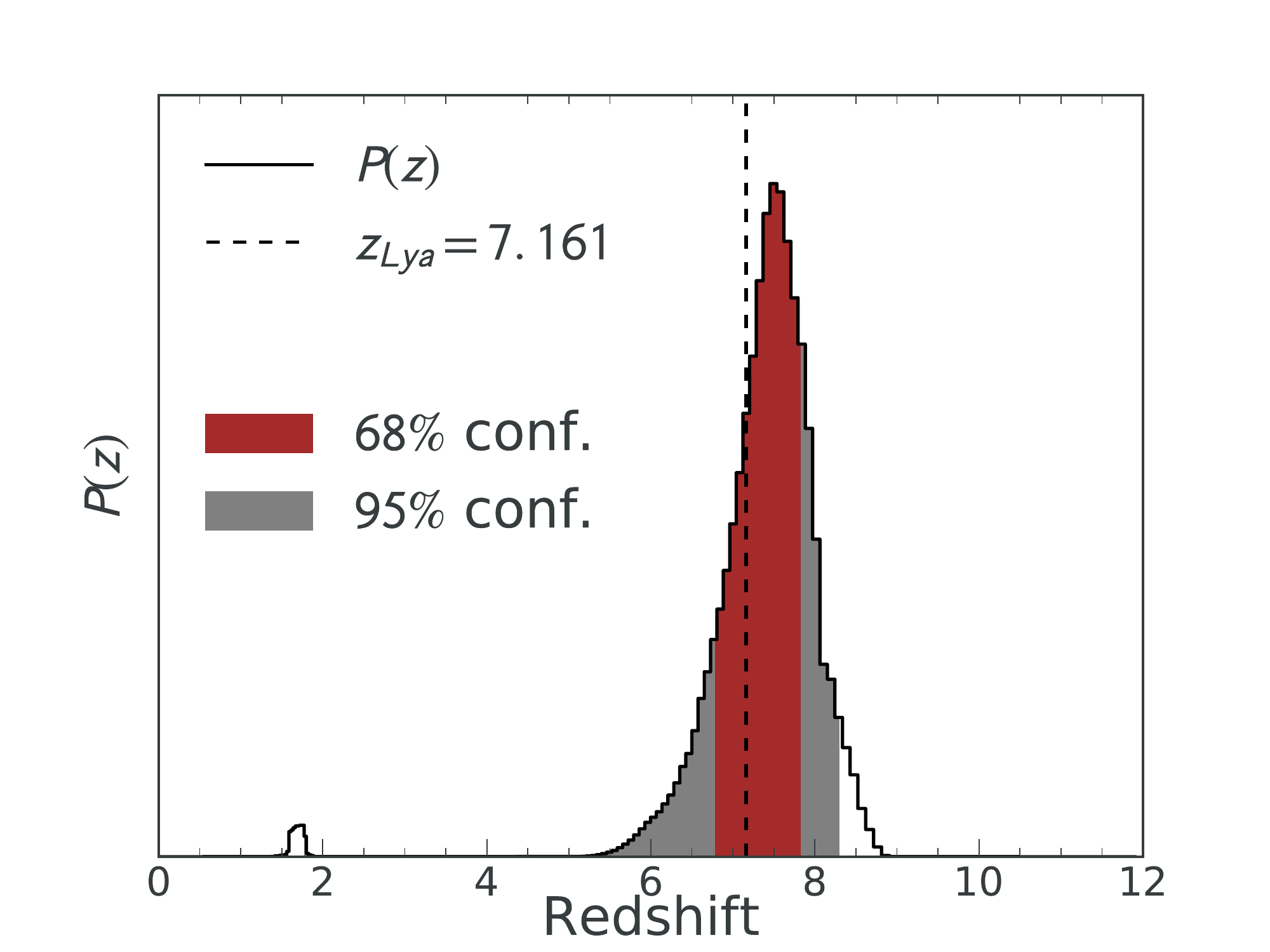}
    \caption{\textbf{Left}: Best-fit \citet{BC03} spectral energy distribution of the target \rxjsampleid{} fixed at the \lya{} redshift of $z=7.161$. The layout is the same as in Figure~\ref{fig:macs0744sedpz}, except that here we also show the best-fit SED using only the HST photometry in blue. The IRAC photometry is contaminated by a bright neighbor (Figure~\ref{fig:rxj1347postage}), so we consider the SED with and without the IRAC constraints. \textbf{Right}: The photometric redshift distribution, $P(z)$ of the \rxjsampleid{} obtained from EA$z$Y (using the IRAC constraints). The \lya{} spectroscopic redshift is in good agreement with the photometric redshift, and there is a very small probability of a lower-$z$ solution. This is still the case when using only the HST photometry to infer the $P(z)$. }
 \label{fig:rxj1347sedpz}
 \end{figure*}

We also show the spectral energy distributions (SEDs) of both galaxies in Figures~\ref{fig:macs0744sedpz} and ~\ref{fig:rxj1347sedpz}. We perform SED-fitting using the method described by \citet{Huang+16a}. In brief, we use EA$z$Y with the \citet{BC03} stellar population synthesis models, adopting a \citet{Chabrier03} initial mass function between $0.1$ and $100 M_{\odot}$, a \citet{Calzetti+00} dust attenuation law, and an exponentially declining star-formation history with a fixed metallicity of $Z=0.2Z_{\odot}$. Nebular emission lines are added to the templates as described by \citet{Huang+16a}.

We find that \macssampleid{} has a considerably bluer rest-frame UV slope than \rxjsampleid{}. This results in a stark difference in some of their physical properties, which are provided in Tables~\ref{tab:m0744_tab} and~\ref{tab:rxj1347_tab}. In particular, \macssampleid{} is significantly younger (age = $17.38^{+5.53}_{-6.91}~\mathrm{Myr}$) than \rxjsampleid{} (age = $640.47^{+78.15}_{-131.73}~\mathrm{Myr}$). Younger stellar ages ($\sim10-100~\mathrm{Myr}$) are expected at $z\sim7$ given the relatively brief amount of time at this redshift for star-formation to have taken place since the Big Bang ($\sim750~\mathrm{Myr}$). With this in mind,  \rxjsampleid{} is surprisingly old. However, relatively evolved stellar populations have possibly been observed at similar redshifts and beyond \citep[e.g.][Strait et al. 2019, in prep.]{Richard+11,Zheng+12,Bradac+14,Hashimoto+18}, suggesting the onset of star formation within the first $\sim300~\mathrm{Myr}$ after the Big Bang. \citet{Katz+18} showed that the SEDs of these rare, evolved stellar populations can be reproduced by cosmological simulations, but only in the most massive halos in their simulation.

\begin{figure*}
    \centering
    \includegraphics[width=\textwidth]{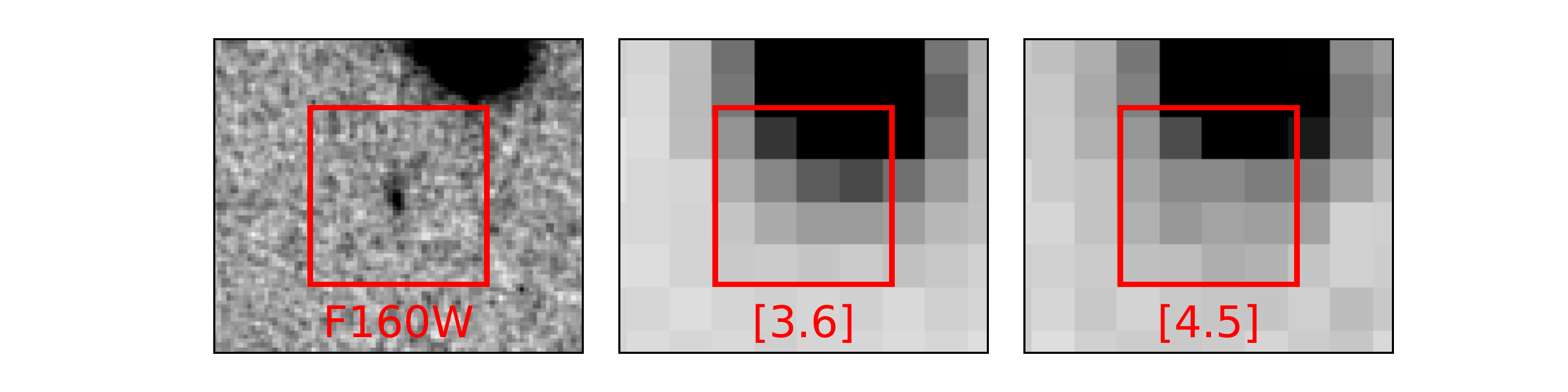}
    \caption{\HST{} and \spitzer{}/IRAC images of \rxjsampleid{}. Shown are the \HST{} F160W (left), IRAC $[3.6]~\mu m$ (middle) and IRAC $[4.5]~\mu m$ images. While the galaxy is well resolved in \HST{}, it is heavily blended with the much brighter neighbor galaxy in the two IRAC bands due to the larger PSF. As a result, the detections in the two IRAC bands are not trustworthy. The red box is $2\farcs5$ on a side. }
 \label{fig:rxj1347postage}
 \end{figure*}

The red UV slope of \rxjsampleid{} is partially driven by the detections in the IRAC $[3.6]~\mu m$ and $[4.5]~\mu m$ bands. For example, in Figure~\ref{fig:rxj1347sedpz} we show our best-fit SED for \rxjsampleid{} using only the \HST{} constraints. The UV slope is bluer then when we include the IRAC constraints. The IRAC detections are questionable due to the presence of a bright neighboring galaxy. In Figure~\ref{fig:rxj1347postage}, we show a cutout of the \HST{} F160W image as well as the two IRAC bands centered on \rxjsampleid{}. The galaxy is resolved in \HST{}, but it is heavily blended with the neighboring galaxy in IRAC due to the much larger PSF. While we attempted to subtract the neighbor using T-PHOT when performing photometry on \rxjsampleid{}, the proximity to \rxjsampleid{} makes the fluxes untrustworthy. As a result, we show the SED with and without the IRAC constraints in Figure~\ref{fig:rxj1347sedpz}, and we report the physical properties of the galaxy in both cases in Table~\ref{tab:rxj1347_tab}. Using \HST{} photometry only, the age of the galaxy is very poorly constrained (age = $321.00^{+397.62}_{-310.52}~\mathrm{Myr}$) and the stellar mass is an order of magnitude smaller than when we include the IRAC constraints.  

In some cases, age and stellar mass inferences can be biased due to strong unmodeled rest-frame UV and optical emission lines falling in the near-IR \HST{} and mid-IR IRAC filters \citep[e.g.][]{Schaerer+09,Labbe+13,Smit+14}. The strongest of such emission lines are the \oii{}, \oiii{}, H$\beta$ and H$\alpha$ lines. At the redshift of \rxjsampleid{}, $z=7.161$, none of these strong lines fall in  $[3.6]$, so they could not explain the bright flux in this filter even if it were real. While H$\beta$ and \oiii{} fall in the $[4.5]$ band at $z=7.161$ and could potentially explain this flux, this would not explain the flux in the $[3.6]$ band.

\citet{Molino+17} performed photometry and SED-fitting for the CLASH clusters, which include MACS0744 and RXJ1347. Both \macssampleid{} and \rxjsampleid{} are present in their publicly available photometric catalogs\footnote{\url{https://archive.stsci.edu/prepds/clash/}}. We compare to the only two fitted quantities present in their catalog, which are the photo-$z$ and the stellar mass. For \rxjsampleid{} (CLASHID$=0911$), they obtained a best-fit photometric redshift of $z=7.61^{+0.13}_{-0.81}$ (95\% conf.), in agreement with the \lya{} spectroscopic redshift (and the photo-$z$) that we measured. They report a stellar mass for \rxjsampleid{} (uncertainties are not provided) of: $3.89\times10^{9} M_{\odot}$ before accounting for magnification. Adopting our measurement of $\mu_{best}=21.4$, this results in a stellar mass of $1.82\times10^{8} M_{\odot}$. We note that \citet{Molino+17} did not use \Spitzer{}/IRAC data, so comparing to our stellar mass measurement of $M_{\star}=1.45^{+4.14}_{-1.19} M_{\odot}$ derived without  IRAC, the two results are in agreement. 
 
For \macssampleid{} (CLASHID$=1176$), the catalog photometric redshift is: $z=4.14^{+1.42}_{-3.70}$ (95\% conf.). This is significantly different than our photo-$z$, which does not show any probability near the peak of their distribution, i.e. at $z\sim4$ (Figure~\ref{fig:macs0744sedpz}). We investigated different origins for this discrepancy. As for \rxjsampleid{}, \citet{Molino+17} do not use the IRAC $[3.6]$ and $[4.5]$ measurements as constraints on the photo-$z$. However, this is not the primary reason for the photo-$z$ discrepancy. We re-ran EA$z$Y on our photometry without the two IRAC bands, and found approximately the same photo-$z$ as when we included the IRAC data. We also checked to see if the different photo-$z$s arose due to differences in \HST{} photometry. We re-ran EA$z$Y using the \citet{Molino+17} photometry and found $z=6.39^{+0.51}_{-5.27}$, similar to the result using our own photometry. This means that the difference most likely arises from the different codes used to derive the redshifts, rather than the photometry. \citet{Molino+17} use Bayesian photometric redshifts \citep[BPZ;][]{Benitez+00,Coe+06}, whereas we use EA$z$Y. We note that the brightness of this galaxy is near the detection limit of the \HST{} images; the $S/N$ in the four bands in which it is detected above $S/N=3$ are F160W: 3.8, F140W: 4.0, F125W: 3.4, F110W: 5.3. \rxjsampleid{} is detected with much higher $S/N$ in \HST{}, and there the photo-$z$s are in much better agreement. As a result, we expect that the different templates and other slight implementation differences between the two codes are amplified when the signal to noise ratio of the data is low, giving rise to the different photometric redshift result.

The \lya{} properties of the two galaxies are also shown in Tables~\ref{tab:m0744_tab} and~\ref{tab:rxj1347_tab}. We calculated rest-frame \lya{} equivalent widths ($\mathrm{EW_{Ly\alpha}}$) by dividing the line flux by the continuum flux density measured in F160W and then by the scale factor ($1+z$). \macssampleid{} has $\mathrm{EW_{Ly\alpha}}=59.4$ \AA{}, and \rxjsampleid{} has $\mathrm{EW_{Ly\alpha}}=27.8$ \AA{}. These are within the range of values for $\mathrm{EW_{Ly\alpha}}$ among the other known \lya{} emitters at $z=7-7.5$ \citep[see Table 2 of ][]{Stark+16}. However, our measurements are for much fainter galaxies than are typically observed at $z\sim7$ due to the lensing magnification. \macssampleid{} and \rxjsampleid{} have intrinsic absolute magnitudes of $M_{UV} - 2.5 \log_{10}( \mu / \mu_{\mathrm{best}}) =-18.5\pm0.4$ mag ($L= 0.12^{+0.06}_{-0.04}L_{\star}$) and $M_{UV} - 2.5\log_{10}( \mu / \mu_{\mathrm{best}}) =-17.2\pm0.2$ mag ($L=0.03\pm0.01L_{\star}$), respectively\footnote{We adopt $M_{\star}=-20.87\pm0.26$ mag from \citet{Bouwens+15a} to obtain $L_{\star}$}. Galaxies this faint far outnumber those observed in the field at luminosities of $L\sim L_{\star}$, which likely live in rare and massive halos \citep{RobertsBorsani+16,Mason+18b}. As a result, we can probe the neutral hydrogen fraction from more typical halo masses at $z\sim7-8$, as opposed to fewer halos at the high mass end. An ideal approach, and one we hope to adopt in future work, is to combine data sets to probe the neutral fraction using the entire available range of halo masses.

%%%%%%% Figure 4: MACS0744 detection %%%%%%%

 %%%%%%% Figure 5: RXJ1347 detection %%%%%%%

 \subsection{Non-detections}
 \label{sec:nondetections}
The remaining 20/22 candidate lines that were flagged by the automatic line detector are much less secure than the two lines presented above. Most fail multiple of the visual inspection tests. However, 2 of these correspond to real emission lines from a single lower-$z$ galaxy that we did not target, but which happened to fall in the same slit as one of our $z\gtrsim7$ targets, MACS1149-014. The two emission lines are confidently identified as the \oiii{}$\lambda \lambda 4959,5007$ emission line pair at $z=1.225$, and are present at a spatial offset from the central trace of our actual target consistent with that expected from the position of the slit relative to our target and the lower-$z$ contaminating galaxy in the \HST{} image.

We note that while \citet{Hoag+17} presented a detection of \lya{} in ID=MACS1423-040, an object in our MOSFIRE sample, it does not pass the above automatic line detection criteria. This is primarily due to our re-scaling (increasing) of the noise as described in Section~\ref{sec:specdata}. If we perform the same extraction of the line as done by \citet{Hoag+17}, but use the re-scaled noise, we find $S/N_{int} = 3.4$. Because the \lya{} EW of this object is so low \citep[$9$\AA{};][]{Hoag+17}, it has an insignificant impact on our neutral fraction results. We demonstrate this in section~\ref{sec:neutral_fraction} by inferring the neutral fraction both with this target as a detection and as a non-detection.

\begin{figure*}[htb]
    \centering
    \includegraphics[width=0.8\textwidth]{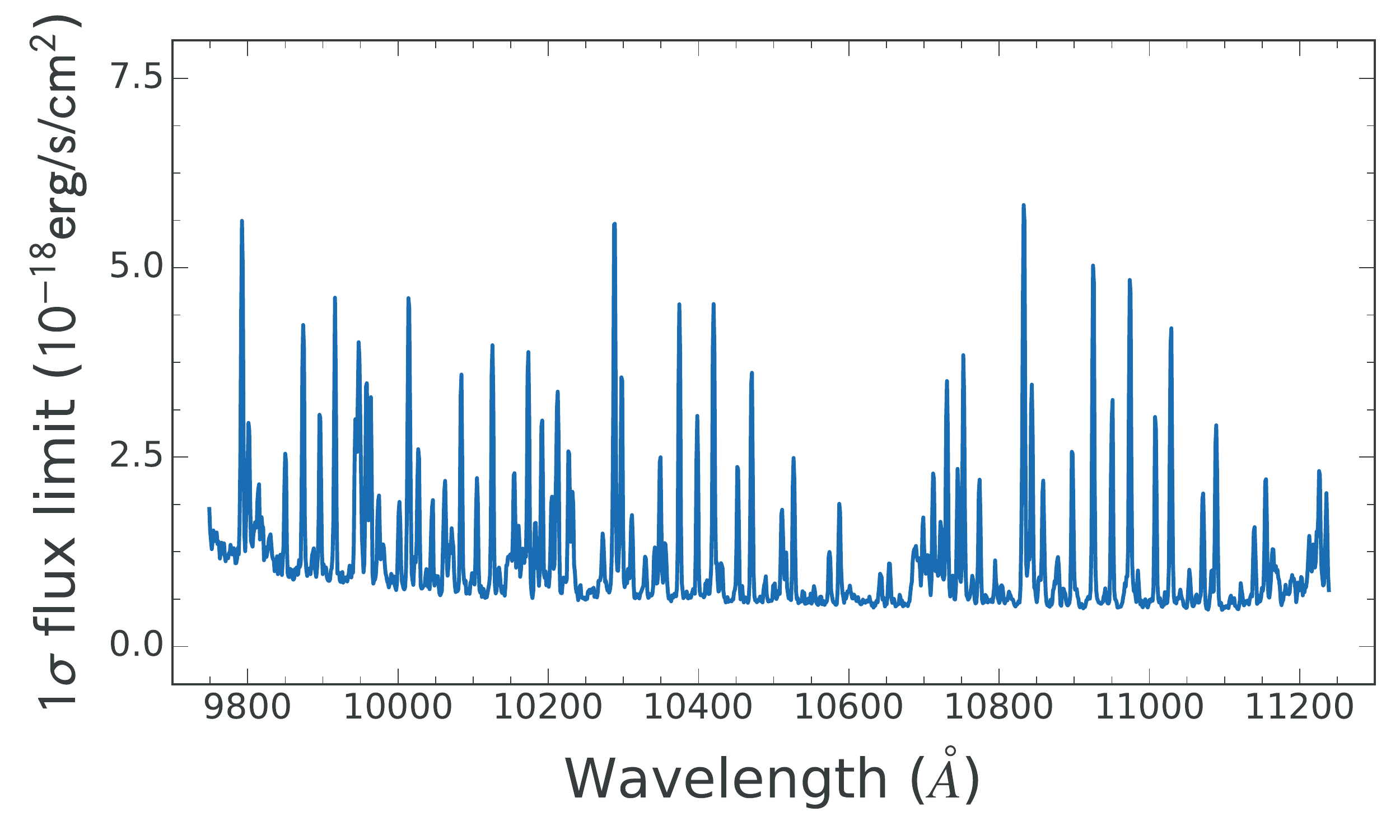}
    \caption{Median $1\sigma$ flux limit of all \nhighz{} targets in the MOSFIRE sample. The limits shown here are calculated assuming unresolved lines using a spectral bandpass of twice the FWHM of the MOSFIRE Y-band grating. We use resolved lines when computing the likelihood on the neutral fraction for each object, which slightly decreases the sensitivity. The spikes in sensitivity at certain wavelength are due to the sky emission, and the variation is accounted for in the Bayesian framework from which we infer the neutral hydrogen fraction (Mason et al. 2019, submitted). The median flux limits vary from $\sim1-5 \times 10^{-18} \mathrm{erg/s/cm^2}$ among the \nhighz{} targets, assuming unresolved lines.}
 \label{fig:fluxlimit}
 \end{figure*}

While we likely only detect \lya{} in 2 of the \nhighz{} galaxies in our sample, our non-detections are useful constraints on the neutral fraction. An individual $1\sigma$ flux limit spectrum, $f_{\mathrm{lim}} (\lambda)$, is obtained from the scaled noise spectrum of each object, $\sigma(\lambda)_{\mathrm{MOS, scaled}}$, via:

\begin{equation} f_{\mathrm{lim}} (\lambda) =  \Delta \lambda \times \sqrt{2*\mathrm{FWHM} / \Delta \lambda} \times \sigma(\lambda)_{\mathrm{MOS, scaled}}, \end{equation}
where $\Delta \lambda = 1.086$\AA{} is the spectral pixel size and $\mathrm{FWHM}$ is the full-width at half maximum of the MOSFIRE Y-band spectral resolution, $\sim3$\AA{}. 
We show the median flux limit spectrum of the sample in Figure~\ref{fig:fluxlimit}. An individual $1\sigma$ rest-frame \lya{} equivalent width spectrum is obtained from the flux limit spectrum via:

\begin{equation}  \mathrm{EW_{Ly\alpha}} =  f_{\mathrm{lim}} (\lambda) / f_{\mathrm{cont}} / (1+z) \end{equation}
where $f_{\mathrm{cont}}$ is the continuum flux density measured in F160W. We note that F160W samples the rest-UV of galaxies in our redshift range without encompassing the \lya{} line. The neutral fraction inference uses the entire flux density spectrum in the calculation of the likelihood for each object, which takes into account the variable sensitivity of the spectrum (Mason et al. 2019, submitted). 

 %%%%%%% Figure 6: SED and P(z) of LAE in MACS0744 %%%%%%%

 %%%%%%% Figure 7: SED and P(z) of LAE in RXJ1347 %%%%%%%

 %%%%%%% Figure 8: F160W + IRAC Postage Stamps of RXJ1347 LAE %%%%%%%

%%%%%%%%%%%%%%%%%%%%%%%%%%%%%%%%
\begin{deluxetable}{ccc}
				\tablecaption{Properties of \macssampleid{}}
				\tablecolumns{3}
				\tablewidth{0pt}
				\scriptsize
				\tablehead{Quantity & Unit & Value}
\startdata
$\alpha_{\mathrm{J2000}}$ & (degree) & 116.24648 \\
$\delta_{\mathrm{J2000}}$ & (degree) & 39.46042 \\
$F160W$ & mag & $27.17\pm0.38$ \\
$[3.6]$ & mag & $<25.54$ \\
$[4.5]$ & mag & $<25.37$ \\
$\mu_{\mathrm{best}}$ & & $3.2\pm0.1$ \\
$M_{UV} - 2.5 \mathrm{log}_{10}(\mu / \mu_{\mathrm{best}})$ & mag & $-18.5\pm0.4$ \\
$L_{UV} \times \mu / \mu_{\mathrm{best}} $ & $L_{\star}$ & $0.12^{+0.06}_{-0.04}$ \\
$z_{\mathrm{phot}}$ & & $6.9^{+0.5}_{-0.2}$ \\
$z_{Ly\alpha}$ & & $7.148\pm0.001$ \\
$\mathrm{SFR} \times \mu/\mu_{\mathrm{best}}$ & $M_{\odot}/\mathrm{yr}$ & $0.66^{+0.18}_{-0.11}$ \\
$M_{\star} \times \mu/\mu_{\mathrm{best}}$ & $10^7 M_{\odot}$ & $1.11^{+0.35}_{-0.28}$ \\
Age & $\mathrm{Myr}$ & $17.38^{+5.53}_{-6.91}$ \\
E(B-V) & mag & $<0.01$ \\
\hline
$f_{,Ly\alpha}$ & $10^{-18} \mathrm{erg\, s^{-1}\, cm^{-2}}$ & $7.1\pm0.7$ \\
$W_{0,Ly\alpha}$ & \AA{} & $58.3\pm25.1$ \\
$\mathrm{FWHM_{Ly\alpha}}$ & \AA{} & $5.4\pm1.3$ \\
$S/N_{int,Ly\alpha}$ & & 10.3
\enddata
\label{tab:m0744_tab}
\tablecomments{IRAC $[3.6]$ and $[4.5]$ magnitude limits are $3\sigma$. The factor $\mu / \mu_{\mathrm{best}}$ is introduced to show how the magnification factor enters into some properities, where $\mu_{\mathrm{best}}$ is the best-fit magnification value measured in this work. This factor allows one to re-calculate the value of a given property provided with a new magnification factor. $M_{UV}$ and $L_{UV}$ are calculated using the $F160W$ magnitude to approximate the rest-frame $UV$ luminosity. $L_{\star}$ is calculated from $M_{UV}^{\star}=20.87\pm0.26$ measured by \citet{Bouwens+15a} at $z=6.8$.}
\end{deluxetable}
\begin{deluxetable}{ccc}
				\tablecaption{Properties of \rxjsampleid{}}
				\tablecolumns{3}
				\tablewidth{0pt}
				\scriptsize
				\tablehead{Quantity & Unit & Value}
\startdata
$\alpha_{\mathrm{J2000}}$ & (degree) & 206.89124 \\
$\delta_{\mathrm{J2000}}$ & (degree) & -11.75261 \\
$F160W$ & mag & $26.43 \pm 0.14$ \\
$[3.6]$ & mag & $24.47 \pm 0.17$ \\
$[4.5]$ & mag & $24.07 \pm 0.11$ \\
$\mu_{\mathrm{best}}$ & & $21.4^{+1.7}_{-1.3}$ \\
$M_{UV} - 2.5 \mathrm{log}_{10}(\mu / \mu_{\mathrm{best}})$ & mag & $-17.2\pm0.2$ \\
$L_{UV} \times \mu/\mu_{\mathrm{best}}$ & $L_{\star}$ & $0.03\pm 0.01$ \\
$z_{\mathrm{phot}}$ & & $7.5^{+0.3}_{-0.7}$ \\
$z_{Ly\alpha}$ & & $7.161\pm0.001$ \\
$\mathrm{SFR} \times \mu/\mu_{\mathrm{best}}$ & $M_{\odot}/\mathrm{yr}$ & $3.40^{+1.45}_{-0.59}$ \\
$M_{\star} \times \mu/\mu_{\mathrm{best}}$ & $10^7 M_{\odot}$ & $143.34^{+26.38}_{-35.18}$ \\
Age & $\mathrm{Myr}$ & $640.47^{+78.15}_{-131.73}$ \\
$(\mathrm{SFR} \times \mu/\mu_{\mathrm{best}})_{\mathrm{HST}}$ & $M_{\odot}/\mathrm{yr}$ & $1.59^{+3.63}_{-1.05}$ \\
$(M_{\star} \times \mu/\mu_{\mathrm{best}})_{\mathrm{HST}} $ & $10^7 M_{\odot}$ & $14.53^{+41.37}_{-11.88}$ \\
$\mathrm{(Age)_{HST}}$ & $\mathrm{Myr}$ & $321.00^{+397.62}_{-310.52}$ \\
E(B-V) & mag & $<0.35$ \\
\hline 
$f_{,Ly\alpha}$ & $10^{-18} \mathrm{erg\, s^{-1}\, cm^{-2}}$ & $6.6\pm1.3$ \\
$W_{0,Ly\alpha}$ & \AA{} & $27.2 \pm 6.5$ \\
$\mathrm{FWHM_{Ly\alpha}}$ & \AA{} & $7.6\pm1.3$ \\
$S/N_{int,Ly\alpha}$ & & 5.1
\enddata
\label{tab:rxj1347_tab}
\tablecomments{Quantities with subscript $\mathrm{HST}$ are derived from the best-fit SED to HST photometry only, whereas the other quantities include the IRAC photometry. The uncertainty on $\mathrm{FWHM_{Ly\alpha}}$ does not include additional uncertainty introduced due to the sky emission line blueward of the emission peak. }
\end{deluxetable}

%%%%%%% Figure 6: Flux limits %%%%%%%

\subsection{Neutral fraction inference}
\label{sec:neutral_fraction}

Here we use the formalism introduced by \citet{Treu+12} and extended by \citet[][Mason et al. 2019, submitted]{Mason+18a} including models by \cite{Mesinger+15} to infer the volume-averaged neutral hydrogen fraction in the IGM at $z\sim7.5$ using our \lya{} spectroscopy from MOSFIRE. We adopt $z=7.6$ as our fiducial redshift as it is the average \lya{} redshift probed by the MOSFIRE Y-band. The formalism uses the Evolution of 21cm Structure cosmological simulations \citep{Mesinger+15,Mesinger+16}, i.e. cosmological-scale IGM simulations of inhomogeneous reionization, to obtain the global neutral fraction distribution as well as a realistic treatment of the impact of the ISM and CGM on the emerging \lya{} line. 

The model yields a likelihood that can be used to interpret
observables of LBGs in terms of neutral fraction. The input
observables are the rest-frame \lya{} equivalent width measurements
and limits (for non-detections) as well as galaxy luminosities. For
non-detections, we use the entire flux density and noise spectra,
taking into account the varying sensitivity as a function of
wavelength (see Figure~\ref{fig:fluxlimit}). 

Our ignorance of the underlying \lya{} FWHM distribution of the objects without \lya{} detections introduces uncertainty into the inferred neutral fraction. Furthermore, larger assumed FWHMs result in weaker constraints due to the line flux being spread out over more detector pixels. To parameterize our ignorance on the FWHM, we marginalize over it when calculating the posterior for each object. We explored three different priors for the FWHM distribution: 1) A uniform prior between $100-400~\mathrm{km/s}$ for each object, 2) a log-normal $p(\mathrm{FWHM} | M_{UV})$ that uses the \citet{Mason+18a} scaling relation between \lya{} velocity offset and $M_{UV}$ and the \citet{Verhamme+18} scaling relation between velocity offset and FWHM, where we adopt the median $M_{UV}$ of the sample for each object, 3) the same functional form and scaling relations as in (2), but where we use the individual $M_{UV}$ for each galaxy rather than the median of the sample in the scaling relations. We found that the result was robust against the choice of prior, so we adopted case (3) as it is the most physically motivated. The log-normal tends to peak around $50-100~\mathrm{km/s}$ for the $M_{UV}$ values in the MOSFIRE sample. For more details and the form of the full likelihood, see \citet{Mason+18a}. 

We also use the photometric redshift
probability distributions $P(z)$ as priors in the likelihood. Effectively, we weight each galaxy in the inference, where the weight applied is given by the probability that \lya{} falls within the MOSFIRE
Y-band, i.e. $P(7<z<8.2)$. Doing so accounts for the
photometric redshift impurity of our sample. Galaxy intrinsic luminosities
enter into the inference via the halo-mass dependence (and hence
luminosity dependence) of the local neutral fraction in the IGM
simulations. Furthermore, the model assumes an intrinsic emitted EW distribution at $z\sim6$ before any attenuation. This model is based off of real observations at $z\sim6$ compiled by \citet{deBarros+17}. For more details see \citet{Mason+18a} and Mason et al. (2019, submitted).

After applying the prior, we obtain the posterior, $p(\overline{x}_{\mathrm{HI}} | \{\mathrm{f}(\lambda),m,\mu\}$) where the data consist of the set of flux density spectra, $\mathrm{f}(\lambda)$, apparent magnitudes, $m$, and magnifications, $\mu$. For objects where there are detections, we use the measured equivalent width of the line rather than the entire flux spectrum to construct the posterior.

\begin{figure}[htb]
    \centering
    \includegraphics[width=\linewidth]{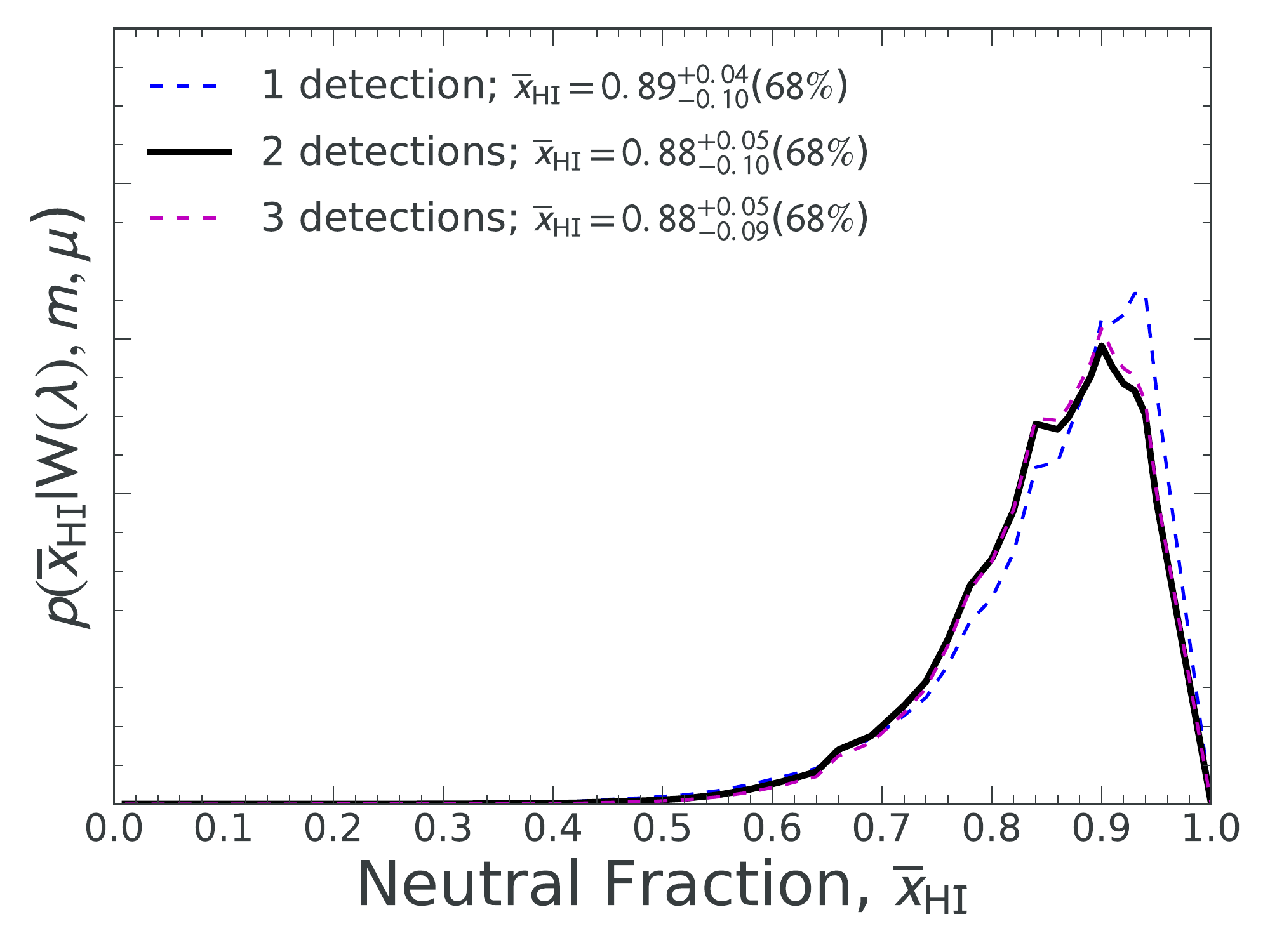}
    \caption{Inference on the volume-averaged neutral hydrogen fraction, \xhi{}, at $z=7.6\pm0.6$ using the MOSFIRE sample of \nhighz{} galaxies. We infer a neutral fraction of \xhibest{} in our fiducial case (2/\nhighz{} \lya{} detections; solid black). We show the posterior distributions for the neutral fraction for 2 other scenarios (see Section~\ref{sec:lya_search}):  1/\nhighz{} \lya{} detection (blue dashed), and 3/\nhighz{} \lya{} detections (magenta dashed). All cases have one detected object in common: \macssampleid{}, the \lya{} detection in which we are most confident ($S/N_{int}\sim10$). The fact that the difference between the scenarios is very small illustrates the influence that \macssampleid{} has on the inference, and the unimportance of low $S/N$ detections. }
 \label{fig:posterior}
 \end{figure}

We show our posterior on the volume-averaged neutral fraction \xhi{} in Figure~\ref{fig:posterior}. Using our fiducial sample of 2 \lya{} detections (see Section~\ref{sec:lya_search}), we infer a neutral fraction of \xhibest{}. Our result is robust against the set of \lya{} detections we choose, as long as the \macssampleid{} detection is included. Excluding the \rxjsampleid{} detection  (the 1 detection case), we infer a statistically consistent neutral fraction to the fiducial case: \xhione{}. If instead we include the additional \lya{} detection presented by \citet{Hoag+17} (the 3 detection case), we also infer a statistically consistent neutral fraction: \xhithree{}. The difference between the three cases shown is small because in all cases the \macssampleid{} detection is included. This detection is $S/N\sim10$, so it has much larger statistical weight in the inference than the other detections. While we find the 0 detection case unlikely, we cannot rule it out because there is a small chance that the single emission line in each of the three spectra is not \lya{}. If this were the case, we would put a lower limit on neutral fraction. 

\section{Discussion}
\label{sec:discussion}
\begin{figure}[htb]
    \centering
    \includegraphics[width=\linewidth]{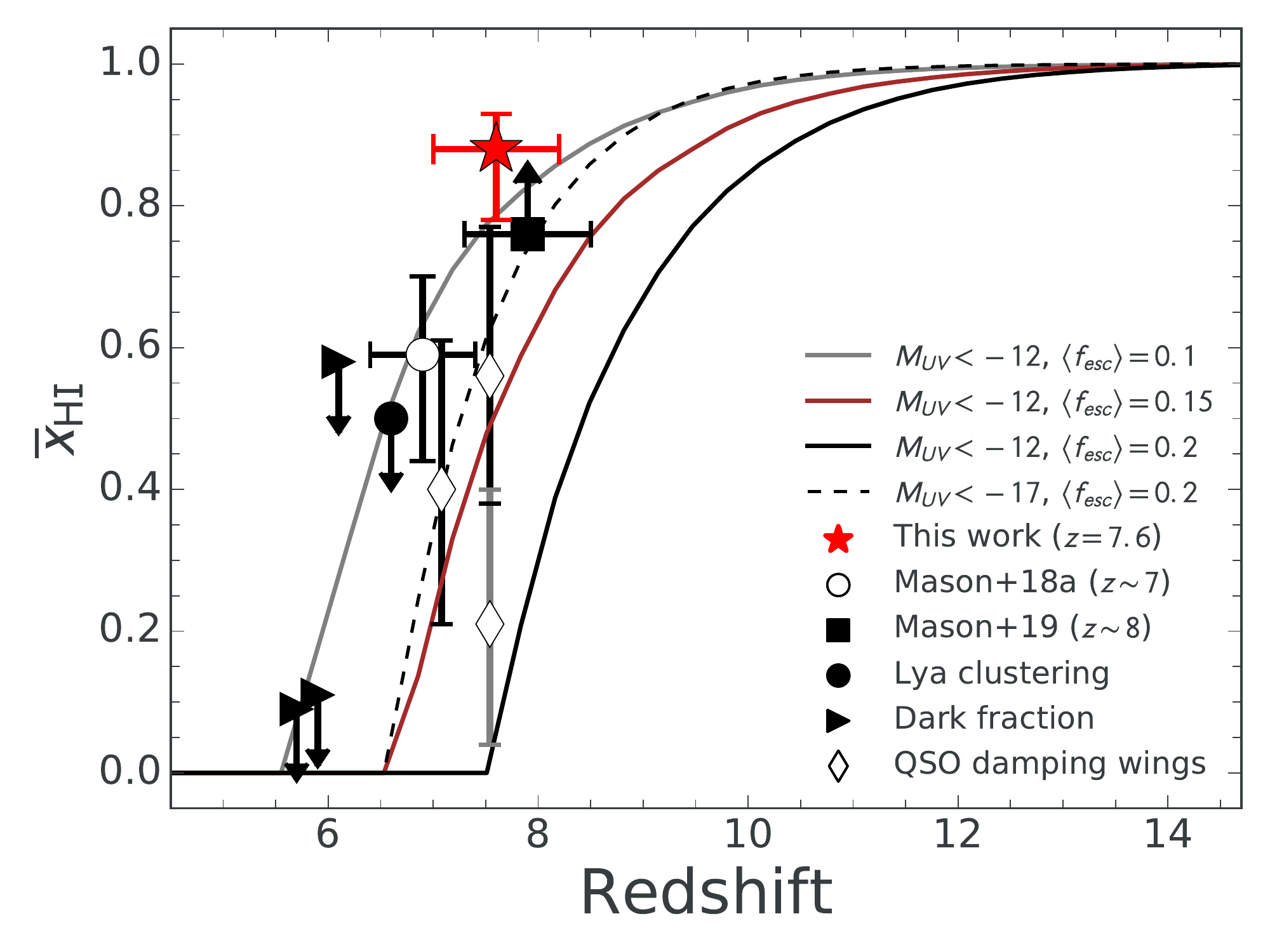}
    \caption{Constraints on the volume-averaged neutral hydrogen fraction, \xhi{}, during reionization, including our constraint at $z=7.6$ of \xhibest{} (red star). The four curves correspond to different reionization histories derived from the \citet{Mason+15} luminosity function (LF) models. The two parameters of these models are the mean Lyman-continuum escape fraction, $\langle f_{\mathrm{esc}} \rangle$, and the faint-end cut-off luminosity when extrapolating the LF. The data collectively exclude reionization histories with both large mean escape fractions ($\langle f_{\mathrm{esc}} \rangle \gtrsim 0.2$ with faint cutoff luminosities $M_{UV} \lesssim -12$, which would result in an earlier reionization. The QSO damping wing measurement of $\overline{x}_{\mathrm{HI}} = 0.56^{+0.21}_{-0.18}$ at $z=7.54$ by \citet{Banados+18} (black open diamond) and the re-measurement from the same QSO by \citet{Greig+18} (gray open diamond) are both shown.  }
 \label{fig:neutral_fraction}
 \end{figure}

Our fiducial result of \xhibest{} implies a universe at $z\sim7.5$ that is mostly neutral. This result is the most precise constraint on \xhi{} at $z\gtrsim7$, and it is consistent with the emerging picture from other probes of reionization implying a ``late'' and rapid reionization scenario. In Figure~\ref{fig:neutral_fraction}, we show published constraints on the neutral fraction from other authors. These constraints come from various independent approaches: 1) the ``dark fraction'' in quasar spectra, which provide upper limits on \xhi{} at $z=5.7, 5.9, 6.1$ \citep{Mcgreer+15}, 2) an upper limit from \lya{} clustering measurements at $z=6.6$ \citep{Ouchi+10},  3) measurements of the \lya{} damping wings of QSOs at $z=7.08$ \citep{Greig+17a} and $z=7.54$ \citep{Banados+18}, and 4) the same approach as in this work done by \citet{Mason+18a} at $z=6.9\pm0.5$ and Mason et al. (2019, submitted) at $z=7.9\pm0.6$. Specifically, \citet{Mason+18a} used \lya{} spectroscopy of 68 LBGs at $z\sim7$ compiled by \cite{Pentericci+14}, spanning a wide range in intrinsic luminosity ($-22.75\lesssim M_{UV} \lesssim -17.8$) over multiple sight lines. Mason et al. (2019, submitted) used the KMOS IFU to target 53 lensed galaxies at $z\sim8$, providing an excellent consistency check with our work.

The closest measurement in redshift to ours is by \citet{Banados+18}, who discovered the bright QSO ULASJ1342+0928 at $z=7.54$, inferring a neutral fraction of $\overline{x}_{\mathrm{HI}} = 0.56^{+0.21}_{-0.18}$ from the quasar's \lya{} damping wing. \citet{Greig+18} performed an independent analysis of this object using a complementary technique and found a lower fraction: $\overline{x}_{\mathrm{HI}} = 0.21^{+0.17}_{-0.19}$. Taken together, these two inferences from the same object span a large range in the neutral fraction at $z\sim7.5$, which are lower than our neutral fraction constraint, albeit with large uncertainties. At this point, the uncertainties are large enough that the results are not in significant statistical disagreement with our result. We also point out that their measurement is derived from a single QSO, which probes only a single line of sight through the IGM. 

Because the hosts of QSOs such as the ones detected by \citet{Greig+17a} and \citet{Banados+18} are probably very massive halos, one might expect them to reside in more ionized regions of the universe at $z>7$. The authors attempt to account for this bias by linking their observations to halos in a global reionization simulation, as done in this work.  Similarly, \citet{Mason+18a} targeted a higher luminosity sample of galaxies, and rely on the same reionization simulations in this work to infer the global neutral fraction. We stress that using a large sample of low luminosity galaxies, as achieved in this work with gravitational lensing, allows us to sample the neutral fraction from more typical-luminosity halos directly, rather than relying on the model to extrapolate using only the rarer, massive halos. In future work, we plan to use the full range of halo masses probed by the data in a consistent manner to infer the neutral fraction.

We also compare our result to the inferences by \citet{Mason+18a} at $z\sim7$ and Mason et al. (2019, submitted) at $z\sim8$, who used the same framework as us to infer the neutral fraction. It is reassuring that we find a more neutral universe at $z\sim7.5$ than their result at $z\sim7$. Similarly, their lower limit at $z\sim8$ of $>0.76$ ($68\%$ confidence) is consistent with our inference. 

We note that the range in redshift probed by our result is large, covering $z=7-8.2$, a period over which the neutral fraction is likely rapidly evolving. In fact, both of our \lya{} detections are at $z<7.2$, hinting at evolution within our own data set. We opted to not bin our data in redshift more finely due to the already large uncertainty in a single redshift bin. Combining our data with other similar data sets in the future may allow us to meaningfully separate the data in redshift.

In Figure~\ref{fig:neutral_fraction}, we show various reionization histories derived from the \citet{Mason+15} luminosity function (LF) models. The two parameters of these models that we vary are the mean Lyman-continuum escape fraction, $\langle f_{\mathrm{esc} }\rangle$, and the cut-off luminosity when extrapolating the LF. To construct these histories, we also assume a log-normal ionizing efficiency, $\xi_{\mathrm{ion}}$ with mean 25.2 and standard deviation of 0.15 dex, as well as a uniform distribution for the clumping factor over the range $C=1-6$. Estimates from our data and those from similar works exclude earlier reionization histories which require large mean escape fractions ($\langle f_{\mathrm{esc}} \rangle \gtrsim 0.2$) and faint cutoff luminosities ($M_{UV} \lesssim -12$). Such scenarios result in an earlier reionization due to a higher abundance of ionizing photons at higher redshift. The escape fraction and cutoff luminosity are somewhat degenerate from the allowed reionization histories. For example, a brighter cutoff luminosity with large $\langle f_{\mathrm{esc}} \rangle$ has a similar reionization history to one with a fainter cutoff luminosity but smaller $\langle f_{\mathrm{esc}} \rangle$. With current data, the two scenarios cannot be distinguished. In future work, we hope to constrain the allowed parameter space of $\langle f_{\mathrm{esc}} \rangle$ and cutoff luminosity using all available constraints on the neutral fraction. 

The escape fraction is notoriously difficult to measure, even at lower redshift. At redshifts during reionization, ionizing radiation is efficiently absorbed by the neutral hydrogen in the IGM. The cutoff luminosity, on the other hand, will be more tightly constrained via extremely deep surveys with the \emph{James Webb Space Telescope} (JWST), though the exact value is likely deeper than even JWST will probe. Currently, there is no strong evidence that the $z>6$ LF deviates from a very steep faint-end slope \citep[$\alpha \lesssim -2$;][]{Finkelstein+15,Livermore+16,Bouwens+17}. A hopeful scenario is if the cutoff luminosity can be well-constrained by future observations and uncertainties on measurements of the neutral fraction such as the one presented here are reduced. In this case, the escape fraction could be inferred. This is one example in which constraining the neutral fraction can inform the properties of galaxies at the earliest epochs.

Our neutral fraction constraint is among only a handful at $z>7$, each with large uncertainties. To better constrain the reionization timeline, and ultimately the sources of the reionization, more constraints at $z\sim7-8$ and constraints at higher redshifts ($z\gtrsim8$) are needed. From the current data, it is clear that reionization is ongoing at $z\sim7-8$, so prospects of constraining the timeline with future surveys are bright. The method adopted in this work is readily deployed with JWST at higher redshift. In particular, the NIRSPEC instrument is equipped with a multi-object spectrograph similar to MOSFIRE that is able to probe to longer wavelengths ($\sim1-5~\mu m$) with higher efficiency. In analogy to the GLASS \lya{} survey \citep{Schmidt+16}, JWST NIRISS observations will also potentially allow a more complete follow-up of clusters via sensitive grism spectroscopy due to its simultaneous full field coverage and lack of slit losses. While the number of photometric targets at $z\gtrsim8$ per cluster is currently small ($<5$/cluster), JWST NIRCAM will find larger samples with its superior sensitivity to \HST{}, making an analysis akin to this work at $z\gtrsim8$ a possibility. 

%%%%%%%% Figure 8: Neutral fraction posterior %%%%%%%%

%%%%%%%%%%%%%%%%%%%%%%%%%%%%% 

%%%%%%%% Figure 9: Neutral fraction vs. z and literature %%%%%%%%

%%%%%%%%%%%%%%%%%%%%%%%%%%%%% 

\section{Summary}
\label{sec:summary}
We presented a spectroscopic campaign targeting lensed LBGs in \ncluster{} galaxy cluster fields using Keck MOSFIRE. Using a consistent photometric selection across the \ncluster{} fields, we obtained a sample of 70 LBGs. 2 of these were confirmed to be at $6<z<7$ by a different spectroscopic campaign running simultaneously with our campaign, so we do not include those in our inference on the neutral fraction. Among the remaining \nhighz{} targets, we identified 2 probable \lya{} emitters, one at $z=7.148\pm0.001$ in MACS0744 and the other at $7.161\pm0.001$ in RXJ1347. Both \lya{} emitters are faint ($M_{UV}=-18.5\pm0.4$ and $M_{UV}=-17.2\pm0.2$, respectively). We also obtained sensitive \lya{} flux and equivalent width limits for the targets from which we did not detect \lya{}.

We used the Bayesian framework of \citet[][Mason et al. 2019, submitted]{Mason+18a} to infer the volume-averaged neutral hydrogen fraction, \xhi{}, from our \lya{} spectroscopy, taking into account the luminosity of each galaxy when linking to simulated halos in a realistic large-scale reionization simulation. We inferred a neutral hydrogen fraction of \xhibest{} at $z=7.6\pm0.6$. Our result favors a late reionization scenario, consistent more so with the \citet{Planck+16} optical depth to reionization than the higher optical depth inferred from the full-depth WMAP dataset. With larger samples on current and future telescopes, this method holds promise to precisely constrain the evolution of the hydrogen neutral fraction and ultimately the timeline of cosmic reionization.

\vspace*{0.5cm}

\acknowledgments{The data presented herein were obtained at the W.M. Keck Observatory, which is operated as a scientific partnership among the California Institute of Technology, the University of California and NASA. The observatory was made possible by the generous financial support of the W.M. Keck Foundation. The authors thank L. Rizzi, M. Kassis, and S. Yeh for help with the Multi-Object Spectrometer
for Infra-Red Exploration (MOSFIRE) observations and data reduction. The authors recognize and acknowledge the very significant cultural role and reverence that the summit of Maunakea has always had within the indigenous Hawaiian community.
We are most fortunate to have the opportunity to conduct observations from this mountain. Support for this work was provided by NASA through an award issued by JPL/Caltech (for SURFS UP project) and by {\HST}/STScI: {\HST}-AR-13235, {\HST}-AR-14280, {\HST}-GO-13177, and HST-GO-15212. Support was also provided by the National Science Foundation through grant: NSF-AAG-1815458. This research was also supported in part by the Australian Research Council Centre of Excellence for All Sky Astrophysics in 3 Dimensions (ASTRO 3D), through project number CE170100013. This work utilizes gravitational lensing models produced by PIs Brada\v{c}, Richard, Natarajan \& Kneib, Sharon, Williams, Keeton, Bernstein and Diego, and Oguri. This lens modeling was partially funded by the HST Frontier Fields program conducted by STScI. STScI is operated by the Association of Universities for Research in Astronomy, Inc. under NASA contract NAS 5-26555. The lens models were obtained from the Mikulski Archive for Space Telescopes (MAST). CM acknowledges support provided by NASA through the NASA Hubble Fellowship grant HST-HF2-51413.001-A awarded by the Space Telescope Science Institute, which is operated by the Association of Universities for Research in Astronomy, Inc., for NASA, under contract NAS5-26555. }

% \bibliography{ms.bib}

\begin{thebibliography}{}
\expandafter\ifx\csname natexlab\endcsname\relax\def\natexlab#1{#1}\fi

\bibitem[{{Ba{\~n}ados} {et~al.}(2018){Ba{\~n}ados}, {Venemans},
  {Mazzucchelli}, {Farina}, {Walter}, {Wang}, {Decarli}, {Stern}, {Fan},
  {Davies}, {Hennawi}, {Simcoe}, {Turner}, {Rix}, {Yang}, {Kelson}, {Rudie}, \&
  {Winters}}]{Banados+18}
{Ba{\~n}ados}, E., {Venemans}, B.~P., {Mazzucchelli}, C., {et~al.} 2018, \nat,
  553, 473

\bibitem[{{Becker} {et~al.}(2001){Becker}, {Fan}, {White}, {Strauss},
  {Narayanan}, {Lupton}, {Gunn}, {Annis}, {Bahcall}, {Brinkmann}, {Connolly},
  {Csabai}, {Czarapata}, {Doi}, {Heckman}, {Hennessy}, {Ivezi{\'c}}, {Knapp},
  {Lamb}, {McKay}, {Munn}, {Nash}, {Nichol}, {Pier}, {Richards}, {Schneider},
  {Stoughton}, {Szalay}, {Thakar}, \& {York}}]{Becker+01}
{Becker}, R.~H., {Fan}, X., {White}, R.~L., {et~al.} 2001, \aj, 122, 2850

\bibitem[{{Ben{\'{\i}}tez}(2000)}]{Benitez+00}
{Ben{\'{\i}}tez}, N. 2000, \apj, 536, 571

\bibitem[{{Bertin} \& {Arnouts}(1996)}]{Bertin+96}
{Bertin}, E., \& {Arnouts}, S. 1996, \aaps, 117, 393

\bibitem[{{Bouwens} {et~al.}(2015{\natexlab{a}}){Bouwens}, {Illingworth},
  {Oesch}, {Caruana}, {Holwerda}, {Smit}, \& {Wilkins}}]{Bouwens+15b}
{Bouwens}, R.~J., {Illingworth}, G.~D., {Oesch}, P.~A., {et~al.}
  2015{\natexlab{a}}, \apj, 811, 140

\bibitem[{{Bouwens} {et~al.}(2017){Bouwens}, {Oesch}, {Illingworth}, {Ellis},
  \& {Stefanon}}]{Bouwens+17}
{Bouwens}, R.~J., {Oesch}, P.~A., {Illingworth}, G.~D., {Ellis}, R.~S., \&
  {Stefanon}, M. 2017, \apj, 843, 129

\bibitem[{{Bouwens} {et~al.}(2015{\natexlab{b}}){Bouwens}, {Illingworth},
  {Oesch}, {Trenti}, {Labb{\'e}}, {Bradley}, {Carollo}, {van Dokkum},
  {Gonzalez}, {Holwerda}, {Franx}, {Spitler}, {Smit}, \& {Magee}}]{Bouwens+15a}
{Bouwens}, R.~J., {Illingworth}, G.~D., {Oesch}, P.~A., {et~al.}
  2015{\natexlab{b}}, \apj, 803, 34

\bibitem[{{Brada{\v c}} {et~al.}(2005){Brada{\v c}}, {Schneider}, {Lombardi},
  \& {Erben}}]{Bradac+05}
{Brada{\v c}}, M., {Schneider}, P., {Lombardi}, M., \& {Erben}, T. 2005, \aap,
  437, 39

\bibitem[{{Brada{\v c}} {et~al.}(2009){Brada{\v c}}, {Treu}, {Applegate},
  {Gonzalez}, {Clowe}, {Forman}, {Jones}, {Marshall}, {Schneider}, \&
  {Zaritsky}}]{Bradac+09}
{Brada{\v c}}, M., {Treu}, T., {Applegate}, D., {et~al.} 2009, \apj, 706, 1201

\bibitem[{{Brada{\v c}} {et~al.}(2014){Brada{\v c}}, {Ryan}, {Casertano},
  {Huang}, {Lemaux}, {Schrabback}, {Gonzalez}, {Allen}, {Cain}, {Gladders},
  {Hall}, {Hildebrandt}, {Hinz}, {von der Linden}, {Lubin}, {Treu}, \&
  {Zaritsky}}]{Bradac+14}
{Brada{\v c}}, M., {Ryan}, R., {Casertano}, S., {et~al.} 2014, \apj, 785, 108

\bibitem[{{Brammer} {et~al.}(2008){Brammer}, {van Dokkum}, \&
  {Coppi}}]{Brammer+08}
{Brammer}, G.~B., {van Dokkum}, P.~G., \& {Coppi}, P. 2008, \apj, 686, 1503

\bibitem[{{Brammer} {et~al.}(2016){Brammer}, {Marchesini}, {Labb{\'e}},
  {Spitler}, {Lange-Vagle}, {Barker}, {Tanaka}, {Fontana}, {Galametz},
  {Ferr{\'e}-Mateu}, {Kodama}, {Lundgren}, {Martis}, {Muzzin}, {Stefanon},
  {Toft}, {van der Wel}, {Vulcani}, \& {Whitaker}}]{Brammer+16}
{Brammer}, G.~B., {Marchesini}, D., {Labb{\'e}}, I., {et~al.} 2016, \apjs, 226,
  6

\bibitem[{{Bruzual} \& {Charlot}(2003)}]{BC03}
{Bruzual}, G., \& {Charlot}, S. 2003, \mnras, 344, 1000

\bibitem[{{Calzetti} {et~al.}(2000){Calzetti}, {Armus}, {Bohlin}, {Kinney},
  {Koornneef}, \& {Storchi-Bergmann}}]{Calzetti+00}
{Calzetti}, D., {Armus}, L., {Bohlin}, R.~C., {et~al.} 2000, \apj, 533, 682

\bibitem[{{Castellano} {et~al.}(2016){Castellano}, {Amor{\'{\i}}n}, {Merlin},
  {Fontana}, {McLure}, {M{\'a}rmol-Queralt{\'o}}, {Mortlock}, {Parsa},
  {Dunlop}, {Elbaz}, {Balestra}, {Boucaud}, {Bourne}, {Boutsia}, {Brammer},
  {Bruce}, {Buitrago}, {Capak}, {Cappelluti}, {Ciesla}, {Comastri}, {Cullen},
  {Derriere}, {Faber}, {Giallongo}, {Grazian}, {Grillo}, {Mercurio},
  {Micha{\l}owski}, {Nonino}, {Paris}, {Pentericci}, {Pilo}, {Rosati},
  {Santini}, {Schreiber}, {Shu}, \& {Wang}}]{Castellano+16b}
{Castellano}, M., {Amor{\'{\i}}n}, R., {Merlin}, E., {et~al.} 2016, \aap, 590,
  A31

\bibitem[{{Chabrier}(2003)}]{Chabrier03}
{Chabrier}, G. 2003, \pasp, 115, 763

\bibitem[{{Coe} {et~al.}(2006){Coe}, {Ben{\'{\i}}tez}, {S{\'a}nchez}, {Jee},
  {Bouwens}, \& {Ford}}]{Coe+06}
{Coe}, D., {Ben{\'{\i}}tez}, N., {S{\'a}nchez}, S.~F., {et~al.} 2006, \aj, 132,
  926

\bibitem[{{De Barros} {et~al.}(2017){De Barros}, {Pentericci}, {Vanzella},
  {Castellano}, {Fontana}, {Grazian}, {Conselice}, {Yan}, {Koekemoer},
  {Cristiani}, {Dickinson}, {Finkelstein}, \& {Maiolino}}]{deBarros+17}
{De Barros}, S., {Pentericci}, L., {Vanzella}, E., {et~al.} 2017, \aap, 608,
  A123

\bibitem[{{Fan} {et~al.}(2006){Fan}, {Strauss}, {Becker}, {White}, {Gunn},
  {Knapp}, {Richards}, {Schneider}, {Brinkmann}, \& {Fukugita}}]{Fan+06}
{Fan}, X., {Strauss}, M.~A., {Becker}, R.~H., {et~al.} 2006, \aj, 132, 117

\bibitem[{{Finkelstein} {et~al.}(2015){Finkelstein}, {Ryan}, {Papovich},
  {Dickinson}, {Song}, {Somerville}, {Ferguson}, {Salmon}, {Giavalisco},
  {Koekemoer}, {Ashby}, {Behroozi}, {Castellano}, {Dunlop}, {Faber}, {Fazio},
  {Fontana}, {Grogin}, {Hathi}, {Jaacks}, {Kocevski}, {Livermore}, {McLure},
  {Merlin}, {Mobasher}, {Newman}, {Rafelski}, {Tilvi}, \&
  {Willner}}]{Finkelstein+15}
{Finkelstein}, S.~L., {Ryan}, Jr., R.~E., {Papovich}, C., {et~al.} 2015, \apj,
  810, 71

\bibitem[{{Finney} {et~al.}(2018){Finney}, {Brada{\v c}}, {Huang}, {Hoag},
  {Morishita}, {Schrabback}, {Treu}, {Borello Schmidt}, {Lemaux}, {Wang}, \&
  {Mason}}]{Finney+18}
{Finney}, E.~Q., {Brada{\v c}}, M., {Huang}, K.-H., {et~al.} 2018, \apj, 859,
  58

\bibitem[{{Fontana} {et~al.}(2010){Fontana}, {Vanzella}, {Pentericci},
  {Castellano}, {Giavalisco}, {Grazian}, {Boutsia}, {Cristiani}, {Dickinson},
  {Giallongo}, {Maiolino}, {Moorwood}, \& {Santini}}]{Fontana+10}
{Fontana}, A., {Vanzella}, E., {Pentericci}, L., {et~al.} 2010, \apjl, 725,
  L205

\bibitem[{{Greig} {et~al.}(2018){Greig}, {Mesinger}, \&
  {Ba{\~n}ados}}]{Greig+18}
{Greig}, B., {Mesinger}, A., \& {Ba{\~n}ados}, E. 2018, ArXiv e-prints,
  arXiv:1807.01593

\bibitem[{{Greig} {et~al.}(2017){Greig}, {Mesinger}, {Haiman}, \&
  {Simcoe}}]{Greig+17a}
{Greig}, B., {Mesinger}, A., {Haiman}, Z., \& {Simcoe}, R.~A. 2017, \mnras,
  466, 4239

\bibitem[{{Grillo} {et~al.}(2016){Grillo}, {Karman}, {Suyu}, {Rosati},
  {Balestra}, {Mercurio}, {Lombardi}, {Treu}, {Caminha}, {Halkola}, {Rodney},
  {Gavazzi}, \& {Caputi}}]{Grillo+16}
{Grillo}, C., {Karman}, W., {Suyu}, S.~H., {et~al.} 2016, \apj, 822, 78

\bibitem[{{Gunn} \& {Peterson}(1965)}]{Gunn+1965}
{Gunn}, J.~E., \& {Peterson}, B.~A. 1965, \apj, 142, 1633

\bibitem[{{Haiman} \& {Loeb}(1998)}]{Haiman+98}
{Haiman}, Z., \& {Loeb}, A. 1998, \apj, 503, 505

\bibitem[{{Hashimoto} {et~al.}(2018){Hashimoto}, {Laporte}, {Mawatari},
  {Ellis}, {Inoue}, {Zackrisson}, {Roberts-Borsani}, {Zheng}, {Tamura},
  {Bauer}, {Fletcher}, {Harikane}, {Hatsukade}, {Hayatsu}, {Matsuda}, {Matsuo},
  {Okamoto}, {Ouchi}, {Pell{\'o}}, {Rydberg}, {Shimizu}, {Taniguchi},
  {Umehata}, \& {Yoshida}}]{Hashimoto+18}
{Hashimoto}, T., {Laporte}, N., {Mawatari}, K., {et~al.} 2018, \nat, 557, 392

\bibitem[{{Hinshaw} {et~al.}(2013){Hinshaw}, {Larson}, {Komatsu}, {Spergel},
  {Bennett}, {Dunkley}, {Nolta}, {Halpern}, {Hill}, {Odegard}, {Page}, {Smith},
  {Weiland}, {Gold}, {Jarosik}, {Kogut}, {Limon}, {Meyer}, {Tucker}, {Wollack},
  \& {Wright}}]{Hinshaw+13}
{Hinshaw}, G., {Larson}, D., {Komatsu}, E., {et~al.} 2013, \apjs, 208, 19

\bibitem[{{Hoag} {et~al.}(2015){Hoag}, {Brada{\v c}}, {Huang}, {Ryan},
  {Sharon}, {Schrabback}, {Schmidt}, {Cain}, {Gonzalez}, {Hildebrandt}, {Hinz},
  {Lemaux}, {von der Linden}, {Lubin}, {Treu}, \& {Zaritsky}}]{Hoag+15}
{Hoag}, A., {Brada{\v c}}, M., {Huang}, K.~H., {et~al.} 2015, \apj, 813, 37

\bibitem[{{Hoag} {et~al.}(2016){Hoag}, {Huang}, {Treu}, {Brada{\v c}},
  {Schmidt}, {Wang}, {Brammer}, {Broussard}, {Amorin}, {Castellano}, {Fontana},
  {Merlin}, {Schrabback}, {Trenti}, \& {Vulcani}}]{Hoag+16}
{Hoag}, A., {Huang}, K.-H., {Treu}, T., {et~al.} 2016, ArXiv e-prints,
  arXiv:1603.00505

\bibitem[{{Hoag} {et~al.}(2017){Hoag}, {Brada{\v c}}, {Trenti}, {Treu},
  {Schmidt}, {Huang}, {Lemaux}, {He}, {Bernard}, {Abramson}, {Mason},
  {Morishita}, {Pentericci}, \& {Schrabback}}]{Hoag+17}
{Hoag}, A., {Brada{\v c}}, M., {Trenti}, M., {et~al.} 2017, Nature Astronomy,
  1, 0091

\bibitem[{{Huang} {et~al.}(2016{\natexlab{a}}){Huang}, {Lemaux}, {Schmidt},
  {Hoag}, {Brada{\v c}}, {Treu}, {Dijkstra}, {Fontana}, {Henry}, {Malkan},
  {Mason}, {Morishita}, {Pentericci}, {Ryan}, {Trenti}, \& {Wang}}]{Huang+16b}
{Huang}, K.-H., {Lemaux}, B.~C., {Schmidt}, K.~B., {et~al.} 2016{\natexlab{a}},
  \apjl, 823, L14

\bibitem[{{Huang} {et~al.}(2016{\natexlab{b}}){Huang}, {Brada{\v c}}, {Lemaux},
  {Ryan}, {Hoag}, {Castellano}, {Amor{\'{\i}}n}, {Fontana}, {Brammer}, {Cain},
  {Lubin}, {Merlin}, {Schmidt}, {Schrabback}, {Treu}, {Gonzalez}, {von der
  Linden}, \& {Knight}}]{Huang+16a}
{Huang}, K.-H., {Brada{\v c}}, M., {Lemaux}, B.~C., {et~al.}
  2016{\natexlab{b}}, \apj, 817, 11

\bibitem[{{Johnson} {et~al.}(2014){Johnson}, {Sharon}, {Bayliss}, {Gladders},
  {Coe}, \& {Ebeling}}]{Johnson+14}
{Johnson}, T.~L., {Sharon}, K., {Bayliss}, M.~B., {et~al.} 2014, \apj, 797, 48

\bibitem[{{Katz} {et~al.}(2018){Katz}, {Laporte}, {Ellis}, {Devriendt}, \&
  {Slyz}}]{Katz+18}
{Katz}, H., {Laporte}, N., {Ellis}, R.~S., {Devriendt}, J., \& {Slyz}, A. 2018,
  ArXiv e-prints, arXiv:1809.07210

\bibitem[{{Labb{\'e}} {et~al.}(2013){Labb{\'e}}, {Oesch}, {Bouwens},
  {Illingworth}, {Magee}, {Gonz{\'a}lez}, {Carollo}, {Franx}, {Trenti}, {van
  Dokkum}, \& {Stiavelli}}]{Labbe+13}
{Labb{\'e}}, I., {Oesch}, P.~A., {Bouwens}, R.~J., {et~al.} 2013, \apjl, 777,
  L19

\bibitem[{{Lagattuta} {et~al.}(2017){Lagattuta}, {Richard}, {Cl{\'e}ment},
  {Mahler}, {Patr{\'{\i}}cio}, {Pell{\'o}}, {Soucail}, {Schmidt}, {Wisotzki},
  {Martinez}, \& {Bina}}]{Lagattuta+17}
{Lagattuta}, D.~J., {Richard}, J., {Cl{\'e}ment}, B., {et~al.} 2017, \mnras,
  469, 3946

\bibitem[{{Le F{\`e}vre} {et~al.}(2017){Le F{\`e}vre}, {Lemaux}, {Nakajima},
  {Schaerer}, {Talia}, {Zamorani}, {Cassata}, {Garilli}, {Maccagni},
  {Pentericci}, {Tasca}, {Zucca}, {Amorin}, {Bardelli}, {Cimatti},
  {Giavalisco}, {Guaita}, {Hathi}, {Marchi}, {Vanzella}, {Vergani}, \&
  {Dunlop}}]{LeFevre+17}
{Le F{\`e}vre}, O., {Lemaux}, B.~C., {Nakajima}, K., {et~al.} 2017, arXiv
  e-prints, arXiv:1710.10715

\bibitem[{{Leclercq} {et~al.}(2017){Leclercq}, {Bacon}, {Wisotzki}, {Mitchell},
  {Garel}, {Verhamme}, {Blaizot}, {Hashimoto}, {Herenz}, {Conseil},
  {Cantalupo}, {Inami}, {Contini}, {Richard}, {Maseda}, {Schaye}, {Marino},
  {Akhlaghi}, {Brinchmann}, \& {Carollo}}]{Leclercq+17}
{Leclercq}, F., {Bacon}, R., {Wisotzki}, L., {et~al.} 2017, \aap, 608, A8

\bibitem[{{Limousin} {et~al.}(2012){Limousin}, {Ebeling}, {Richard},
  {Swinbank}, {Smith}, {Jauzac}, {Rodionov}, {Ma}, {Smail}, {Edge}, {Jullo}, \&
  {Kneib}}]{Limousin+12}
{Limousin}, M., {Ebeling}, H., {Richard}, J., {et~al.} 2012, \aap, 544, A71

\bibitem[{{Livermore} {et~al.}(2017){Livermore}, {Finkelstein}, \&
  {Lotz}}]{Livermore+16}
{Livermore}, R.~C., {Finkelstein}, S.~L., \& {Lotz}, J.~M. 2017, \apj, 835, 113

\bibitem[{{Lotz} {et~al.}(2016){Lotz}, {Koekemoer}, {Coe}, {Grogin}, {Capak},
  {Mack}, {Anderson}, {Avila}, {Barker}, {Borncamp}, {Brammer}, {Durbin},
  {Gunning}, {Hilbert}, {Jenkner}, {Khandrika}, {Levay}, {Lucas}, {MacKenty},
  {Ogaz}, {Porterfield}, {Reid}, {Robberto}, {Royle}, {Smith},
  {Storrie-Lombardi}, {Sunnquist}, {Surace}, {Taylor}, {Williams}, {Bullock},
  {Dickinson}, {Finkelstein}, {Natarajan}, {Richard}, {Robertson}, {Tumlinson},
  {Zitrin}, {Flanagan}, {Sembach}, {Soifer}, \& {Mountain}}]{Lotz+16}
{Lotz}, J.~M., {Koekemoer}, A., {Coe}, D., {et~al.} 2016, ArXiv e-prints,
  arXiv:1605.06567

\bibitem[{{Madau} {et~al.}(1999){Madau}, {Haardt}, \& {Rees}}]{Madau+99}
{Madau}, P., {Haardt}, F., \& {Rees}, M.~J. 1999, \apj, 514, 648

\bibitem[{{Mason} {et~al.}(2015){Mason}, {Trenti}, \& {Treu}}]{Mason+15}
{Mason}, C.~A., {Trenti}, M., \& {Treu}, T. 2015, \apj, 813, 21

\bibitem[{{Mason} {et~al.}(2018{\natexlab{a}}){Mason}, {Treu}, {Dijkstra},
  {Mesinger}, {Trenti}, {Pentericci}, {de Barros}, \& {Vanzella}}]{Mason+18a}
{Mason}, C.~A., {Treu}, T., {Dijkstra}, M., {et~al.} 2018{\natexlab{a}}, \apj,
  856, 2

\bibitem[{{Mason} {et~al.}(2018{\natexlab{b}}){Mason}, {Treu}, {de Barros},
  {Dijkstra}, {Fontana}, {Mesinger}, {Pentericci}, {Trenti}, \&
  {Vanzella}}]{Mason+18b}
{Mason}, C.~A., {Treu}, T., {de Barros}, S., {et~al.} 2018{\natexlab{b}},
  \apjl, 857, L11

\bibitem[{{McGreer} {et~al.}(2018){McGreer}, {Fan}, {Jiang}, \&
  {Cai}}]{McGreer+18}
{McGreer}, I.~D., {Fan}, X., {Jiang}, L., \& {Cai}, Z. 2018, \aj, 155, 131

\bibitem[{{McGreer} {et~al.}(2015){McGreer}, {Mesinger}, \&
  {D'Odorico}}]{Mcgreer+15}
{McGreer}, I.~D., {Mesinger}, A., \& {D'Odorico}, V. 2015, \mnras, 447, 499

\bibitem[{{McLean} {et~al.}(2010){McLean}, {Steidel}, {Epps}, {Matthews},
  {Adkins}, {Konidaris}, {Weber}, {Aliado}, {Brims}, {Canfield}, {Cromer},
  {Fucik}, {Kulas}, {Mace}, {Magnone}, {Rodriguez}, {Wang}, \&
  {Weiss}}]{mosfire}
{McLean}, I.~S., {Steidel}, C.~C., {Epps}, H., {et~al.} 2010, 7735,
  doi:10.1117/12.856715

\bibitem[{{Merlin} {et~al.}(2015){Merlin}, {Fontana}, {Ferguson}, {Dunlop},
  {Elbaz}, {Bourne}, {Bruce}, {Buitrago}, {Castellano}, {Schreiber}, {Grazian},
  {McLure}, {Okumura}, {Shu}, {Wang}, {Amor{\'{\i}}n}, {Boutsia}, {Cappelluti},
  {Comastri}, {Derriere}, {Faber}, \& {Santini}}]{Merlin+15}
{Merlin}, E., {Fontana}, A., {Ferguson}, H.~C., {et~al.} 2015, \aap, 582, A15

\bibitem[{{Merlin} {et~al.}(2016){Merlin}, {Amor{\'{\i}}n}, {Castellano},
  {Fontana}, {Buitrago}, {Dunlop}, {Elbaz}, {Boucaud}, {Bourne}, {Boutsia},
  {Brammer}, {Bruce}, {Capak}, {Cappelluti}, {Ciesla}, {Comastri}, {Cullen},
  {Derriere}, {Faber}, {Ferguson}, {Giallongo}, {Grazian}, {Lotz},
  {Micha{\l}owski}, {Paris}, {Pentericci}, {Pilo}, {Santini}, {Schreiber},
  {Shu}, \& {Wang}}]{Merlin+16}
{Merlin}, E., {Amor{\'{\i}}n}, R., {Castellano}, M., {et~al.} 2016, \aap, 590,
  A30

\bibitem[{{Mesinger} {et~al.}(2015){Mesinger}, {Aykutalp}, {Vanzella},
  {Pentericci}, {Ferrara}, \& {Dijkstra}}]{Mesinger+15}
{Mesinger}, A., {Aykutalp}, A., {Vanzella}, E., {et~al.} 2015, \mnras, 446, 566

\bibitem[{{Mesinger} {et~al.}(2016){Mesinger}, {Greig}, \&
  {Sobacchi}}]{Mesinger+16}
{Mesinger}, A., {Greig}, B., \& {Sobacchi}, E. 2016, \mnras, 459, 2342

\bibitem[{{Molino} {et~al.}(2017){Molino}, {Ben{\'{\i}}tez}, {Ascaso}, {Coe},
  {Postman}, {Jouvel}, {Host}, {Lahav}, {Seitz}, {Medezinski}, {Rosati},
  {Schoenell}, {Koekemoer}, {Jimenez-Teja}, {Broadhurst}, {Melchior},
  {Balestra}, {Bartelmann}, {Bouwens}, {Bradley}, {Czakon}, {Donahue}, {Ford},
  {Graur}, {Graves}, {Grillo}, {Infante}, {Jha}, {Kelson}, {Lazkoz}, {Lemze},
  {Maoz}, {Mercurio}, {Meneghetti}, {Merten}, {Moustakas}, {Nonino}, {Orgaz},
  {Riess}, {Rodney}, {Sayers}, {Umetsu}, {Zheng}, \& {Zitrin}}]{Molino+17}
{Molino}, A., {Ben{\'{\i}}tez}, N., {Ascaso}, B., {et~al.} 2017, \mnras, 470,
  95

\bibitem[{{Monna} {et~al.}(2017){Monna}, {Seitz}, {Balestra}, {Rosati},
  {Grillo}, {Halkola}, {Suyu}, {Coe}, {Caminha}, {Frye}, {Koekemoer},
  {Mercurio}, {Nonino}, {Postman}, \& {Zitrin}}]{Monna+17}
{Monna}, A., {Seitz}, S., {Balestra}, I., {et~al.} 2017, \mnras, 466, 4094

\bibitem[{{Ono} {et~al.}(2012){Ono}, {Ouchi}, {Mobasher}, {Dickinson},
  {Penner}, {Shimasaku}, {Weiner}, {Kartaltepe}, {Nakajima}, {Nayyeri},
  {Stern}, {Kashikawa}, \& {Spinrad}}]{Ono+12}
{Ono}, Y., {Ouchi}, M., {Mobasher}, B., {et~al.} 2012, \apj, 744, 83

\bibitem[{{Onoue} {et~al.}(2017){Onoue}, {Kashikawa}, {Willott}, {Hibon}, {Im},
  {Furusawa}, {Harikane}, {Imanishi}, {Ishikawa}, {Kikuta}, {Matsuoka},
  {Nagao}, {Niino}, {Ono}, {Ouchi}, {Tanaka}, {Tang}, {Toshikawa}, \&
  {Uchiyama}}]{Onoue+17}
{Onoue}, M., {Kashikawa}, N., {Willott}, C.~J., {et~al.} 2017, \apjl, 847, L15

\bibitem[{{Ouchi} {et~al.}(2010){Ouchi}, {Shimasaku}, {Furusawa}, {Saito},
  {Yoshida}, {Akiyama}, {Ono}, {Yamada}, {Ota}, {Kashikawa}, {Iye}, {Kodama},
  {Okamura}, {Simpson}, \& {Yoshida}}]{Ouchi+10}
{Ouchi}, M., {Shimasaku}, K., {Furusawa}, H., {et~al.} 2010, \apj, 723, 869

\bibitem[{{Pentericci} {et~al.}(2011){Pentericci}, {Fontana}, {Vanzella},
  {Castellano}, {Grazian}, {Dijkstra}, {Boutsia}, {Cristiani}, {Dickinson},
  {Giallongo}, {Giavalisco}, {Maiolino}, {Moorwood}, {Paris}, \&
  {Santini}}]{Pentericci+11}
{Pentericci}, L., {Fontana}, A., {Vanzella}, E., {et~al.} 2011, \apj, 743, 132

\bibitem[{{Pentericci} {et~al.}(2014){Pentericci}, {Vanzella}, {Fontana},
  {Castellano}, {Treu}, {Mesinger}, {Dijkstra}, {Grazian}, {Brada{\v c}},
  {Conselice}, {Cristiani}, {Dunlop}, {Galametz}, {Giavalisco}, {Giallongo},
  {Koekemoer}, {McLure}, {Maiolino}, {Paris}, \& {Santini}}]{Pentericci+14}
{Pentericci}, L., {Vanzella}, E., {Fontana}, A., {et~al.} 2014, \apj, 793, 113

\bibitem[{{Planck Collaboration} {et~al.}(2016{\natexlab{a}}){Planck
  Collaboration}, {Ade}, {Aghanim}, {Arnaud}, {Ashdown}, {Aumont},
  {Baccigalupi}, {Banday}, {Barreiro}, {Bartlett}, \& et~al.}]{Planck2015}
{Planck Collaboration}, {Ade}, P.~A.~R., {Aghanim}, N., {et~al.}
  2016{\natexlab{a}}, \aap, 594, A13

\bibitem[{{Planck Collaboration} {et~al.}(2016{\natexlab{b}}){Planck
  Collaboration}, {Adam}, {Aghanim}, {Ashdown}, {Aumont}, {Baccigalupi},
  {Ballardini}, {Banday}, {Barreiro}, {Bartolo}, {Basak}, {Battye}, {Benabed},
  {Bernard}, {Bersanelli}, {Bielewicz}, {Bock}, {Bonaldi}, {Bonavera}, {Bond},
  {Borrill}, {Bouchet}, {Boulanger}, {Bucher}, {Burigana}, {Calabrese},
  {Cardoso}, {Carron}, {Chiang}, {Colombo}, {Combet}, {Comis}, {Couchot},
  {Coulais}, {Crill}, {Curto}, {Cuttaia}, {Davis}, {de Bernardis}, {de Rosa},
  {de Zotti}, {Delabrouille}, {Di Valentino}, {Dickinson}, {Diego}, {Dor{\'e}},
  {Douspis}, {Ducout}, {Dupac}, {Elsner}, {En{\ss}lin}, {Eriksen}, {Falgarone},
  {Fantaye}, {Finelli}, {Forastieri}, {Frailis}, {Fraisse}, {Franceschi},
  {Frolov}, {Galeotta}, {Galli}, {Ganga}, {G{\'e}nova-Santos}, {Gerbino},
  {Ghosh}, {Gonz{\'a}lez-Nuevo}, {G{\'o}rski}, {Gruppuso}, {Gudmundsson},
  {Hansen}, {Helou}, {Henrot-Versill{\'e}}, {Herranz}, {Hivon}, {Huang},
  {Ili{\'c}}, {Jaffe}, {Jones}, {Keih{\"a}nen}, {Keskitalo}, {Kisner}, {Knox},
  {Krachmalnicoff}, {Kunz}, {Kurki-Suonio}, {Lagache}, {L{\"a}hteenm{\"a}ki},
  {Lamarre}, {Langer}, {Lasenby}, {Lattanzi}, {Lawrence}, {Le Jeune},
  {Levrier}, {Lewis}, {Liguori}, {Lilje}, {L{\'o}pez-Caniego}, {Ma},
  {Mac{\'{\i}}as-P{\'e}rez}, {Maggio}, {Mangilli}, {Maris}, {Martin},
  {Mart{\'{\i}}nez-Gonz{\'a}lez}, {Matarrese}, {Mauri}, {McEwen}, {Meinhold},
  {Melchiorri}, {Mennella}, {Migliaccio}, {Miville-Desch{\^e}nes}, {Molinari},
  {Moneti}, {Montier}, {Morgante}, {Moss}, {Naselsky}, {Natoli}, {Oxborrow},
  {Pagano}, {Paoletti}, {Partridge}, {Patanchon}, {Patrizii}, {Perdereau},
  {Perotto}, {Pettorino}, {Piacentini}, {Plaszczynski}, {Polastri}, {Polenta},
  {Puget}, {Rachen}, {Racine}, {Reinecke}, {Remazeilles}, {Renzi}, {Rocha},
  {Rossetti}, {Roudier}, {Rubi{\~n}o-Mart{\'{\i}}n}, {Ruiz-Granados},
  {Salvati}, {Sandri}, {Savelainen}, {Scott}, {Sirri}, {Sunyaev}, {Suur-Uski},
  {Tauber}, {Tenti}, {Toffolatti}, {Tomasi}, {Tristram}, {Trombetti},
  {Valiviita}, {Van Tent}, {Vielva}, {Villa}, {Vittorio}, {Wandelt}, {Wehus},
  {White}, {Zacchei}, \& {Zonca}}]{Planck+16}
{Planck Collaboration}, {Adam}, R., {Aghanim}, N., {et~al.} 2016{\natexlab{b}},
  \aap, 596, A108

\bibitem[{{Postman} {et~al.}(2012){Postman}, {Coe}, {Ben{\'{\i}}tez},
  {Bradley}, {Broadhurst}, {Donahue}, {Ford}, {Graur}, {Graves}, {Jouvel},
  {Koekemoer}, {Lemze}, {Medezinski}, {Molino}, {Moustakas}, {Ogaz}, {Riess},
  {Rodney}, {Rosati}, {Umetsu}, {Zheng}, {Zitrin}, {Bartelmann}, {Bouwens},
  {Czakon}, {Golwala}, {Host}, {Infante}, {Jha}, {Jimenez-Teja}, {Kelson},
  {Lahav}, {Lazkoz}, {Maoz}, {McCully}, {Melchior}, {Meneghetti}, {Merten},
  {Moustakas}, {Nonino}, {Patel}, {Reg{\"o}s}, {Sayers}, {Seitz}, \& {Van der
  Wel}}]{Postman+12}
{Postman}, M., {Coe}, D., {Ben{\'{\i}}tez}, N., {et~al.} 2012, \apjs, 199, 25

\bibitem[{{Richard} {et~al.}(2011){Richard}, {Kneib}, {Ebeling}, {Stark},
  {Egami}, \& {Fiedler}}]{Richard+11}
{Richard}, J., {Kneib}, J.-P., {Ebeling}, H., {et~al.} 2011, \mnras, 414, L31

\bibitem[{{Roberts-Borsani} {et~al.}(2016){Roberts-Borsani}, {Bouwens},
  {Oesch}, {Labbe}, {Smit}, {Illingworth}, {van Dokkum}, {Holden}, {Gonzalez},
  {Stefanon}, {Holwerda}, \& {Wilkins}}]{RobertsBorsani+16}
{Roberts-Borsani}, G.~W., {Bouwens}, R.~J., {Oesch}, P.~A., {et~al.} 2016,
  \apj, 823, 143

\bibitem[{{Rosati} {et~al.}(2014){Rosati}, {Balestra}, {Grillo}, {Mercurio},
  {Nonino}, {Biviano}, {Girardi}, {Vanzella}, \& {Clash-VLT Team}}]{Rosati+14}
{Rosati}, P., {Balestra}, I., {Grillo}, C., {et~al.} 2014, The Messenger, 158,
  48

\bibitem[{{Ryan} {et~al.}(2014){Ryan}, {Gonzalez}, {Lemaux}, {Brada{\v c}},
  {Casertano}, {Allen}, {Cain}, {Gladders}, {Hall}, {Hildebradt}, {Hinz},
  {Huang}, {Lubin}, {Schrabback}, {Stiavelli}, {Treu}, {von der Linden}, \&
  {Zaritsky}}]{Ryan+14}
{Ryan}, Jr., R.~E., {Gonzalez}, A.~H., {Lemaux}, B.~C., {et~al.} 2014, \apjl,
  786, L4

\bibitem[{{Schaerer} \& {de Barros}(2009)}]{Schaerer+09}
{Schaerer}, D., \& {de Barros}, S. 2009, \aap, 502, 423

\bibitem[{{Schenker} {et~al.}(2014){Schenker}, {Ellis}, {Konidaris}, \&
  {Stark}}]{Schenker+14}
{Schenker}, M.~A., {Ellis}, R.~S., {Konidaris}, N.~P., \& {Stark}, D.~P. 2014,
  \apj, 795, 20

\bibitem[{{Schenker} {et~al.}(2012){Schenker}, {Stark}, {Ellis}, {Robertson},
  {Dunlop}, {McLure}, {Kneib}, \& {Richard}}]{Schenker+12}
{Schenker}, M.~A., {Stark}, D.~P., {Ellis}, R.~S., {et~al.} 2012, \apj, 744,
  179

\bibitem[{{Schlegel} {et~al.}(1998){Schlegel}, {Finkbeiner}, \&
  {Davis}}]{Schlegel+98}
{Schlegel}, D.~J., {Finkbeiner}, D.~P., \& {Davis}, M. 1998, \apj, 500, 525

\bibitem[{Schmidt {et~al.}(2014)Schmidt, Treu, Brammer, Brada{\v c}, Wang,
  Dijkstra, Dressler, Fontana, Gavazzi, Henry, Hoag, Jones, Kelly, Malkan,
  Mason, Pentericci, Poggianti, Stiavelli, Trenti, von~der Linden, \&
  Vulcani}]{Schmidt+14}
Schmidt, K.~B., Treu, T., Brammer, G.~B., {et~al.} 2014, The Astrophysical
  Journal Letters, 782, L36

\bibitem[{{Schmidt} {et~al.}(2016){Schmidt}, {Treu}, {Brada{\v c}}, {Vulcani},
  {Huang}, {Hoag}, {Maseda}, {Guaita}, {Pentericci}, {Brammer}, {Dijkstra},
  {Dressler}, {Fontana}, {Henry}, {Jones}, {Mason}, {Trenti}, \&
  {Wang}}]{Schmidt+16}
{Schmidt}, K.~B., {Treu}, T., {Brada{\v c}}, M., {et~al.} 2016, \apj, 818, 38

\bibitem[{{Shapley} {et~al.}(2003){Shapley}, {Steidel}, {Pettini}, \&
  {Adelberger}}]{Shapley+03}
{Shapley}, A.~E., {Steidel}, C.~C., {Pettini}, M., \& {Adelberger}, K.~L. 2003,
  \apj, 588, 65

\bibitem[{{Smit} {et~al.}(2014){Smit}, {Bouwens}, {Labb{\'e}}, {Zheng},
  {Bradley}, {Donahue}, {Lemze}, {Moustakas}, {Umetsu}, {Zitrin}, {Coe},
  {Postman}, {Gonzalez}, {Bartelmann}, {Ben{\'{\i}}tez}, {Broadhurst}, {Ford},
  {Grillo}, {Infante}, {Jimenez-Teja}, {Jouvel}, {Kelson}, {Lahav}, {Maoz},
  {Medezinski}, {Melchior}, {Meneghetti}, {Merten}, {Molino}, {Moustakas},
  {Nonino}, {Rosati}, \& {Seitz}}]{Smit+14}
{Smit}, R., {Bouwens}, R.~J., {Labb{\'e}}, I., {et~al.} 2014, \apj, 784, 58

\bibitem[{{Stark} {et~al.}(2010){Stark}, {Ellis}, {Chiu}, {Ouchi}, \&
  {Bunker}}]{Stark+10}
{Stark}, D.~P., {Ellis}, R.~S., {Chiu}, K., {Ouchi}, M., \& {Bunker}, A. 2010,
  \mnras, 408, 1628

\bibitem[{{Stark} {et~al.}(2011){Stark}, {Ellis}, \& {Ouchi}}]{Stark+11}
{Stark}, D.~P., {Ellis}, R.~S., \& {Ouchi}, M. 2011, \apjl, 728, L2

\bibitem[{{Stark} {et~al.}(2014){Stark}, {Richard}, {Siana}, {Charlot},
  {Freeman}, {Gutkin}, {Wofford}, {Robertson}, {Amanullah}, {Watson}, \&
  {Milvang-Jensen}}]{Stark+14}
{Stark}, D.~P., {Richard}, J., {Siana}, B., {et~al.} 2014, \mnras, 445, 3200

\bibitem[{{Stark} {et~al.}(2015){Stark}, {Walth}, {Charlot}, {Cl{\'e}ment},
  {Feltre}, {Gutkin}, {Richard}, {Mainali}, {Robertson}, {Siana}, {Tang}, \&
  {Schenker}}]{Stark+15}
{Stark}, D.~P., {Walth}, G., {Charlot}, S., {et~al.} 2015, \mnras, 454, 1393

\bibitem[{{Stark} {et~al.}(2016){Stark}, {Ellis}, {Charlot}, {Chevallard},
  {Tang}, {Belli}, {Zitrin}, {Mainali}, {Gutkin}, {Vidal-Garcia}, {Bouwens}, \&
  {Oesch}}]{Stark+16}
{Stark}, D.~P., {Ellis}, R.~S., {Charlot}, S., {et~al.} 2016, ArXiv e-prints,
  arXiv:1606.01304

\bibitem[{{Steidel} {et~al.}(2011){Steidel}, {Bogosavljevi{\'c}}, {Shapley},
  {Kollmeier}, {Reddy}, {Erb}, \& {Pettini}}]{Steidel+11}
{Steidel}, C.~C., {Bogosavljevi{\'c}}, M., {Shapley}, A.~E., {et~al.} 2011,
  \apj, 736, 160

\bibitem[{{Strait} {et~al.}(2018){Strait}, {Bradac}, {Hoag}, {Huang}, {Treu},
  {Wang}, {Amorin}, {Castellano}, {Fontana}, {Lemaux}, {Merlin}, {Schmidt},
  {Schrabback}, {Tomczack}, {Trenti}, \& {Vulcani}}]{Strait+18}
{Strait}, V., {Bradac}, M., {Hoag}, A., {et~al.} 2018, ArXiv e-prints,
  arXiv:1805.08789

\bibitem[{{Tilvi} {et~al.}(2014){Tilvi}, {Papovich}, {Finkelstein}, {Long},
  {Song}, {Dickinson}, {Ferguson}, {Koekemoer}, {Giavalisco}, \&
  {Mobasher}}]{Tilvi+14}
{Tilvi}, V., {Papovich}, C., {Finkelstein}, S.~L., {et~al.} 2014, \apj, 794, 5

\bibitem[{{Trenti} {et~al.}(2011){Trenti}, {Bradley}, {Stiavelli}, {Oesch},
  {Treu}, {Bouwens}, {Shull}, {MacKenty}, {Carollo}, \&
  {Illingworth}}]{Trenti+11}
{Trenti}, M., {Bradley}, L.~D., {Stiavelli}, M., {et~al.} 2011, \apjl, 727, L39

\bibitem[{{Treu} {et~al.}(2013){Treu}, {Schmidt}, {Trenti}, {Bradley}, \&
  {Stiavelli}}]{Treu+13}
{Treu}, T., {Schmidt}, K.~B., {Trenti}, M., {Bradley}, L.~D., \& {Stiavelli},
  M. 2013, \apjl, 775, L29

\bibitem[{{Treu} {et~al.}(2012){Treu}, {Trenti}, {Stiavelli}, {Auger}, \&
  {Bradley}}]{Treu+12}
{Treu}, T., {Trenti}, M., {Stiavelli}, M., {Auger}, M.~W., \& {Bradley}, L.~D.
  2012, \apj, 747, 27

\bibitem[{{Treu} {et~al.}(2015){Treu}, {Schmidt}, {Brammer}, {Vulcani}, {Wang},
  {Brada{\v c}}, {Dijkstra}, {Dressler}, {Fontana}, {Gavazzi}, {Henry}, {Hoag},
  {Huang}, {Jones}, {Kelly}, {Malkan}, {Mason}, {Pentericci}, {Poggianti},
  {Stiavelli}, {Trenti}, \& {von der Linden}}]{Treu+15a}
{Treu}, T., {Schmidt}, K.~B., {Brammer}, G.~B., {et~al.} 2015, \apj, 812, 114

\bibitem[{{Treu} {et~al.}(2016){Treu}, {Brammer}, {Diego}, {Grillo}, {Kelly},
  {Oguri}, {Rodney}, {Rosati}, {Sharon}, {Zitrin}, {Balestra}, {Brada{\v c}},
  {Broadhurst}, {Caminha}, {Halkola}, {Hoag}, {Ishigaki}, {Johnson}, {Karman},
  {Kawamata}, {Mercurio}, {Schmidt}, {Strolger}, {Suyu}, {Filippenko}, {Foley},
  {Jha}, \& {Patel}}]{Treu+16}
{Treu}, T., {Brammer}, G., {Diego}, J.~M., {et~al.} 2016, \apj, 817, 60

\bibitem[{{Verhamme} {et~al.}(2018){Verhamme}, {Garel}, {Ventou}, {Contini},
  {Bouch{\'e}}, {Herenz}, {Richard}, {Bacon}, {Schmidt}, {Maseda}, {Marino},
  {Brinchmann}, {Cantalupo}, {Caruana}, {Cl{\'e}ment}, {Diener}, {Drake},
  {Hashimoto}, {Inami}, {Kerutt}, {Kollatschny}, {Leclercq}, {Patr{\'{\i}}cio},
  {Schaye}, {Wisotzki}, \& {Zabl}}]{Verhamme+18}
{Verhamme}, A., {Garel}, T., {Ventou}, E., {et~al.} 2018, \mnras, 478, L60

\bibitem[{{Wang} {et~al.}(2015){Wang}, {Hoag}, {Huang}, {Treu}, {Brada{\v c}},
  {Schmidt}, {Brammer}, {Vulcani}, {Jones}, {Ryan}, {Amor{\'{\i}}n},
  {Castellano}, {Fontana}, {Merlin}, \& {Trenti}}]{Wang+15}
{Wang}, X., {Hoag}, A., {Huang}, K.-H., {et~al.} 2015, \apj, 811, 29

\bibitem[{{Wisotzki} {et~al.}(2016){Wisotzki}, {Bacon}, {Blaizot},
  {Brinchmann}, {Herenz}, {Schaye}, {Bouch{\'e}}, {Cantalupo}, {Contini},
  {Carollo}, {Caruana}, {Courbot}, {Emsellem}, {Kamann}, {Kerutt}, {Leclercq},
  {Lilly}, {Patr{\'{\i}}cio}, {Sandin}, {Steinmetz}, {Straka}, {Urrutia},
  {Verhamme}, {Weilbacher}, \& {Wendt}}]{Wisotzki+16}
{Wisotzki}, L., {Bacon}, R., {Blaizot}, J., {et~al.} 2016, \aap, 587, A98

\bibitem[{{Zheng} {et~al.}(2012){Zheng}, {Postman}, {Zitrin}, {Moustakas},
  {Shu}, {Jouvel}, {H{\o}st}, {Molino}, {Bradley}, {Coe}, {Moustakas},
  {Carrasco}, {Ford}, {Ben{\'{\i}}tez}, {Lauer}, {Seitz}, {Bouwens},
  {Koekemoer}, {Medezinski}, {Bartelmann}, {Broadhurst}, {Donahue}, {Grillo},
  {Infante}, {Jha}, {Kelson}, {Lahav}, {Lemze}, {Melchior}, {Meneghetti},
  {Merten}, {Nonino}, {Ogaz}, {Rosati}, {Umetsu}, \& {van der Wel}}]{Zheng+12}
{Zheng}, W., {Postman}, M., {Zitrin}, A., {et~al.} 2012, \nat, 489, 406

\bibitem[{{Zitrin} {et~al.}(2015){Zitrin}, {Fabris}, {Merten}, {Melchior},
  {Meneghetti}, {Koekemoer}, {Coe}, {Maturi}, {Bartelmann}, {Postman},
  {Umetsu}, {Seidel}, {Sendra}, {Broadhurst}, {Balestra}, {Biviano}, {Grillo},
  {Mercurio}, {Nonino}, {Rosati}, {Bradley}, {Carrasco}, {Donahue}, {Ford},
  {Frye}, \& {Moustakas}}]{Zitrin+15a}
{Zitrin}, A., {Fabris}, A., {Merten}, J., {et~al.} 2015, \apj, 801, 44

\end{thebibliography}
% \bibliographystyle{apj}

\appendix

\section*{List of spectroscopic targets}
\label{sec:appendixA}

\renewcommand\thetable{A\arabic{table}} 
\setcounter{table}{0}

%\clearpage
\LongTables
\tabletypesize{\tiny} \tabcolsep=0.11cm
				\begin{deluxetable*}{| ccccccccccc |}
				\tablecaption{Keck/MOSFIRE spectroscopic targets}
				\tablecolumns{11}
				\tablehead{Cluster & ID & RA & DEC & $H_{160}$ & $\mu$ & $M_{UV} $ & $z_{\mathrm{phot}}$ & 				dates of observation & $t_{\mathrm{exp}}$ & $\sigma_{\mathrm{EW}}$ \\ & & (deg.) & (deg.) & (mag.) & & (mag.) & & (UTC) & (s) & (\AA{}) }

	\startdata
A370 & 0 & 39.982354 & -1.581067 & $26.47 \pm 0.09$ & $8.3^{+0.1}_{-0.1}$ & $-18.4^{+0.1}_{-0.1}$ & $8.2^{+0.4}_{-0.3}$ & 2013Dec16/2013Dec18 & 6660 & 6.2 \\ 
A370 & 1 & 39.963746 & -1.569358 & $26.00 \pm 0.06$ & $18.9^{+1.0}_{-0.7}$ & $-17.9^{+0.1}_{-0.1}$ & $7.8^{+0.2}_{-0.2}$ & 2013Dec16/2013Dec18/2017Oct01 & 21240 & 2.5 \\ 
A370 & 2 & 39.960679 & -1.574164 & $25.57 \pm 0.05$ & $16.2^{+0.1}_{-0.2}$ & $-18.5^{+0.1}_{-0.1}$ & $7.8^{+0.1}_{-0.2}$ & 2013Dec16/2013Dec18 & 6660 & 2.6 \\ 
A370 & 3 & 39.972946 & -1.569983 & $28.04 \pm 0.26$ & $12.5^{+0.3}_{-0.3}$ & $-13.3^{+0.3}_{-0.3}$ & $1.4^{+2.3}_{-1.0}$ & 2017Oct01 & 14580 & 9.4 \\ 
A370 & 4 & 39.975808 & -1.587256 & $26.70 \pm 0.11$ & $7.2^{+0.1}_{-0.1}$ & $-18.3^{+0.1}_{-0.1}$ & $8.2^{+0.2}_{-6.4}$ & 2017Oct01 & 14580 & 3.9 \\ 
A370 & 5 & 39.955425 & -1.572589 & $27.33 \pm 0.15$ & $5.2^{+0.2}_{-0.1}$ & $-18.1^{+0.2}_{-0.1}$ & $8.2^{+0.1}_{-6.8}$ & 2017Oct01 & 14580 & 5.9 \\ 
A370 & 6 & 39.966071 & -1.594767 & $27.56 \pm 0.19$ & $8.6^{+1.3}_{-1.0}$ & $-17.3^{+0.2}_{-0.2}$ & $8.1^{+0.3}_{-0.9}$ & 2017Oct01 & 14580 & 7.2 \\ 
A370 & 7 & 39.948246 & -1.586531 & $27.56 \pm 0.25$ & $2.3^{+0.1}_{-0.1}$ & $-18.5^{+0.2}_{-0.2}$ & $7.2^{+0.6}_{-5.5}$ & 2017Oct01 & 14580 & 7.0 \\ 
A370\tablenotemark{$a$} & 8 & 39.964604 & -1.613703 & $26.10 \pm 0.22$ & \nodata & $-21.6^{+0.2}_{-0.2}$ & $11.3^{-2.8}_{-10.2}$ & 2017Oct01 & 14580 & 2.9 \\ 
A370 & 9 & 39.949317 & -1.605278 & $24.40 \pm 0.05$ & $2.0^{+0.0}_{-0.0}$ & $-21.8^{+0.1}_{-0.1}$ & $7.1^{+0.1}_{-0.1}$ & 2017Oct01 & 14580 & 0.4 \\ 
\hline 
MACS1149 & 10 & 177.394529 & 22.382308 & $28.35 \pm 0.18$ & $1.7^{+0.0}_{-0.0}$ & $-18.3^{+0.2}_{-0.2}$ & $8.5^{0.0}_{-7.3}$ & 2016Feb22 & 10800 & 32.5 \\ 
MACS1149 & 11 & 177.392725 & 22.384714 & $28.64 \pm 0.28$ & $1.7^{+0.0}_{-0.0}$ & $-16.5^{+0.3}_{-0.3}$ & $3.3^{+1.2}_{-2.5}$ & 2016Feb22 & 10800 & 34.9 \\ 
MACS1149 & 12 & 177.412083 & 22.389050 & $27.92 \pm 0.13$ & $5.6^{+0.1}_{-0.1}$ & $-17.1^{+0.1}_{-0.1}$ & $6.8^{+0.2}_{-0.5}$ & 2016Feb22 & 10800 & 18.8 \\ 
MACS1149 & 13 & 177.404417 & 22.412400 & $27.70 \pm 0.12$ & $2.0^{+0.0}_{-0.0}$ & $-18.5^{+0.1}_{-0.1}$ & $7.3^{0.0}_{-6.9}$ & 2016Feb22 & 10800 & 16.2 \\ 
MACS1149 & 14 & 177.417746 & 22.417442 & $25.08 \pm 0.03$ & $1.6^{+0.0}_{-0.0}$ & $-21.5^{+0.0}_{-0.0}$ & $7.8^{+0.1}_{-0.2}$ & 2016Feb22 & 10800 & 1.7 \\ 
\hline 
RXJ1347 & 15 & 206.867204 & -11.756654 & $27.72 \pm 0.27$ & $34.6^{+8.2}_{-5.7}$ & $-14.6^{+0.3}_{-0.3}$ & $4.4^{+1.1}_{-1.5}$ & 2018Jun01 & 11700 & 15.9 \\ 
RXJ1347 & 16 & 206.882308 & -11.742173 & $27.00 \pm 0.19$ & $20.3^{+1.5}_{-1.1}$ & $-16.6^{+0.2}_{-0.2}$ & $6.8^{+0.1}_{-0.2}$ & 2018Jun01 & 11700 & 8.2 \\ 
RXJ1347 & 17 & 206.887116 & -11.745009 & $25.66 \pm 0.08$ & $17.3^{+0.9}_{-0.8}$ & $-18.1^{+0.1}_{-0.1}$ & $6.7^{+0.3}_{-0.2}$ & 2018Jun01 & 11700 & 2.5 \\ 
RXJ1347\tablenotemark{$b$} & 18 & 206.891246 & -11.752606 & $26.43 \pm 0.14$ & $21.4^{+1.7}_{-1.3}$ & $-17.3^{+0.2}_{-0.2}$ & $7.5^{+0.3}_{-0.7}$ & 2018Jun01 & 11700 & \nodata \\ 
RXJ1347 & 19 & 206.893075 & -11.760237 & $27.92 \pm 0.34$ & $15.3^{+1.0}_{-0.9}$ & $-16.0^{+0.4}_{-0.3}$ & $6.9^{+0.6}_{-1.8}$ & 2018Jun01 & 11700 & 19.4 \\ 
\hline 
RCS2327 & 20 & 351.852258 & -2.062694 & $26.61 \pm 0.29$ & $5.2^{+0.9}_{-0.5}$ & $-18.6^{+0.3}_{-0.3}$ & $7.4^{+0.4}_{-7.1}$ & 2013Dec15/2013Dec17/2013Dec18 & 10620 & 4.7 \\ 
RCS2327 & 21 & 351.856092 & -2.093228 & $26.38 \pm 0.27$ & $15.7^{+4.1}_{-3.0}$ & $-14.1^{+0.4}_{-0.4}$ & $1.1^{+5.3}_{-0.1}$ & 2013Dec15/2013Dec17/2013Dec18 & 10620 & 3.9 \\ 
RCS2327 & 22 & 351.880600 & -2.076275 & $24.83 \pm 0.05$ & $4.9^{+0.4}_{-0.3}$ & $-20.5^{+0.1}_{-0.1}$ & $7.4^{+0.1}_{-0.1}$ & 2013Dec15/2013Dec17/2013Dec18 & 10620 & 1.0 \\ 
\hline 
MACS2129 & 23 & 322.345250 & -7.671411 & $25.65 \pm 0.15$ & $1.5^{+0.1}_{-0.1}$ & $-18.6^{+0.2}_{-0.2}$ & $2.0^{+0.3}_{-1.7}$ & 2015May28/2015Nov07/2015Nov08 & 25560 & 1.1 \\ 
MACS2129 & 24 & 322.350850 & -7.675244 & $26.38 \pm 0.13$ & $1.7^{+0.0}_{-0.0}$ & $-19.9^{+0.1}_{-0.1}$ & $6.5^{0.0}_{-6.0}$ & 2015May28/2015Nov07/2015Nov08 & 25560 & 2.2 \\ 
MACS2129 & 25 & 322.348408 & -7.680228 & $27.17 \pm 0.18$ & $2.1^{+0.0}_{-0.0}$ & $-15.9^{+0.2}_{-0.2}$ & $1.3^{+5.2}_{-0.9}$ & 2015May28/2015Nov07/2015Nov08 & 25560 & 4.5 \\ 
MACS2129\tablenotemark{$c$} & 26 & 322.353242 & -7.697442 & $26.79 \pm 0.19$ & \nodata & \nodata & \nodata & \nodata & \nodata & \nodata \\ 
MACS2129 & 27 & 322.364192 & -7.701939 & $27.40 \pm 0.28$ & $2.7^{+0.2}_{-0.2}$ & $-18.6^{+0.3}_{-0.3}$ & $7.7^{-0.6}_{-7.1}$ & 2015May28/2015Nov07/2015Nov08 & 25560 & 5.5 \\ 
\hline 
MACS0416 & 28 & 64.030963 & -24.059686 & $27.68 \pm 0.16$ & $2.7^{+0.1}_{-0.1}$ & $-15.6^{+0.2}_{-0.2}$ & $1.6^{+0.1}_{-1.3}$ & 2015Nov07/2015Nov08 & 16200 & 8.6 \\ 
MACS0416 & 29 & 64.043133 & -24.057911 & $28.09 \pm 0.16$ & $5.6^{+0.1}_{-0.1}$ & $-17.4^{+0.1}_{-0.2}$ & $9.5^{+0.2}_{-7.5}$ & 2015Nov07/2015Nov08 & 16200 & 15.1 \\ 
MACS0416 & 30 & 64.049583 & -24.064589 & $27.56 \pm 0.17$ & $144.8^{+399.3}_{-73.1}$ & $-14.2^{+1.4}_{-0.8}$ & $8.1^{+0.4}_{-0.6}$ & 2015Nov07/2015Nov08 & 16200 & 8.3 \\ 
MACS0416 & 31 & 64.060329 & -24.064961 & $28.16 \pm 0.16$ & $4.6^{+0.1}_{-0.1}$ & $-17.4^{+0.2}_{-0.2}$ & $8.1^{0.0}_{-0.9}$ & 2015Nov07/2015Nov08 & 16200 & 14.0 \\ 
MACS0416 & 32 & 64.027421 & -24.089975 & $25.18 \pm 0.03$ & $14.9^{+8.7}_{-2.4}$ & $-19.0^{+0.5}_{-0.2}$ & $7.8^{0.0}_{-0.3}$ & 2015Nov07/2015Nov08 & 16200 & 1.0 \\ 
MACS0416 & 33 & 64.047992 & -24.081669 & $26.76 \pm 0.06$ & $2.5^{+0.1}_{-0.1}$ & $-19.5^{+0.1}_{-0.1}$ & $8.8^{-0.1}_{-0.4}$ & 2015Nov07/2015Nov08 & 16200 & 4.1 \\ 
MACS0416 & 34 & 64.046212 & -24.091339 & $28.00 \pm 0.24$ & $2.2^{+0.1}_{-0.0}$ & $-18.4^{+0.2}_{-0.3}$ & $8.7^{+0.7}_{-6.7}$ & 2015Nov07/2015Nov08 & 16200 & 13.5 \\ 
MACS0416 & 35 & 64.046063 & -24.094239 & $28.03 \pm 0.15$ & $2.4^{+0.1}_{-0.1}$ & $-17.9^{+0.2}_{-0.1}$ & $6.8^{+0.3}_{-0.5}$ & 2015Nov07/2015Nov08 & 16200 & 11.2 \\ 
\hline 
MACS2214 & 36 & 333.716917 & -14.002444 & $26.26 \pm 0.45$ & $1.5^{+0.0}_{-0.0}$ & $-20.5^{+0.5}_{-0.4}$ & $7.9^{+0.2}_{-0.8}$ & 2017Oct01 & 14400 & 2.9 \\ 
MACS2214 & 37 & 333.739317 & -14.026907 & $26.35 \pm 0.33$ & $1.4^{+0.1}_{-0.0}$ & $-20.1^{+0.4}_{-0.3}$ & $6.3^{+0.1}_{-5.3}$ & 2017Oct01 & 14400 & 2.6 \\ 
\hline 
MACS1423 & 38 & 215.958129 & 24.077017 & $27.20 \pm 0.22$ & $3.0^{+0.0}_{-0.0}$ & $-16.1^{+0.2}_{-0.2}$ & $1.8^{+4.7}_{-0.7}$ & 2015May28/2016Mar20/2015Apr27 & 31500 & 4.1 \\ 
MACS1423 & 39 & 215.942100 & 24.079403 & $26.01 \pm 0.13$ & $1.6^{+0.1}_{-0.0}$ & $-20.3^{+0.1}_{-0.1}$ & $6.8^{+0.2}_{-5.6}$ & 2015May28/2016Mar20/2015Apr27 & 31500 & 1.5 \\ 
MACS1423\tablenotemark{$b$} & 40 & 215.942400 & 24.069656 & $25.03 \pm 0.08$ & $10.0^{+1.9}_{-1.8}$ & $-19.7^{+0.2}_{-0.2}$ & $8.3^{+0.3}_{-0.2}$ & 2015May28/2016Mar20 & 23940 & 0.8 \\ 
MACS1423 & 41 & 215.933929 & 24.079950 & $27.19 \pm 0.28$ & $2.0^{+0.1}_{-0.1}$ & $-15.4^{+0.3}_{-0.3}$ & $1.0^{+4.2}_{-0.1}$ & 2015May28/2016Mar20 & 23940 & 4.7 \\ 
MACS1423 & 42 & 215.928800 & 24.083906 & $25.70 \pm 0.15$ & $1.7^{+0.1}_{-0.1}$ & $-20.8^{+0.2}_{-0.1}$ & $7.7^{+0.6}_{-0.7}$ & 2015May28/2016Mar20/2015Apr27 & 31500 & 1.3 \\ 
MACS1423 & 43 & 215.933379 & 24.070978 & $26.33 \pm 0.19$ & $2.0^{+0.1}_{-0.1}$ & $-17.0^{+0.2}_{-0.2}$ & $1.4^{+0.4}_{-0.4}$ & 2015May28/2016Mar20/2013Jun13 & 28980 & 2.2 \\ 
MACS1423 & 44 & 215.934721 & 24.063594 & $25.24 \pm 0.14$ & $2.5^{+0.1}_{-0.1}$ & $-20.9^{+0.2}_{-0.1}$ & $8.0^{+0.3}_{-7.6}$ & 2015May28/2016Mar20 & 23940 & 0.9 \\ 
MACS1423 & 45 & 215.947900 & 24.082450 & $26.79 \pm 0.22$ & $12.5^{+0.7}_{-0.9}$ & $-17.4^{+0.2}_{-0.2}$ & $6.8^{+0.1}_{-5.4}$ & 2015Apr27 & 7560 & 5.2 \\ 
MACS1423 & 46 & 215.945529 & 24.072431 & $25.93 \pm 0.11$ & $5.1^{+0.4}_{-0.3}$ & $-19.2^{+0.1}_{-0.1}$ & $6.9^{+0.2}_{-0.5}$ & 2013Jun13 & 5040 & 3.1 \\ 
MACS1423\tablenotemark{$c$} & 47 & 215.935858 & 24.078411 & $26.37 \pm 0.14$ & \nodata & \nodata & \nodata & \nodata & \nodata & \nodata \\ 
\hline 
MACS0454 & 48 & 73.551792 & -3.001011 & $26.39 \pm 0.19$ & $2.5^{+0.2}_{-0.2}$ & $-19.5^{+0.2}_{-0.2}$ & $6.7^{+0.1}_{-0.4}$ & 2013Dec15 & 3240 & 7.2 \\ 
MACS0454 & 49 & 73.551321 & -3.004286 & $26.37 \pm 0.32$ & $2.5^{+0.2}_{-0.2}$ & $-19.6^{+0.3}_{-0.4}$ & $7.5^{+0.7}_{-0.6}$ & 2013Dec15 & 3240 & 7.5 \\ 
MACS0454 & 50 & 73.535988 & -2.997678 & $26.42 \pm 0.23$ & $5.7^{+1.1}_{-0.9}$ & $-18.6^{+0.3}_{-0.3}$ & $6.8^{0.0}_{-1.1}$ & 2013Dec15 & 3240 & 7.7 \\ 
\hline 
A2744 & 51 & 3.604512 & -30.380475 & $25.84 \pm 0.05$ & $2.8^{+0.1}_{-0.1}$ & $-20.2^{+0.1}_{-0.1}$ & $8.2^{+0.2}_{-0.2}$ & 2015Nov07/2015Nov08 & 6840 & 3.2 \\ 
A2744 & 52 & 3.588983 & -30.378661 & $27.27 \pm 0.12$ & $3.1^{+0.1}_{-0.1}$ & $-15.8^{+0.1}_{-0.1}$ & $1.6^{+0.3}_{-0.3}$ & 2015Nov07/2015Nov08 & 6840 & 9.9 \\ 
A2744 & 53 & 3.596100 & -30.385833 & $26.95 \pm 0.07$ & $11.7^{+2.2}_{-1.5}$ & $-17.6^{+0.2}_{-0.2}$ & $8.6^{0.0}_{-0.4}$ & 2015Nov07/2015Nov08 & 6840 & 9.2 \\ 
A2744 & 54 & 3.596892 & -30.390453 & $28.90 \pm 0.23$ & $2.8^{+0.2}_{-0.1}$ & $-16.8^{+0.2}_{-0.2}$ & $6.4^{+0.5}_{-5.4}$ & 2015Nov07/2015Nov08 & 6840 & 42.9 \\ 
A2744 & 55 & 3.597833 & -30.395967 & $27.29 \pm 0.11$ & $3.2^{+0.1}_{-0.1}$ & $-18.5^{+0.1}_{-0.1}$ & $7.4^{0.0}_{-0.6}$ & 2015Nov07/2015Nov08 & 6840 & 11.2 \\ 
A2744 & 56 & 3.586250 & -30.392708 & $28.94 \pm 0.39$ & $4.1^{+0.3}_{-0.3}$ & $-16.3^{+0.4}_{-0.4}$ & $6.2^{-0.1}_{-5.5}$ & 2015Nov07/2015Nov08 & 6840 & 52.3 \\ 
A2744 & 57 & 3.604558 & -30.409364 & $28.67 \pm 0.22$ & $7.0^{+0.6}_{-0.3}$ & $-15.9^{+0.2}_{-0.2}$ & $6.0^{+0.4}_{-5.1}$ & 2015Nov07/2015Nov08 & 6840 & 39.5 \\ 
A2744 & 58 & 3.579846 & -30.401594 & $28.21 \pm 0.14$ & $8.0^{+1.1}_{-0.9}$ & $-16.6^{+0.2}_{-0.2}$ & $7.4^{+0.1}_{-6.0}$ & 2015Nov07/2015Nov08 & 6840 & 26.2 \\ 
A2744 & 59 & 3.580454 & -30.405044 & $27.24 \pm 0.08$ & $4.9^{+0.3}_{-0.4}$ & $-18.0^{+0.1}_{-0.1}$ & $7.2^{0.0}_{-0.6}$ & 2015Nov07/2015Nov08 & 6840 & 10.8 \\ 
A2744 & 60 & 3.567771 & -30.401283 & $27.11 \pm 0.21$ & $2.4^{+0.0}_{-0.0}$ & $-18.9^{+0.2}_{-0.2}$ & $6.9^{+0.3}_{-0.5}$ & 2015Nov07/2015Nov08 & 6840 & 8.3 \\ 
A2744 & 61 & 3.572542 & -30.413272 & $28.61 \pm 0.24$ & $2.8^{+0.1}_{-0.1}$ & $-17.5^{+0.2}_{-0.2}$ & $8.8^{-0.3}_{-7.8}$ & 2015Nov07/2015Nov08 & 6840 & 40.5 \\ 
\hline 
MACS0744 & 62 & 116.212767 & 39.468133 & $26.57 \pm 0.33$ & $4.3^{+0.1}_{-0.1}$ & $-18.7^{+0.4}_{-0.4}$ & $6.7^{+0.2}_{-1.3}$ & 2016Mar20 & 5040 & 5.5 \\ 
MACS0744 & 63 & 116.234258 & 39.472594 & $27.34 \pm 0.46$ & $4.9^{+0.5}_{-0.3}$ & $-14.9^{+0.4}_{-0.5}$ & $1.4^{+4.7}_{-1.0}$ & 2016Mar20/2016Feb22/2016Feb23 & 32940 & 5.2 \\ 
MACS0744\tablenotemark{$b$} & 64 & 116.246483 & 39.460414 & $27.17 \pm 0.38$ & $3.2^{+0.1}_{-0.1}$ & $-18.5^{+0.3}_{-0.4}$ & $6.9^{+0.5}_{-0.2}$ & 2016Mar20/2016Feb22/2016Feb23 & 5040 & \nodata \\ 
MACS0744 & 65 & 116.214337 & 39.472056 & $26.71 \pm 0.41$ & $3.8^{+0.0}_{-0.1}$ & $-15.9^{+0.5}_{-0.4}$ & $1.4^{+1.6}_{-0.6}$ & 2016Feb22/2016Feb23 & 27900 & 3.0 \\ 
MACS0744 & 66 & 116.230142 & 39.476083 & $27.26 \pm 0.88$ & $4.0^{+0.3}_{-0.2}$ & $-18.0^{+0.9}_{-1.0}$ & $6.7^{+0.6}_{-5.8}$ & 2016Feb22/2016Feb23 & 27900 & 4.7 \\ 
MACS0744 & 67 & 116.223754 & 39.450433 & $28.05 \pm 1.03$ & $10.3^{+1.3}_{-1.1}$ & $-16.5^{+0.9}_{-1.0}$ & $8.1^{+0.1}_{-6.4}$ & 2016Feb22/2016Feb23 & 27900 & 11.8 \\ 
MACS0744 & 68 & 116.250412 & 39.453011 & $25.58 \pm 0.16$ & $2.4^{+0.1}_{-0.1}$ & $-20.4^{+0.2}_{-0.2}$ & $6.8^{+0.2}_{-0.2}$ & 2016Feb22/2016Feb23/2015Nov08 & 35100 & 0.9 \\ 
MACS0744 & 69 & 116.220858 & 39.473664 & $26.62 \pm 0.28$ & $4.7^{+0.2}_{-0.2}$ & $-18.7^{+0.3}_{-0.2}$ & $7.2^{+0.6}_{-6.0}$ & 2015Nov08 & 7200 & 4.6 
\enddata
\label{tab:targets}
\tablenotetext{$a$}{ID=8 fell outside of the lensing field of view, so we adopted $\mu=1$ for this target when deriving its absolute magnitude.}\tablenotetext{$b$}{\lya{} was detected in the spectrum of this target.}\tablenotetext{$c$}{Spectroscopically confirmed to be at lower redshift ($z<7$) by other surveys. } 
\end{deluxetable*}

%%%%%%%%%%%%%%%%%%%%%%%%%%%%%%%%%%

\end{document}